\documentclass[aps,prl,amsmath,superscriptaddress,twocolumn,longbibliography]{revtex4-1}

\usepackage[T1]{fontenc}
\usepackage{amsmath,amsfonts,amssymb,amsthm,bbm,bm, braket}
\usepackage{wasysym}
\usepackage{comment}
\usepackage{scalerel}
\usepackage{enumerate}
\usepackage{graphicx}
\usepackage{mathtools}
\usepackage{ifthen}
\usepackage{tensor}
\usepackage{lipsum}
\usepackage[T1]{fontenc}
\usepackage{array}
\usepackage{booktabs}
  
\usepackage{longtable,array,booktabs}
\usepackage{supertabular}
\usepackage{longtable}
\usepackage{graphicx}
\usepackage{amsmath}
\usepackage{amsfonts}
\usepackage{amssymb}
\usepackage{braket}
\usepackage{bm}
\usepackage{multirow}
\usepackage{color}
\usepackage[normalem]{ulem}
\usepackage{mathrsfs}
\usepackage{mathtools}
\usepackage{float}
\usepackage{graphicx}
\usepackage{amsfonts}
\usepackage{epsfig}
\usepackage{dcolumn}
\usepackage{amsthm}
\usepackage{color}

\usepackage{latexsym}
\usepackage{amscd}
\usepackage{hyperref}
\usepackage[mathscr]{eucal}
\usepackage{physics}

\usepackage{bm}      
\usepackage{ifpdf}

\makeatletter

\newcommand{\Rmnum}[1]{\expandafter\@slowromancap\romannumeral #1@}
\makeatother

\hypersetup{colorlinks=true, linkcolor=blue, citecolor=blue, filecolor=blue, urlcolor=blue}
\AtBeginDocument{%
    \newwrite\bibnotes
    \def\bibnotesext{Notes.bib}
    \immediate\openout\bibnotes=\jobname\bibnotesext
    \immediate\write\bibnotes{@CONTROL{REVTEX41Control}}
    \immediate\write\bibnotes{@CONTROL{%
    apsrev41Control,author="08",editor="1",pages="1",title="0",year="1"}}
     \if@filesw
     \immediate\write\@auxout{\string\citation{apsrev41Control}}%
    \fi
}%

\begin{document}
\title{Exactly solvable dynamics and signatures of integrability in an infinite-range many-body Floquet spin system}

\author{Harshit Sharma}
\email{ds21phy007@students.vnit.ac.in}
\affiliation{Department of Physics, National Institute of Technology, Nagpur 440010, India}
 
\author{Udaysinh T. Bhosale}
\email{udaysinhbhosale@phy.vnit.ac.in}
\affiliation{Department of Physics, National Institute of Technology, Nagpur 440010, India}

\date{\today}

\begin{abstract}
We study $N$ qubits having infinite-range Ising interaction and subjected to periodic pulse of external magnetic field. 
We solve the cases of $N=5$ to $11$ qubits analytically, finding its eigensystem, the dynamics of the entanglement 
for various initial states, and the unitary evolution operator. These quantities shows signatures of quantum integrability.
For the general case of $N>11$ qubits, we provide a conjecture on quantum integrability based on the numerical evidences like  
degenerate spectrum, and the exact periodic nature of the time-evolved unitary evolution operator and the entanglement dynamics.
Using linear entropy we show that for class of initial unentangled state the entanglement
displays periodically maximum and zero values.

\end{abstract}

\maketitle
{\it Introduction.---}
Classical and quantum systems with long-range interactions have played an important role in our understanding 
\cite{milotti2001exactly,milotti2002exactly,campa2009statistical,henkel2010three,ciani2010long,campa2014physics,zeiher2016many,monroe2021programmable,das2022multiple,defenu2023long,roberts2023exact,agrawal2023ordering,mohdeb2023entanglement,diessel2023generalized}. 
They have found to be useful for quantum technology applications 
like quantum heat engine \cite{solfanelli2023quantum}, quantum computing \cite{lewis2021optimal,lewis2023ion}, ion traps 
\cite{gambetta2020long}, etc. Generalized Higgs mechanism is used recently to understand such systems \cite{diessel2023generalized}.
There are several experimental systems where these interactions are present, for example cold atoms 
in cavities \cite{baumann2010dicke}, polar molecules \cite{yan2013observation}, dipolar quantum gases 
\cite{lahaye2009physics,chomaz2022dipolar}, Rydberg atoms \cite{saffman2010quantum}, etc.
% These systems have beed studied from the perspective of quantum information.
They have found to be efficient for quantum computing and quantum simulation tasks as they can realize
highly entangled states \cite{avellino2006quantum,richerme2014non,jurcevic2014quasiparticle,eldredge2017fast}.
These systems have given rise to new phases of matter \cite{kastner2011diverging} and 
measurement-induced phase-transition \cite{muller2022measurement,block2022measurement,minato2022fate}.
Studies have shown that one can view them as short-range interacting systems in higher dimensions 
\cite{gong2016kaleidoscope,maghrebi2017continuous}. Recent work has studied quantum many-body scars in them \cite{lerose2023theory}.
Various studies have also addressed entanglement in such systems \cite{dur2005entanglement,kuzmak2018entanglement,amghar2023geometrical}.
Propagation of multipartite entanglement and quantum scrambling in them is also addressed \cite{pappalardi2018scrambling}, which can be 
measured in experiments \cite{garttner2017measuring}.
Bound on the scrambling in such all-to-all interaction models is well understood \cite{yin2020bound,li2020fast,belyansky2020minimal}.
It is known that the semiclassical limit for these systems can be obtained 
\cite{swingle2016measuring,Hakke87}. Thus, one can study effects of the underlying classical correlations on the quantum 
correlations 
\cite{neill2016ergodic,Arjendu2017,pappalardi2018scrambling,madhok2018quantum,BandyoArulPRE,kumari2019untangling,
kumari2018quantum,munoz2021nonlinear,Udaysinhbifurcation2017}.

The interaction in the long-range system decay as power-law with distance ($1/r^{\alpha}$), where $\alpha$ characterizes the given 
system. For van der Waals interactions in Rydberg atoms $\alpha=6$, for polar molecules and magnetic atoms (dipole-dipole interaction) 
$\alpha=3$, monopole-dipole $\alpha=2$, monopole-monopole or Coulomb-like $\alpha=1$, whereas for atoms coupled to cavities $\alpha=0$
\cite{britton2012engineered,schauss2012observation,peter2012anomalous,yan2013observation,gil2016nonequilibrium,jurcevic2014quasiparticle,hazzard2014many,richerme2014non,douglas2015quantum}.
The case $\alpha=0$ corresponds to the class of infinite-range or all-to-all interaction physical systems.
This kind of interaction is very useful for the construction of robust and high-fidelity geometric quantum logic gates \cite{leibfried2003experimental}.

In this work, we consider a system of qubits in magnetic field with periodic application (or kicking) of uniform infinite-range 
Ising interaction ($\alpha=0$). Thus, the energy is not conserved due to periodic kicking. In such models the Hamiltonian can also be reduced to 
total spin operators 
\cite{milburn1999simulating,Wang04,VaibhavMadhok2015,neill2016ergodic,Arjendu2017,pappalardi2018scrambling,madhok2018quantum,
ExactDogra2019,lerose2020bridging,sreeram2023witnessing}. In fact there is a one-to-one correspondence between all-to-all Ising 
interacting qubits and total spin operators. It is used depending on the problem under consideration. Specifically, the qubit 
representation is useful for studying quantum correlations whereas the total spin operators helps in understanding the semiclassical 
limit. 
% When energy is conserved in a Hamiltonian 
Such Hamiltonians with and without energy conservation are studied from the direction of integrability-chaos transition in 
Refs.\cite{bentsen2019integrable} and \cite{bulchandani2022onset} respectively.

In this Letter, we study a model consisting of $N$ qubits kept in an external magnetic field and are subjected to periodic global 
pulses of uniform infinite-range Ising interaction. We show that the system shows quantum integrability 
\cite{owusu2008link,doikou2010introduction,SantosLea12,yuzbashyan2013quantum}. Its signatures can be seen in spectral statistics being 
Poissonian or degenerate spectrum or level crossings, exact periodicity of the time-evolution operator, etc. 
\cite{LevelCrossingsYuzbashyan2008,yuzbashyan2013quantum,BogomolnyDistribution2013,naik2019controlled,patoary2023chaotic,retore2022introduction,doikou2010introduction,faddeev1995algebraic}.
It should be noted that the Refs. \cite{owusu2008link,doikou2010introduction,yuzbashyan2013quantum} considers a parameter dependent 
family of quantum integrable Hamiltonians, whereas we get integrability only at a special case. And we expect that these general 
signatures to remain intact for our case too. Our model has connection to nearest neighbor Ising interaction model and 
a special case of the quantum chaotic kicked top (QKT) (to be discussed in later part of the paper). The QKT has been implemented in 
various 
experimental beds for values of $N$ up to six \cite{Chaudhury09,neill2016ergodic,KrithikaKickedTop2019}. For larger values of $N$ 
of the order of $100$, use of ion traps \cite{monroe2021programmable,defenu2023long} has been proposed in the Ref.\cite{sieberer2019digital}. 
Once our model is mapped to the QKT, it is shown to display integrability {\it only} upto four qubits \cite{ExactDogra2019}.
Whereas for large $N$ it is not understood whether the integrability persists or not.

In our present work, we first analytically solve the model for the cases of five 
to eleven qubits. We analytically obtain the eigenvalues, eigenvectors and, entanglement and operator dynamics exactly. 
We show the time-periodic nature of both, the entanglement and the operator itself.

This nature is in accordance with results from the 
Refs.\cite{ArulLakshminarayan2005,naik2019controlled,mishra2015protocol,pal2018entangling}, where quantum integrable kicked Ising 
model for a range of parameters are considered \cite{briegel2001persistent,prosen2004ruelle}. This is an important observation as our 
model falls in the same category of integrability class as that of these models (refer Eq.(\ref{uni})) \cite{caux2011remarks}. 
Thus, we can use these signatures to quantify integrability in our model too.

In Refs.\cite{ArulLakshminarayan2005,naik2019controlled,mishra2015protocol,pal2018entangling}, using numerical and 
analytical studies it is found that the integrable periodically-kicked spin chains with Ising interaction show time-periodic nature 
of the entanglement for various initial states. In the Ref.\cite{pal2018entangling}, it is shown that the integrable systems display 
the time-periodic nature of the entangling power of the Floquet operator. Whereas in Ref.\cite{naik2019controlled}, the time-periodic nature 
of the Floquet operator itself is observed, which implies its time-periodic entangling power. In our work, we show a result for the 
general value of $N$, analytically for five to eleven qubits and numerically for $N>11$, that the Floquet operator itself is 
time-periodic implying the time-periodic nature of the entangling power. Another signature of integrability can be observed from the 
degeneracy in the spectra of the Hamiltonian \cite{LevelCrossingsYuzbashyan2008,yuzbashyan2013quantum}. It means the system 
eigenvalues are lacking the repulsion among themselves. Similarly, in the Ref.\cite{naik2019controlled} a highly degenerate spectra 
is observed in the periodically driven Ising system which is also integrable. Thus, we can conclude that a system with 
Ising-like interaction shows quantum integrability if its spectra is degenerate; the entanglement and the evolution operator 
dynamics are time-periodic.

For the general case of $N>11$ qubits, we provide numerical evidence of the integrability  
using the degeneracy in the spectra, the periodic nature of entanglement dynamics as well as the corresponding Floquet operator itself.

{\it Model.---} The model Hamiltonian is given as follows:
\begin{equation}
\label{Eq:QKT}
H(t)= \sum_{ l< l'=1}^{N} \sigma^z_{l} \sigma^z_{l'}  + 
\,  \sum_{n = -\infty}^{ \infty} \delta(n-t/\tau) \sum_{l=1}^{N}\sigma^y_l, 
\end{equation}
where $\tau$ is the period with which magnetic field along $y$-axis is applied periodically (second term). The strength of the field is set equal 
to that of the Ising interaction (first term). In our model the Ising interaction is uniform and all-to-all. The model is also
permutation symmetric under the exchange of spins and thus the Hilbert space dimension is $N+1$. 
Its special case, the one with only  
nearest-neighbor (NN) interaction has been studied extensively 
\cite{ArulLakshminarayan2005,mishra2015protocol,apollaro2016entanglement,bertini2019entanglement,naik2019controlled,shukla2022characteristic}.
In the Ref.\cite{mishra2015protocol}, $\tau=\pi/4$ is shown to generate nonlocal Bell pairs (maximum entanglement between two qubits) 
and multiqubit entanglement. Here, we also restrict ourselves to the same $\tau$. In fact, we find that the model is quantum 
integrable {\it only} for $\tau=\pi/4$ whereas for other values it isn't (to be discussed in subsequent part of the paper). 
Scrambling in models similar to ours has been studied in the 
Refs.\cite{belyansky2020minimal,li2020fast,yin2020bound,li2021long,li2022fast,wanisch2023information}, particularly the one in the 
Ref.\cite{li2020fast} which is studied from the perspective of holographic duality is very similar to ours. All these models are 
energy conserving ones whereas our is not.

The Hamiltonian in the Eq.(\ref{Eq:QKT}) also has connection with the model of quantum chaotic kicked top (QKT)
\cite{Hakke87,Haakebook,UdaysinhPeriodicity2018}. Its Hamiltonian is given by 
\begin{equation}
\label{Eq:QKT1}
H_{QKT}(t)= \frac{p}{\tau'} \, {J_y} + \frac{k}{2j}{J_z}^2 \sum_{n = -\infty}^{ \infty} \delta(t-n\tau').
\end{equation}
Its a time-dependent Hamiltonian where the first term represents a rotation and the second one is torsion applied at periodic 
$\delta$ kicks. Here, $J_{x,y,z}$ are components of the angular momentum operator $\mathbf{J}$. For given spin of the top $j$, 
the top can be decomposed in $N=2j$ spin-half particles \cite{milburn1999simulating,Wang04}. Here, time between periodic kicks 
is $\tau'$, $p$ measures rotation about $y$ axis and $k$ controls the degree of chaos in the classical limit. It can be seen that 
$H_{QKT}(t)$ has total spin operators. Various entanglement content has been studied in this model in recent times
\cite{Arjendu2017,UdaysinhPeriodicity2018,Udaysinhbifurcation2017,ShohiniGhose2008,ExactDogra2019,Lombardi2011,Krithika2019,
neill2016ergodic,Haakebook,Miller99,VaibhavMadhok2015,madhok2015comment,piga2019quantum,kumari2019untangling,
BandyoArulPRE,madhok2018quantum,wang2020signatures}. Particularly, its important to note that for a given initial quantum state 
the quantum correlations are periodic in $k$ with period $2 j \pi$ \cite{UdaysinhPeriodicity2018}.

For the special case of parameters $p=\pi/2$, $k=j\pi(=N\pi/2)$, $\tau'=1$ and using many-qubit transformation 
$J_{x,y,z}=\sum_{l=1}^{2j}\sigma_l^{x,y,z}/2$, where $\sigma_l^{x,y,z}$ are the standard Pauli matrices, it can be shown 
that $H_{QKT}(t)$ has close resemblance or mapping with $H(t)$ from the Eq.(\ref{Eq:QKT}). 
Due to this mapping our model can also be shown to have connection with the integrable Lipkin-Meshkov-Glick (LMG) model 
\cite{kumari2022eigenstate}.
The Floquet operator corresponding to the Eq.~(\ref{Eq:QKT}) for $\tau=\pi/4$ is as follows:
\begin{equation}
\label{uni}
{\mathcal U}=\exp\left(-i \frac{\pi}{4}  \sum_{ l< l'=1}^{N} \sigma^z_{l} \sigma^z_{l'}\right)
\exp\left( -i \frac{\pi}{4} \sum_{l=1}^{N}\sigma^y_l \right).
\end{equation}
It gives the evolution of states just before a kick to just before the next one.
In the Ref.\cite{ExactDogra2019} small number of qubits ($N=3$ and $4$) are studied extensively for these parameters.

In this work, our initial states are the standard SU(2) coherent states on the unit sphere with spherical coordinates 
$(\theta_0,\phi_0)$ and given by \cite{Glauber1976,PuriBook}:
\begin{equation}
 |\theta_0,\phi_0\rangle = \otimes^{N} \left(\cos(\theta_0/2) |0\rangle + e^{-i \phi_0} \sin(\theta_0/2) |1\rangle\right).
\end{equation}
These are then evolved using the Floquet operator $\mathcal{U}$ and their entanglement dynamics as a function of time is studied. 
For this study 
we use linear entropy \cite{buscemi2007linear} and concurrence \cite{Wootters01,Wooters} as measures of entanglement.
For the analytical solution, we use the following standard basis permutation symmetric space by generalizing the one given for three 
and four qubits in Ref.\cite{ExactDogra2019}. It is based on the  commutation relation:
$[\mathcal{U},\otimes_{l=1}^{N}\sigma_l^y ]=0$, proved in \cite{supplementry2023}. Thus, the general basis for the odd number of qubits is: 
\begin{equation}\label{Eq:oddBasis1}
\ket{\phi_q^{\pm}}=\frac{1}{\sqrt{2}}\left(\ket{w_q}\pm {i^{\left(N-2q\right)}} \ket{\overline{w_q}}\right),
0\leq q\leq \dfrac{N-1}{2}.
\end{equation}
 For the even case:
\begin{eqnarray}\label{Eq:evenBasis1}
\begin{split}
\ket{\phi_r^{\pm}}&=\frac{1}{\sqrt{2}}\left(\ket{w_r}\pm {(-1)^{\left(N/2-r\right)}} \ket{\overline{w_r}}\right), 0\leq r\leq N/2-1\\
\mbox{and} &\;\;
\ket{\phi_{N/2}^+}=\left({1}/{\sqrt{\binom{N}{N/2}}}\right)\sum_{\mathcal{P}}\left(\otimes^{N/2}\ket{0}\otimes^{N/2}\ket{1}\right)\mathcal{P}
\end{split}
\end{eqnarray}
  where 
$\ket{w_q}=\left({1}/{\sqrt{\binom{N}{q}}}\right)\sum_\mathcal{P}\left(\otimes^q \ket{1} \otimes^{(N-q)}\ket{0}\right)_\mathcal{P}$ and 
$\ket{\overline{w_q}}=\left({1}/{\sqrt{\binom{N}{q}}}\right)\sum_\mathcal{P}\left(\otimes^{q}\ket{0}\otimes^{(N-q)}\ket{1}\right)_\mathcal{P}$,
both being definite particle states \cite{vikram11}. The $\sum_\mathcal{P}$ denotes the sum over all possible permutations.
These basis states are parity symmetric and follows $\otimes_{l=1}^{N}\sigma_l^y\ket{\phi_j^{\pm}}=\pm\ket{\phi_j^{\pm}}$.
In this basis, $\mathcal U$ becomes block-diagonal simplifying further analysis. We derive entanglement dynamics  for the coherent 
states $|\theta_0=0,\phi_0=0\rangle$ and $|\theta_0=\pi/2,\phi_0=-\pi/2\rangle$. In terms of qubit representation, these coherent 
states are given by $\otimes^{N}\ket{0}$ and $\otimes^{N}\ket{+}$ respectively \cite{ExactDogra2019}. These states have importance 
in the QKT. The first state is on the period-4 orbit whereas the second one is on the fixed point in the classical phase space of 
the QKT \cite{ExactDogra2019}.

\begin{figure}[!t]\vspace{0.4cm}
\includegraphics[width=0.450\textwidth]{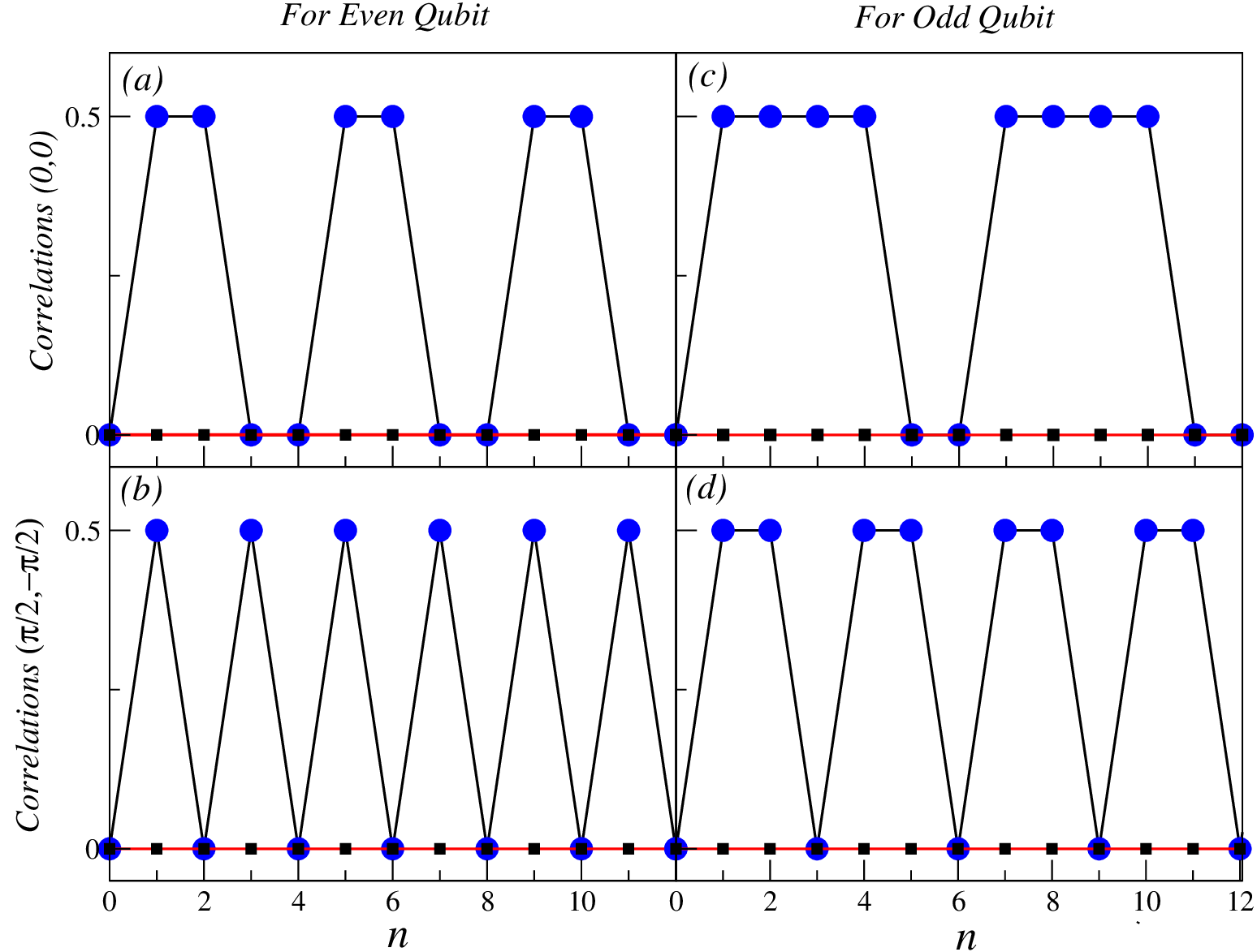}
\caption{Correlations using linear entropy (circles) and concurrence (squares) are plotted for even ((a) and (b)) and 
odd ((c) and (d)) number of qubits for various initial states.}
\label{fig:correlations1} 
\end{figure}

{\it Exact solution for five qubits.---}
Using basis in the Eq.~(\ref{Eq:oddBasis1}) for $N=5$, the unitary operator $\mathcal{U}$ is given by
\begin{equation} 
 \mathcal{U} = \begin{pmatrix}
            \mathcal{U}_{+} & 0 \\ 0 & \mathcal{U}_{-}
            \end{pmatrix}, 
\end{equation}
where $\mathcal{U}_{\pm}$ are $3\times 3$ dimensional matrices and $0$ is a null matrix of same dimension as  
$\mathcal{U}_{\pm}$. The $\mathcal{U}_{+}$ ($\mathcal{U}_{-}$) are written in the positive- (negative-) parity subspaces 
$\left\lbrace \phi_0^{+},\phi_1^{+},\phi_2^{+}\right\rbrace$ $\left(\left\lbrace \phi_0^{-},\phi_1^{-},\phi_2^{-}\right\rbrace\right)$
respectively and are obtained as follows:
\begin{equation}
  \mathcal{U}_{\pm}= \frac{1}{4} e^{\pm{\frac{ i \pi }{4}}}
\begin{pmatrix}
 \mp{1} &  i\sqrt{5}  &  \mp{\sqrt{10}} \\
- i\sqrt{5} & \pm 3 & - i\sqrt{2} \\
  \pm \sqrt{10} & - i\sqrt{2} & \mp 2 \\
\end{pmatrix}.
\end{equation}
The eigenvalues  of $\mathcal{U_+}$ and $\mathcal{U_-} $ are 
\({e^{\frac{i \pi }{4}}\left\{1,e^{\frac{2 i \pi }{3}},e^{-\frac{2 i \pi }{3}}\right\}}\) and 
\({e^{\frac{3i \pi }{4}}\left\{1,e^{-\frac{2 i \pi }{3}},e^{\frac{2 i \pi }{3}}\right\}}\) respectively.
Whereas the eigenvectors of $\mathcal{U_{\pm}}$ are 
$\left[{\pm i}/{\sqrt{5}},1,0\right]^T$, $\left[\pm i\sqrt{{5}/{6}},-{1}/{\sqrt{6}},1\right]^T$ 
and $\left[\mp i\sqrt{{5}/{6}},{1}/{\sqrt{6} },1\right]^T$.
For evolving an initial state we need to get $\mathcal{U}^n$ and therefore $\mathcal{U}_{\pm}^n$, which is given as follows:
\begin{widetext}
\begin{equation}
\mathcal{U}_{\pm}^n =(\pm 1)^n e^{\pm{\frac{ in \pi }{4}}}
{\left[
\begin{array}{ccc}
  \left(1+5 \cos \left({2 n \pi }/{3}\right)\right)/6 &\pm i \sqrt{5} \left(\sin^2\left(n \pi/3\right)\right)/3 & 
  -\sqrt{{5}/{6}}
\sin \left({2 n \pi }/{3}\right) \\
 \mp i \sqrt{5} \left(\sin^2 \left({n \pi }/{3}\right)\right)/3 & \left(5+\ \cos \left({2 n \pi }/{3}\right)\right)/6 &\mp {i
\sin \left({2 n \pi }/{3}\right)}/{\sqrt{6}} \\
 \sqrt{{5}/{6}} \sin \left({2 n \pi }/{3}\right) & \mp {i \sin \left(2 n \pi/3\right)}/{\sqrt{6}} & 
 \cos \left(2n\pi/3\right) \\
\end{array}
\right]}.
\end{equation}
\end{widetext}
It can be shown that the $\mathcal{U}$ is periodic with period $24$, which is a signature of integrability \cite{supplementry2023}.
It is now straightforward to do the time evolution of any initial state. Let us first start with $|00000\rangle$. Its $n$th time 
evolution is given by:
\begin{eqnarray}
\ket{\psi_n}&=&\mathcal{U}^n |00000\rangle=\mathcal{U}^n |w_0\rangle =\mathcal{U}^n \left(|\phi_{0}^{+}\rangle+|\phi_{0}^{-}\rangle\right)/\sqrt{2}\nonumber \\ \nonumber
&=&\left(\mathcal{U}_{+}^n |\phi_{0}^{+} \rangle + \mathcal{U}_{-}^n |\phi_{0}^{-} \rangle \right)/\sqrt{2}\\\nonumber
&=&(1/2) e^{{\frac{ in \pi }{4}}}\left\lbrace(1+ i^n)\left(\alpha_n \ket{w_0}-i\beta_n \ket{\overline{w_1}}+ 
\gamma_n \ket{w_2}\right)\right.  \\\nonumber 
&& \left. +(1-i^n)\left(i\alpha_n \ket{\overline{w_0}}+ \beta_n\ket{w_1}+ i \gamma_n \ket{\overline{w_2}}\right)\right\rbrace,\nonumber
 \end{eqnarray}
where $\alpha_n=\left[1+5 \cos \left(2 n \pi/3\right)\right]/6, \,\beta_n=i \sqrt{5} \sin \left(n \pi/3\right)^2/3$ and 
$\gamma_n=\sqrt{5/6} \sin \left(2 n \pi/3\right)$.
Using $\ket{\psi_n}$ one can obtain the reduced density matrix (RDM) of a single qubit 
$\left(\rho_1(n)=\mbox{Tr}_{\neq1}\left(\ket{\psi_n}\bra{\psi_n}\right)\right)$ and the two qubits 
$\left(\rho_{12}(n)=\mbox{Tr}_{\neq1,2}\left(\ket{\psi_n}\bra{\psi_n}\right)\right)$.
For even time $n=2m$, $\rho_1(2m)$ is diagonal and is given as follows:
\begin{equation}
\rho_1(2m) =
{\left[
\begin{array}{cc}
 \lambda_{2m} & 0 \\
 0 & 1-\lambda_{2m} \\
\end{array}
\right]},
\end{equation}
where $\lambda_{2m}$ and $1-\lambda_{2m}$ are its eigenvalues with
$\lambda_{2m}=\left[6+2\cos\left({4m\pi}/{3}\right)+\cos\left({8m\pi}/{3}\right)\right]/9$.
Whereas for odd time $n=2m-1$ we get:
\begin{equation}
\rho_1(2m-1) =
{\left[
\begin{array}{cc}
 1/2 & h_{2m-1} \\
 h_{2m-1} & 1/2 \\
\end{array}
\right]},
\end{equation}
where $h_{2m-1}=(2/9)\sin^2\left[{(2m-1) \pi }/{3}\right]\{2+\cos\left[{2(2m-1)\pi}/{3}\right]-$
$\sqrt{3}\sin\left[{2(2m-1)\pi}/{3}\right]\}$. Its eigenvalues are $\lambda_{2m-1}$ and $1-\lambda_{2m-1}$
with 
$\lambda_{2m-1} = (1/18) \big\{9- \big[27-17 \cos\left({2 (2m-1) \pi }/{3}\right)-10 
\cos\left({4 (2m-1) \pi }/{3}\right)-17 \sqrt{3} \sin\left({2 (2m-1) \pi }/{3}\right)+
10 \sqrt{3} \sin\left({4(2m-1) \pi }/{3}\right)\big]^{1/2}\big\}$.
These eigenvalues give the linear entropy using $2\lambda_n(1-\lambda_n)$ and is plotted in the Fig.(\ref{fig:correlations1}).
It can be shown from the expressions and the figure that it has a periodic nature with period six.
This periodic nature is also observed in previous works on integrable systems involving periodically-kicked spin chains
Refs.\cite{ArulLakshminarayan2005,mishra2015protocol,naik2019controlled}.
In the context of QKT time-periodicity has been reported earlier for $N=2$ \cite{UdaysinhPeriodicity2018};
$N=3$ and $N=4$ for special values of $k$ in the Eq.~(\ref{Eq:QKT1}) \cite{ExactDogra2019}.
We analytically prove that the entanglement content is the same for consecutive odd and even values of $n$ \cite{supplementry2023}.

Similarly, we evolve the $|+++++\rangle$ state and obtain the linear entropy at the $n$th time.
The eigenvalues of RDM $\rho_1(n)$ are $\lambda_n$ and $1-\lambda_n$ where $\lambda_n=
\left[3-2 \cos\left({2 n \pi}/{3}\right)-\cos\left({4 n \pi}/{3}\right)\right]/9$.
The linear entropy using $2\lambda_n(1-\lambda_n)$ is plotted in the Fig.(\ref{fig:correlations1}).
Its again periodic in time with period three.
Using $\rho_{12}$, it is shown that the pair-wise concurrence is zero all $n$ for both the initial states and is shown in the 
same figure \cite{supplementry2023} (see also references \cite{wootters1998entanglement,TingYu2007} therein). This shows that 
the entanglement is of multipartite nature.\\ 
\begin{table*}[t]
  \centering
 \caption{The linear entropy ($S(n)$) for different initial states and number of qubits $N$.}
 \renewcommand{\arraystretch}{2.4}
 \begin{tabular}{|p{0.35cm}|p{11.7cm}|p{5.2cm}|  }
 \hline 
$N$ &\hspace{3cm} The $S(n)$ for the initial state $\otimes^N\ket{0}$ & \hspace{.2cm}  The $S(n)$ for initial state $\otimes^N\ket{+}$ \\
\hline
6 & 
$S(n) 
=\dfrac{1}{512}\left\lbrace\left[ 1 - \cos\left(\frac{n\pi}{2}\right) + \sin\left(\frac{n\pi}{2}\right)\right]^2 \left[284+ 229\left( \cos\left(\frac{n\pi}{2}\right)-\sin\left(\frac{n\pi}{2}\right)\right)\right.\right.$\newline$  
    \left.\left. \hspace{2cm} -96\sin(n\pi)- 9  \left(\cos\left(\frac{3n\pi}{2}\right)  + 
      \sin\left(\frac{3n\pi}{2}\right)\right)\right]\right\rbrace$
 &\hspace{.2cm} $S(n)= \dfrac{1}{2}\left\lbrace 1-\dfrac{1}{4} \left[1+\cos\left( n \pi \right)\right]^2\right\rbrace $\\
\hline

7 &$S(2m-1)= \dfrac{1}{2}\left\lbrace1-\dfrac{1}{81} \sin^2\left(\frac{\pi  (2m-1)}{3}\right) \left[-4 \sqrt{3} \cos\left[\frac{\pi(2m-1)}{3}\right]+\sqrt{3}\cos(\pi
 (2m-1))\right.\right.$\newline$~~~~~~~~~~~~~~~~~~~~~~\left.\left. +6 \sin\left(\frac{\pi(2m-1)}{3}\right)+\sin\left((2m-1)\pi\right)\right]^2\right\rbrace$
 \newline$S(2m)$=$\dfrac{1}{81}\left\lbrace \left[7+2 \cos\left(\frac{4m\pi }{3}\right)\right] \left[12+5 \cos\left(\frac{4 m \pi }{3}\right)+\cos\left(\frac{2m \pi }{3}\right)\right]\sin^2\left(\frac{2m \pi }{3}\right)\right\rbrace$

  &
 $S(n)=\dfrac{1}{2}\left\lbrace 1-\dfrac{1}{81} \left[3+5 \cos\left(\frac{2 n \pi }{3}\right)\right.\right.$\newline$\left.\left.\hspace{1.65cm}+\cos\left(\frac{4 n \pi }{3}\right)\right]^2 \right\rbrace$
    \\
\hline
8&$S(n)$=$\dfrac{1}{2048}\left\lbrace\left[ 1 - \cos\left(\frac{n\pi}{2}\right) + \sin\left(\frac{n\pi}{2}\right)\right]^2 \left[1212-448\sin(n\pi)+1005\left(\cos\left(\frac{n\pi}{2}\right)\right.\right.\right.$ \newline $\left. \left.\left. \hspace{2cm}    - \sin\left(\frac{n\pi}{2}\right)\right) - 49\left(  \sin\left(\frac{3n\pi}{2}\right)+\cos\left(\frac{3n\pi}{2}\right)\right)\right]\right\rbrace$. &

\hspace{.6cm} $S(n)$=$\dfrac{1}{2}\left[1-\cos^2\left(\frac{n \pi }{2}\right)\right]$\\
\hline
9 &
$S(2m-1)=\left\lbrace \dfrac{1}{2}-\dfrac{8}{81} \sin^4\left(\frac{(2m-1)\pi }{3}\right) \left[2+\cos\left(\frac{2 (2m-1) \pi }{3}\right)-\sqrt{3} \sin\left(\frac{2
(2m-1) \pi }{3}\right)\right]^2\right\rbrace$\newline$S(2m)$=$\dfrac{8}{81}\left\lbrace \left[2+\cos\left(\frac{4 m \pi }{3}\right)\right] \left[6+2 \cos\left(\frac{4 m \pi }{3}\right)+\cos\left(\frac{2
m \pi }{3}\right)\right]\sin^2\left(\frac{2m \pi }{3}\right)\right\rbrace$
& $S(n)$=$\dfrac{1}{2}\left\lbrace 1-\dfrac{1}{81} \left[1+2 \cos\left(\frac{2 n \pi }{3}\right)\right]^4\right\rbrace$  \\
\hline
10 & $S(n)=1-\dfrac{1}{8} \left\lbrace 4+\left[\frac{17}{32} \cos\left(\frac{n \pi }{4}\right)+\cos\left(\frac{5 n \pi }{4}\right)+\frac{15}{32}
\cos\left(\frac{9 n \pi }{4}\right)\right]^2\right\rbrace-\sin^2\left(\frac{n \pi }{2}\right)$
\newline$~~~~~~~~~~~~~  \left[17 \left(\cos\left(\frac{3 n \pi }{4}\right)+\sin\left(\frac{3 n \pi }{4}\right)\right)-15\left(\cos\left(\frac{7 n \pi }{4}\right)+\sin\left(\frac{7 n \pi }{4}\right)\right)\right]^2/{4096}$ & 
\hspace{.2cm} $S(n)=\dfrac{1}{2}\left\lbrace 1-\dfrac{1}{4} [1+\cos(n \pi )]^2\right\rbrace$ \\
\hline
11 &$S(2m-1)=\dfrac{1}{2}\left\lbrace1-\dfrac{1}{144} \sin^2\left(\frac{(2m-1) \pi }{3}\right) \left[-2 \sqrt{3} \cos\left(\frac{(2m-1) \pi }{3}\right)+8 \sin\left(\frac{(2m-1) \pi }{3}\right)\right.\right.$\newline$\left.\left.~~~~~~~~~~~~~~~~~~~~+3\left(\sqrt{3}\cos((2m-1) \pi )+ \sin((2m-1) \pi)\right)\right]^2\right\rbrace$ \newline$S{(2m)}$=$\dfrac{1}{144} \left\lbrace\left[11+6 \cos\left(\frac{4 m \pi }{3}\right)\right] \left[16+5 \cos\left(\frac{4 m \pi }{3}\right)+3 \cos\left(\frac{
2m \pi }{3}\right)\right] \sin^2\left(\frac{2m \pi }{3}\right)\right\rbrace$
& $ S(n)=\dfrac{1}{2}\left\lbrace 1-\dfrac{1}{144} \left[4+5 \cos\left(\frac{2 n\pi }{3}\right)\right.\right.$\newline$\left.\left.\hspace{1.65cm}+3 \cos\left(\frac{4 n \pi }{3}\right)\right]^2\right\rbrace$ \\

\hline
\end{tabular}
  \label{Table:LinearEntropy}
\end{table*}

{\it Exact solutions for six to eleven qubits.---}
Following the similar procedure from the previous part we solve for the cases from $N=6$ to $11$. We tabulate the explicit analytical 
expressions for the linear entropy for initial states:  $\otimes^N|0\rangle$ and $\otimes^N|+\rangle$ (refer 
Table \ref{Table:LinearEntropy}). It must be noted that these expressions including those for the $N=5$ case are true for any 
$n\in\mathcal{R}$. But the signatures of quantum integrability can be obtained by restricting $n$ to positive integers only.
The data points in all the figures of this paper are for $n\in \mathcal{N}^+$.
For these cases also, entropy is same for the consecutive odd and even values of $n$ \cite{ExactDogra2019}.

From these expressions we find them to be periodic in time.
We show that the time period $T$ for odd and even number of qubits for the initial states $\otimes^N\ket{0}$ ($\otimes^N\ket{+}$) 
is $6$($3$) and $4$($2$) respectively (see Fig.\ref{fig:correlations1}). We also show that the dynamics of the corresponding 
Floquet operator and its powers
shows periodic nature in time i.e. ${\mathcal U}^{n+T_1}={\mathcal U}^n $, where $n\geq 1$ and $T_1$ is the period. 
For even $N$, $T_1=8$ while for $N=5$, $7$, $9$ and $11$, $T_1=24$, $12$, $24$ and $12$ respectively.
We have found the eigensystem analytically and observe that the spectrum is degenerate \cite{supplementry2023}.\\
\begin{figure}[!t]\vspace{0.4cm}
\includegraphics[width=0.42\textwidth]{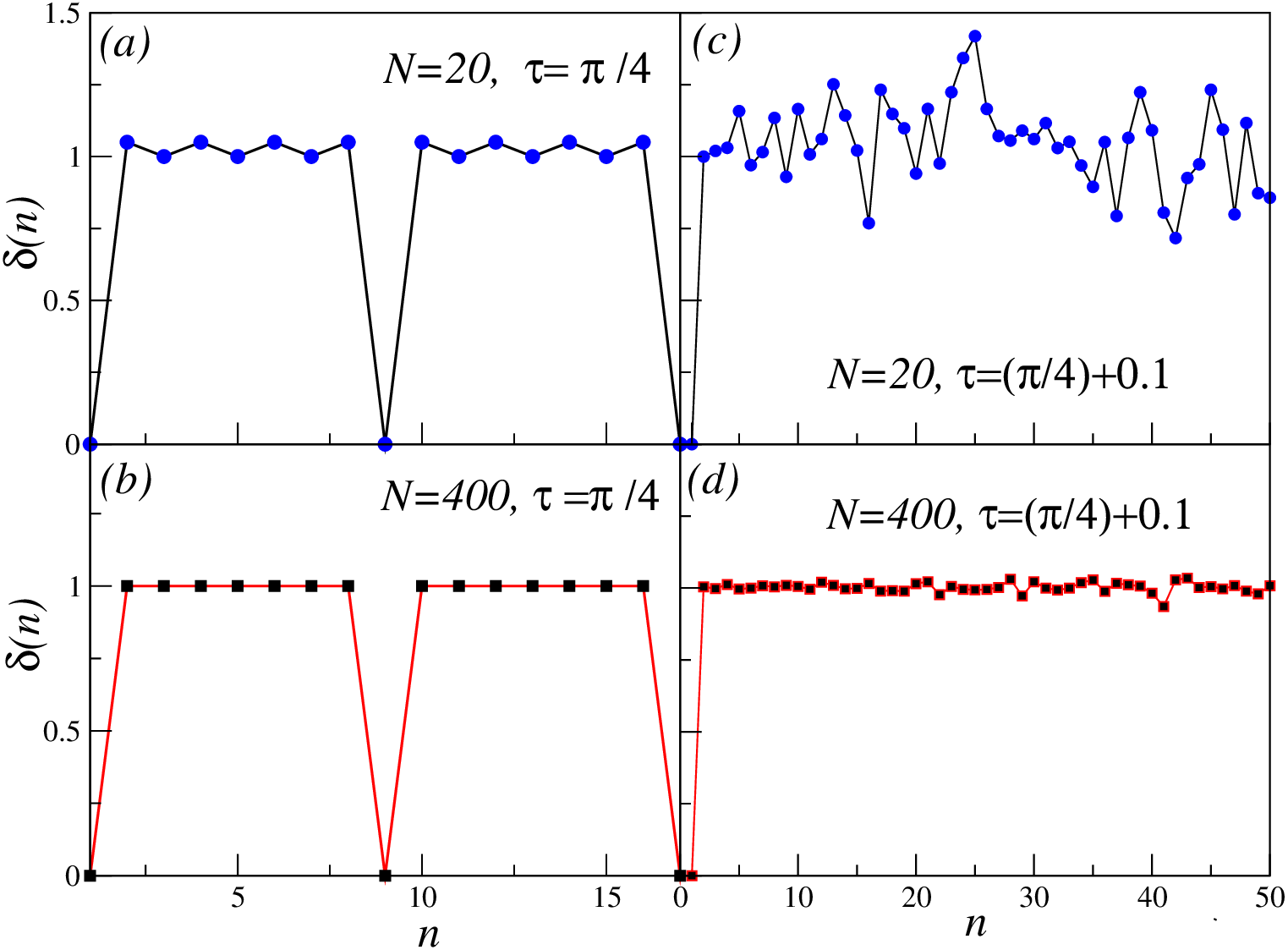}
\caption{Deviation $\delta(n)$ for various values of $N$ and $\tau$.}
\label{fig:periodicUnitary} 
\end{figure}
\begin{figure}[!t]\vspace{0.4cm}
\includegraphics[width=0.42\textwidth]{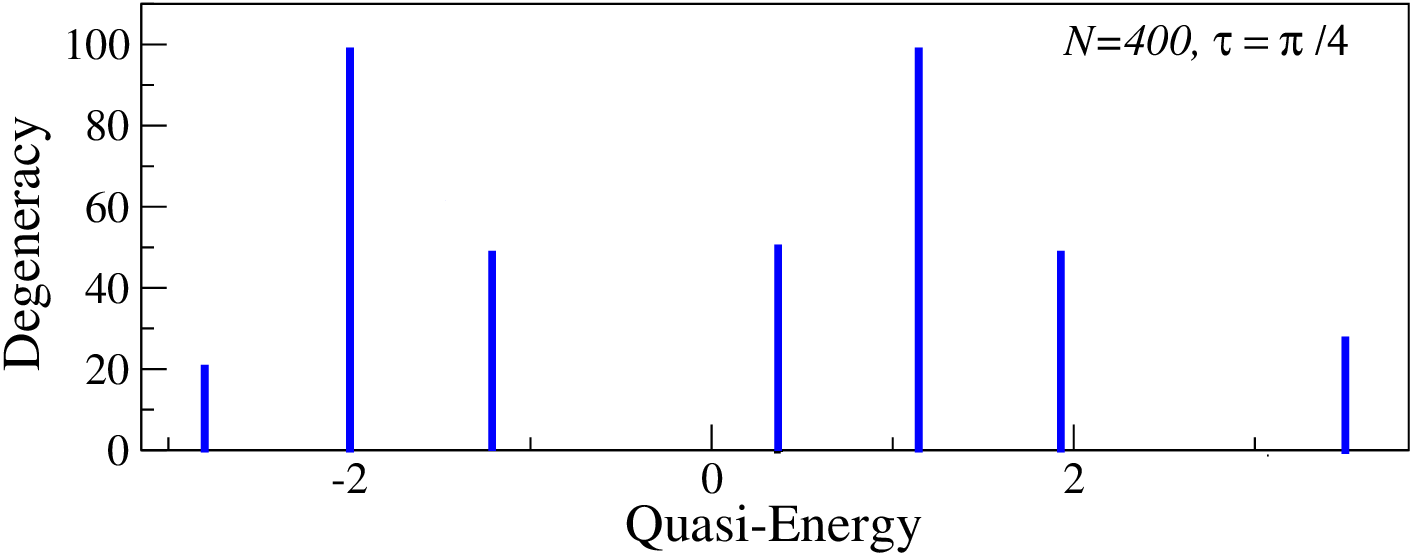}
\caption{Degeneracy of the quasienergies of $\mathcal{U}$.}
\label{fig:degenracy} 
\end{figure}  
{\it Results for general $N$.---}
With our method in principle, one can get the eigensystem and entanglement dynamics analytically for any finite $N$. 
But obtaining a general solution as a function of $N$ is mathematically challenging which can be observed from  the 
Ref. \cite{ArulLakshminarayan2005}, where only the nearest-neighbor interaction is considered.
Thus, we resort to numerical simulations and 
find various signatures of integrability for $N>11$. 
For this purpose we use the same signatures obtained for $N=5$ to $11$ and claim integrability for any $N>11$ and $\tau=\pi/4$.
To our surprise, we find the same signatures (depending only on the parity of $N$) as shown in the Fig.~\ref{fig:correlations1}. 
Similar $N$-independence was observed in the Ref.\cite{ArulLakshminarayan2005} for the case of the integrable kicked-Ising model in 
zero external magnetic field. When field is present in the NN neighbor interaction model the time period shows dependence on $N$ 
\cite{mishra2015protocol,naik2019controlled,pal2018entangling}.
We also study the operator dynamics numerically. To quantify the periodic nature of
time-evolved $\mathcal{U}$, we find its deviation from the original operator itself using: 
$\delta(n)=\sum_{p,q}|{\mathcal U}^n_{p,q}-{\mathcal U}_{p,q}|/2N$. It is zero for any $n>1$ if and only if ${\mathcal U}^n={\mathcal U}$,
thus confirming the time-periodic nature of $\mathcal{U}$.
Numerically it is observed that division by $2N$ ensures the average of $\delta(n)$ to be one.
The results are plotted in Fig.~\ref{fig:periodicUnitary} for even $N$. 
We find that for $\tau=\pi/4$ and $N$ upto $400$, time evolution of $\mathcal U$ is periodic (checked for $n$ as large as $5000$), 
whereas for $\tau\neq \pi/4$ it isn't. For odd $N$ (results not plotted 
here) we find similar periodicity but with different periods $T_1=12$ or $24$. These periods are the same as that of our analytical ones
for $N=5$, $7$, $9$, and $11$. Another signature of integrability is found from the eigenangle spectrum of $\mathcal U$ at $\tau=\pi/4$. 
We find it to be highly degenerate taking values from the set $\{0,\pm \pi/4, \pm \pi/2, \pm 3 \pi/4, \pm \pi \}$ for even $N$ 
(checked for $N$ upto $400$, refer Fig.~\ref{fig:degenracy}). Similarly, for odd $N$, we numerically find a degenerate spectrum. 
Thus, with these signatures, we can very well conjecture that the system is quantum integrable for any $N>11$ number of qubits 
and the time period ($\tau$) of the kick such that $\tau=\pi/4$.\\

{\it Conclusions.---}
Integrable models have played an important role in advancing our understanding of physical systems \cite{BillSutherlandbook}.
 Our model of infinite-range with Ising two-body interaction shows quantum integrability for any number of qubits. 
The cases involving $N=5$ to $11$ qubits are dealt with analytically whereas a conjecture with sufficient evidence is presented for any 
$N>=11$. Previous work as far as integrability in our model (after mapping to QKT) is concerned was limited only up to $4$ qubits. 
We have now generalized it to any $N$.

It must be noted that our conjecture is based on the circumstantial signatures of integrability. A more rigorous proof on quantum 
integrability in our case based on Bethe ansatz \cite{bethe1931theorie,pan1999analytical} and/or obtaining a transfer matrix by 
solving a Yang-Baxter relation in this case is highly warranted \cite{wadati1993quantum,doikou2010introduction}. This transfer matrix 
can then be used to generate infinite number of conserved quantities to prove integrability.
Recent work involved the use of generalized Hubbard-Stratonovich transformation to get an exact solution for quantum strong 
long-range Ising chains \cite{roman2023exact}. Our model can further be investigated in this direction.\\
Our results (for the smaller number of qubits) can be verified experimentally in various setups from NMR \cite{Krithika2019}, 
superconducting qubits \cite{neill2016ergodic} and laser cooled atoms \cite{Chaudhury09}, where the QKT is implemented readily. 
Whereas for the larger number of qubits (of the order of $100$s), one can use ion traps \cite{monroe2021programmable,defenu2023long}. 
Our conjecture can be tested in this setup as well.
With our findings, a search for similar quantum integrable spin systems can be initiated.\\
 {\it Acknowledgement.---}
We are indebted to M.S. Santhanam for valuable comments and suggestions on the manuscript.
We thank anonymous referees for their valuable comments.

\bibliography{reference2023,references,reference22,reference221,reference22013} 
 
% % % % % % % % % % % % % % % % % % % % % % % % % % % 

%%%%%%%%%%%%%%%%%%%%%%%%%%%%%%%%%%%%%%%%%%%%%%%%%%%%%%%%%%%%%%%%%%%%%%%%

\pagebreak
\widetext
\begin{center}
\textbf{\large Supplementary Material for\\ 
``Exactly solvable dynamics and signatures of integrability in an 
infinite-range many-body Floquet spin system''}
\end{center}
%%%%%%%%%% Merge with supplemental materials %%%%%%%%%%
%%%%%%%%%% Prefix a "S" to all equations, figures, tables and reset the counter %%%%%%%%%%
\setcounter{equation}{0}
\setcounter{figure}{0}
\setcounter{table}{0}
\setcounter{page}{1}
\makeatletter
\renewcommand{\theequation}{S\arabic{equation}}
\renewcommand{\thefigure}{S\arabic{figure}}

\onecolumngrid
%\appendix
\section{Exact analytical solution for five-qubits case}

This section presents a detailed description of the analytical solutions that have been discussed in the main text file. As per
Eq. (3), of the main text, unitary operator acting on a system of $5$-qubits, is given by, 
\begin{eqnarray}\label{Eq:5q87}
 \mathcal{U} &=& \exp\left[{-i \frac{\pi}{4} (\sigma_1^z \sigma_2^z +\sigma_1^z \sigma_3^z +\sigma_1^z \sigma_4^z+\sigma_1^z \sigma_5^z+\sigma_2^z \sigma_3^z+\sigma_2^z \sigma_4^z+\sigma_2^z \sigma_5^z+\sigma_3^z \sigma_4^z+\sigma_3^z \sigma_5^z+\sigma_4^z \sigma_5^z)} \right]\times
 \nonumber \\
 &&  \exp \left[{-i \frac{\pi}{4}(\sigma_1^y+\sigma_2^y+\sigma_3^y+\sigma_4^y+\sigma_5^y)} \right], 
\end{eqnarray}
where $\sigma_l^z$ and $\sigma_l^y$  are  Pauli operators. The solution to the $5$-qubit case proceeds from the general observation that (proved in Sec.\ref{Sec:ProofCommutation}),
\begin{equation}
[\mathcal{U},\otimes_{l=1}^{N} \sigma^y_l]=0,
\end{equation} 
 {\it i.e.,} there is
an ``up-down" or parity symmetry \cite{ExactDogra2019}. Here, we are confined to $(N+1=6)$ dimensional permutation symmetric subspace of the total 
$2^{N}$= $2^{5}$= $32$ dimensional Hilbert space. The basis vectors are given as follows:
\begin{eqnarray}
\ket{\phi_0^{\pm}}&=&\frac{1}{\sqrt{2}}(\ket{w_0} \pm i \ket{\overline{w_0}}),\\
\ket{\phi_1^{\pm}}&=&\frac{1}{\sqrt{2}}(\ket{w_1} \mp i \ket{\overline{w_1}}),\\
\ket{\phi_2^{\pm}}&=&\frac{1}{\sqrt{2}}(\ket{w_2} \pm i \ket{\overline{w_2}}).
\end{eqnarray}
These states $\ket{\phi_j^{\pm}}$ are parity eigenstate with eigenvalue $\pm 1$  such that $ \otimes_{l=1}^5 \sigma_l^y \ket{\phi_j^{\pm}}=\pm \ket{\phi_j^{\pm}}$. Where $\ket{w_0}=\ket{00000}$, $\ket{\overline{w_0}}=\ket{11111}$, $\ket{w_1}=\frac{1}{\sqrt{5}} \sum_\mathcal{P} \ket{00001}_\mathcal{P}$, $\ket{\overline {w_1}}=\frac{1}{\sqrt{5}} \sum_\mathcal{P} \ket{01111}_\mathcal{P}$, $\ket{w_2}=\frac{1}{\sqrt{10}} \sum_\mathcal{P} \ket{00011}_\mathcal{P}$ and $\ket{\overline{w_2}}=\frac{1}{\sqrt{10}} \sum_\mathcal{P}\ket{00111}_\mathcal{P}$ and $\sum_\mathcal{P}$  summation over all possible permutations. In this basis, the unitary operator $\mathcal{U}$ is block diagonal  in $\mathcal{U_{\pm}}$ and is given as follows:
\begin{equation} 
 \mathcal{U} = \begin{pmatrix}
            \mathcal{U}_{+} & 0 \\ 0 & \mathcal{U}_{-}
            \end{pmatrix},
\end{equation}
where the  blocks $\mathcal{U}_{\pm}$ are $3\times 3$ dimensional matrices  and $0$ is null matrices of same dimension as of  $\mathcal{U}_{\pm}$. The $\mathcal{U}_{+}\left(\mathcal{U}_{-}\right)$ written in the basis of positive (negative )parity states  $\left\lbrace \phi_0^{+},\phi_1^{+},\phi_2^{+}\right\rbrace\left(\left\lbrace \phi_0^{-},\phi_1^{-},\phi_2^{-}\right\rbrace\right)$ is as follows:
\begin{equation}\label{Eq:5q88}
\mathcal{U}_{\pm} = \begin{pmatrix}
            \bra{\phi_0^{\pm}}\mathcal{U}\ket{\phi_0^{\pm}}& \bra{\phi_0^{\pm}}\mathcal{U}\ket{\phi_1^{\pm}}&\bra{\phi_0^{\pm}}\mathcal{U}\ket{\phi_2^{\pm}} \\ \bra{\phi_1^{\pm}}\mathcal{U}\ket{\phi_0^{\pm}} & \bra{\phi_1^{\pm}}\mathcal{U}\ket{\phi_1^{\pm}}&\bra{\phi_1^{\pm}}\mathcal{U}\ket{\phi_2^{\pm}}\\\bra{\phi_2^{\pm}}\mathcal{U}\ket{\phi_0^{\pm}}&\bra{\phi_2^{\pm}}\mathcal{U}\ket{\phi_1^{\pm}}&\bra{\phi_2^{\pm}}\mathcal{U}\ket{\phi_2^{\pm}}\\
            \end{pmatrix}.
\end{equation}

This block diagonalization makes it easy to take the $n$th power of the unitary operator $\mathcal{U}$,
\begin{equation}
\mathcal{U}^n = \begin{pmatrix}
            \mathcal{U}_{+}^n & 0 \\ 0 & \mathcal{U}_{-}^n
            \end{pmatrix}.
\end{equation}
The block operators  $\mathcal{U}_{\pm}$ are explicitly found by using
Eqs.(\ref{Eq:5q87}) and (\ref{Eq:5q88}), as follows:
\begin{equation}
  \mathcal{U}_{\pm}= \frac{1}{4} e^{\pm{\frac{ i \pi }{4}}}
\begin{pmatrix}
 \mp{1} &  i\sqrt{5}  &  \mp{\sqrt{10}} \\
- i\sqrt{5} & \pm 3 & - i\sqrt{2} \\
  \pm \sqrt{10} & - i\sqrt{2} & \mp 2 \\
\end{pmatrix}.
\end{equation}

The eigenvalues  of $\mathcal{U_+}\left(\mathcal{U_-}\right)$ are ${e^{\frac{i \pi }{4}}}\left\lbrace 1,e^{\frac{2 i \pi }{3}},e^{-\frac{2 i \pi }{3}}\right\rbrace \left(e^{\frac{3i \pi }{4}}\left\lbrace1,e^{-\frac{2 i \pi }{3}},e^{\frac{2 i \pi }{3}}\right\rbrace\right)$ and  the eigenvectors are $\left\lbrace\left[\frac{\pm i}{\sqrt{5}},1,0\right]^T, \left[\pm i\sqrt{\frac{5}{6}},-\frac{1}{\sqrt{6} },1\right]^T,  \left[\mp i\sqrt{\frac{5}{6}},\frac{1}{\sqrt{6} },1\right]^T\right\rbrace$.   To evolve initial states we need $\mathcal{U}^n$ and therefore $\mathcal{U}_{\pm}^n$. Thus,
\begin{equation}
\mathcal{U}_{\pm}^n =(\pm 1)^n e^{\pm{\frac{ in \pi }{4}}}
{\left[
\begin{array}{ccc}
 \frac{1}{6} \left(1+5 \cos \left(\frac{2 n \pi }{3}\right)\right) &\pm \frac{1}{3} i \sqrt{5} \sin^2\left(\frac{n \pi }{3}\right) & -\sqrt{\frac{5}{6}}
\sin \left(\frac{2 n \pi }{3}\right) \\
 \mp \frac{1}{3} i \sqrt{5} \sin^2 \left(\frac{n \pi }{3}\right) & \frac{1}{6} \left(5+\ \cos \left(\frac{2 n\pi }{3}\right)\right) &\mp {i
\sin \left(\frac{2 n \pi }{3}\right)}/{\sqrt{6}} \\
 \sqrt{\frac{5}{6}} \sin \left(\frac{2 n \pi }{3}\right) & \mp {i \sin \left(\frac{2 n \pi }{3}\right)}/{\sqrt{6}} & \cos \left(\frac{2n
 \pi }{3}\right) \\
\end{array}
\right]}.
\end{equation}
Each element of the  $\mathcal{U}_{\pm}^n$  matrix exhibits a time periodicity of $24$. Therefore, the unitary operator 
$\mathcal{U}^n$ for a system of $5$ qubits also possesses a periodicity of $24$. This shows that the unitary operator has a 
periodic nature. 
\subsection{Initial state $\ket{00000}$= $\ket{\theta_0 =0,\phi_0 =0}$} 
The  state $\ket{\psi_n}$ is  obtained by performing $n$ iterations of unitary operator $\mathcal{U}$ on this initial state $\ket{00000}$. Thus,
\begin{equation}
  \begin{split}
\ket{\psi_n}&=\mathcal{U}^n |00000\rangle =\frac{1}{\sqrt{2}} \mathcal{U}^n \left( |\phi_{0}^{+} \rangle + |\phi_{0}^{-} \rangle \right) \\ 
&=\frac{1}{\sqrt{2}} \left( \mathcal{U}_{+}^n |\phi_{0}^{+} \rangle + \mathcal{U}_{-}^n |\phi_{0}^{-} \rangle \right) \\ 
&=\frac{1}{2} e^{{\frac{ in \pi }{4}}}\left\lbrace(1+ i^n)\left(\alpha_n \ket{w_0}-i\beta_n \ket{\overline{w_1}}+ \gamma_n \ket{w_2}\right) \right.  \\ & \left. +(1-i^n)\left(   i\alpha_n \ket{\overline{w_0}}+ \beta_n\ket{w_1}+ i \gamma_n \ket{\overline{w_2}}\right)\right\rbrace,
 \end{split}
\end{equation}
where $\alpha_n=\frac{1}{6} \left[1+5 \cos \left(\frac{2 n \pi }{3}\right)\right]$, $\beta_n=\frac{i \sqrt{5}}{3}  \sin^2 \left(\frac{n \pi }{3}\right)$ and $\gamma_n=\sqrt{\frac{5}{6}} \sin \left(\frac{2 n \pi }{3}\right)$.
By using the state $\ket{\psi_n}$ of system, the one and two-qubit reduced density matrices (RDM) 
$\rho_1(n)=\text{tr}_{2,3,4,5} (|\psi_n \rangle \langle \psi_n |)$ and $\rho_{12}(n)=\text{tr}_{3,4,5} (|\psi_n \rangle \langle 
\psi_n |)$ are obtained. The linear entropy and concurrence are used as measures of entanglement in quantum systems. The linear entropy is 
a measure of the degree of mixedness or impurity of a quantum state. It quantifies the entanglement in a bipartite (two-party) 
quantum system \cite{wootters1998entanglement} which is defined as $1-\text{tr}\left[\rho_1^2(n)\right]$.\\
\subsubsection{\bf{The linear entropy}}
It turns out that for  even $n=2m$ single qubit RDM, $\rho_1(n)$, is diagonal and is given as follows:
\begin{equation}
\rho_1(2m) =
{\left[
\begin{array}{cc}
 \frac{1}{9} \left(6+2 \cos \left(\frac{4m \pi }{3}\right)+\cos\left(\frac{2m \pi }{3}\right)\right) & 0 \\
 0 & \frac{4}{9} \left(2+\cos \left(\frac{4m \pi }{3}\right)\right) \sin^2 \left(\frac{2m \pi }{3}\right) \\
\end{array}
\right]},
\end{equation}
where $m$ is a natural number. The eigenvalues are $\lambda_{2m}$ and $(1-\lambda_{2m})$, where $\lambda_{2m}=\frac{1}{9} \left[6+2 \cos \left(\frac{4m \pi }{3}\right)+\cos\left(\frac{2m\pi }{3}\right)\right.]$. The expression for linear entropy is given as follows:
\begin{eqnarray}\nonumber
\label{Eq:Ap5Q1}
S_{(\theta_0=0,\phi_0=0)}^{(5)}(2m)&=& S_{(0,0)}^{(5)}(2m)=1-\mbox{tr}\left[\rho_1^2(2m)\right]\\ 
&=& 2\lambda_{2m}\left(1-\lambda_{2m}\right) \\ \nonumber
&=&{\frac{8}{81} \left[\left(2+\cos\left(\frac{4m \pi }{3}\right)\right) \left(6+2 \cos\left(\frac{4m \pi }{3}\right)+\cos\left(\frac{2
m \pi }{3}\right)\right) \sin^2\left(\frac{2m \pi }{3}\right)\right]}.
\end{eqnarray}
For odd $n=2m-1$ the single qubit RDM $\rho_1(n)$ is not diagonal and is given as follows:
\begin{equation}
\rho_1(2m-1) =
\frac{1}{2}\left(
\begin{array}{cc}
 1 & P_n \\
 P_n
& 1 \\
\end{array}
\right),
\end{equation}
where 
\begin{equation}\nonumber
 P_n=\frac{4}{9} \sin^2\left[\frac{(2m-1)\pi }{3}\right] \left[2+\cos\left(\frac{2 (2m-1) \pi }{3}\right)-\sqrt{3} \sin\left(\frac{2 (2m-1) \pi }{3}\right)\right].
 \end{equation}
The eigenvalues are $\lambda_{2m-1}$ and $1-\lambda_{2m-1}$ and, \\
$\lambda_{2m-1}=\dfrac{1}{18} \left[9-\sqrt{27-17 \cos\left(\frac{2 (2m-1) \pi }{3}\right)-10 \cos\left(\frac{4 (2m-1) \pi }{3}\right)-17 \sqrt{3} \sin\left(\frac{2 (2m-1) \pi }{3}\right)+10 \sqrt{3} \sin\left(\frac{4(2m-1) \pi }{3}\right)}\right] $.\\

 The expression for the linear entropy is given as follows: 
 \begin{eqnarray}
 \label{Eq:Ap5Q2}
S_{(0,0)}^{(5)}(2m-1)=\frac{1}{2}-\frac{8}{81} \sin^4\left(\frac{(2m-1) \pi }{3}\right) \left[2+\cos\left(\frac{2 (2m-1) \pi }{3}\right)-\sqrt{3} \sin\left(\frac{2
(2m-1) \pi }{3}\right)\right]^2.
\end{eqnarray}
We calculated the linear entropy for 5 qubit  in which we find that it attains the maximum upper bound value i.e. $1/2$. From the Eqs.~(\ref{Eq:Ap5Q1}) and (\ref{Eq:Ap5Q2}), we can see that its entanglement dynamics is periodic in nature with $S_{(0,0)}^{(5)}(n)=S_{(0,0)}^{(5)}(n+6)$ and the   linear entropy for consecutive odd \mbox{and} even values of $n$ are same (Fig.(1)) i.e.
\begin{equation}
S_{(0,0)}^{(5)}(2m-1)=S_{(0,0)}^{(5)}(2m).
\end{equation}

\subsubsection{\bf{Concurrence}}
The linear entropy measures the entanglement in a pure state between one qubit with the rest, while the concurrence quantifies the 
entanglement between any two qubits in the system. Regardless of the specific qubits chosen, there is only one concurrence value due 
to the permutation symmetry in the state. The concurrence is given by,
\begin{equation}
\label{Eq:Ap5Q3}
\mathcal{C}(\rho_{12})=\text{max}\left(0, \sqrt{\lambda_1}-\sqrt{\lambda_2}-\sqrt{\lambda_3}-\sqrt{\lambda_4} \right),
\end{equation}
where $\lambda_l$ are eigenvalues in decreasing order of 
$(\sigma_y \otimes \sigma_y)\rho_{12} (\sigma_y \otimes \sigma_y) \rho_{12}^*$, where $\rho_{12}^*$ is complex conjugation in 
the standard ($\sigma_z$) basis. For even time $n=2m$, the two-qubit RDM $\rho_{12}(2m)$ is an ``$X$ state" \cite{TingYu2007} given as follows:
\begin{equation}
\rho_{12}(2m) ={\left(
\begin{array}{cccc}
 a_n & 0 & 0 & b_n \\
 0 & c_n & c_n & 0 \\
 0 & c_n & c_n & 0 \\
 b_n & 0 & 0 & d_n \\
\end{array}
\right)}.
\end{equation}
Here the coefficients are as follows:
\begin{eqnarray}\nonumber
a_n&=&\frac{1}{18} \left[9+5 \cos\left(\frac{4m \pi }{3}\right)+4 \cos\left(\frac{2m\pi }{3}\right)\right],\;\; b_n=\frac{1}{{6 \sqrt{3}}}\sin\left(\frac{4m\pi }{3}\right)\left[-1+4 \cos\left(\frac{4m \pi }{3}\right)\right] ,\\ \nonumber
c_n&=&\frac{1}{9}\sin^2\left(\frac{2m \pi }{3}\right) \left[5+4 \cos\left(\frac{4m \pi }{3}\right)\right]\;\;\mbox{and}\;\; d_n=\frac{1}{3} \sin^2\left(\frac{2m \pi }{3}\right).
\end{eqnarray}
 Time periodicity of  all the  coefficients is $3$ for even values of $n=2m$ i.e. $\left( a_n(m),b_n(m),c_n(m),d_n(m)\right)=\left( a_{n}(m+3),\right.$\\$\left.b_{n}(m+3),c_{n}(m+3),d_{n}(m+3)\right)$. In terms of these coefficients the eigenvalues $\lambda_l$ of $(\sigma_y \otimes \sigma_y)\rho_{12} (\sigma_y \otimes \sigma_y) \rho_{12}^*$ in decreasing order are $\left\lbrace 0,4c_n^2 ,b_n^2-2b_n \sqrt{a_n d_n }+a_n d_n, b_n^2+ 2b_n \sqrt{a_n d_n }+a_n d_n\right\rbrace$.\\
  For odd time $n=2m-1$, the two qubit RDM $\rho_{12}(2m-1)$ is given as follows:
\begin{equation}
\rho_{12}(2m-1)=
{\left(
\begin{array}{cccc}
 \tilde{a}_n & \tilde{d}_n & \tilde{d}_n & 2\tilde{b}_n \\
 \tilde{d}_n & 2\tilde{c}_n & 2\tilde{c}_n & \tilde{d}_n \\
 \tilde{d}_n & 2\tilde{c}_n & 2\tilde{c}_n & \tilde{d}_n \\
 2\tilde{b}_n & \tilde{d}_n & \tilde{d}_n& \tilde{a}_n \\
\end{array}
\right)}.
\end{equation}
Here the coefficients are as follows:
\begin{eqnarray}\nonumber
 \tilde{d}_n&=&\frac{2}{9} \sin^2\left(\frac{(2m-1) \pi }{3}\right) \left[2+\cos\left(\frac{2 (2m-1) \pi }{3}\right)-\sqrt{3} \sin\left(\frac{2
(2m-1) \pi }{3}\right)\right],  \\ \nonumber
\tilde{a}_n&=&\frac{1}{9} \left[6+\cos\left(\frac{2 (2m-1) \pi }{3}\right)+2 \cos\left(\frac{4 (2m-1) \pi }{3}\right)\right],\\ \nonumber
\tilde{c}_n&=&\frac{1}{9}\sin^2\left(\frac{(2m-1) \pi }{3}\right) \left[5+4 \cos\left(\frac{2(2m-1) \pi }{3}\right)\right] \mbox{and}\\ \nonumber
\tilde{b}_n&=&\frac{1}{{6 \sqrt{3}}}\sin\left(\frac{2(2m-1)\pi }{3}\right)\left[-1+4 \cos\left(\frac{2(2m-1) \pi }{3}\right)\right] .
\end{eqnarray}
Time periodicity of  all the coefficients  is $3$ i.e. $\left( \tilde{a}_n(m),\tilde{d}_n(m),\tilde{c}_n(m),\tilde{b}_n(m)\right)=\left( \tilde{a}_n(m+3),\tilde{d}_n(m+3),\tilde{c}_n(m+3),\right.$\\$\left.\tilde{b}_n(m+3)\right)$. For odd values of $n$ the eigenvalues $\lambda_l$ of $(\sigma_y \otimes \sigma_y)\rho_{12} (\sigma_y \otimes \sigma_y) \rho_{12}^*$  in decreasing order are given as,  \\
$\left\lbrace 0,\frac{1}{8} \left(4 \tilde{b}_n^2+16 \tilde{c}_n^2-8\tilde{d}_n^2 + 4\tilde{b}_n \tilde{a}_n+\tilde{a}_n^2-(2 \tilde{b}_n-4 \tilde{c}_n+\tilde{a}_n) \sqrt{4 \tilde{b}_n^2+16 \tilde{b}_n \tilde{c}_n+16 \tilde{c}_n^2-16 \tilde{d}_n^2+4 \tilde{b}_n \tilde{a}_n+8\tilde{c}_n \tilde{a}_n+\tilde{a}_n^2}\right)\right.$\\
$\left.,\frac{1}{8} \left(4 \tilde{b}_n^2+16 \tilde{c}_n^2-8\tilde{d}_n^2 + 4\tilde{b}_n \tilde{a}_n+\tilde{a}_n^2+(2 \tilde{b}_n-4 \tilde{c}_n+\tilde{a}_n) \sqrt{4 \tilde{b}_n^2+16 \tilde{b}_n \tilde{c}_n+16 \tilde{c}_n^2-16 \tilde{d}_n^2+4 \tilde{b}_n \tilde{a}_n+8\tilde{c}_n \tilde{a}_n+\tilde{a}_n^2}\right)\right.$\\
,$\left.\frac{1}{4} (2 b_n-f_n)^2\right\rbrace$.\\
Due to the periodic nature of the coefficients, we observe that the matrix 
$(\sigma_y \otimes \sigma_y)\rho_{12} (\sigma_y \otimes \sigma_y) \rho_{12}^*$ and its eigenvalues exhibit a pattern with a period 
of $6$ for even and odd values of $n$. The simplified eigenvalues $\lambda_l$ of $(\sigma_y \otimes \sigma_y)\rho_{12} 
(\sigma_y \otimes \sigma_y) \rho_{12}^*$ in decreasing order for $n$ = $0$ to $12$  are $\left\lbrace(0, 0, 0, 0), 
\left(1/4, 1/4, 0, 0\right),\left(1/4, 1/4, 0, 0\right), \left(1/4, 1/4, 0, 0\right),\right.$\\$\left.\left(1/4, 1/4, 0, 0\right), 
(0, 0, 0, 0),(0, 0, 0, 0),\left(1/4, 1/4, 0, 0\right),\left(1/4, 1/4, 0, 0\right), \left(1/4, 1/4, 0, 0\right),\left(1/4, 1/4, 0, 
0\right),  (0, 0, 0, 0),(0, 0, 0, 0)\right\rbrace$. Where we find that these eigenvalues repeat after every $n\rightarrow n+6$, 
which shows that they have a period of $6$ for every value of $n$. By using these $\lambda_l$ in Eq.~(\ref{Eq:Ap5Q3}), we can easily 
see that the concurrence remains zero for every value of $n$. This is shown in the (Fig.(1)).
\subsection{Initial state $\ket{+++++}$= $\ket{\theta_0 =\pi/2,\phi_0 =-\pi/2}$}
We have previously discussed the evolution of the state $\ket{00000}$ in detail. Now, let's examine the case of a five-qubit for state $\ket{+++++}_y$, where $|+\rangle_y=\frac{1}{\sqrt{2}}(|0\rangle+i|1\rangle)$ is an eigenstate of $\sigma_y$ with eigenvalue $+1$. The evolution of this state lies entirely in positive parity subspace and it can also be written as $\otimes ^5  {\ket{+}}_y=\frac{1}{4} \ket{\phi_0^+}+\frac{i\sqrt{5}}{4}\ket{\phi_1^+}- \frac{\sqrt{10}}{4} \ket{\phi_2^+}$. The time evolution of this state is given as,
\begin{eqnarray}
\ket{\psi_n}&=&\mathcal{U}_{+}^n \ket{+++++} \\ \nonumber
            &=&e^{{\frac{ in \pi }{4}}}(\tilde{\alpha}_n \ket{\phi_0^+}+\tilde{\beta}_n \ket{\phi_1^+}+ \tilde{\gamma}_n \ket{\phi_2^+}),
\end{eqnarray}
where the coefficients are as follows:
\begin{eqnarray}\nonumber
\tilde{\alpha}_n &=&\frac{1}{12} \left[-2+5 \cos\left(\frac{2 n \pi }{3}\right)+5 \sqrt{3} \sin\left(\frac{2 n \pi }{3}\right)\right], ~~\tilde{\beta}_n=\frac{i\sqrt{5}}{12} \left[2+\cos\left(\frac{2 n \pi }{3}\right)+\sqrt{3} \cos \left(\frac{2 n \pi }{3}\right)\right]\mbox{and}\\ \nonumber
\tilde{\gamma}_n &=&\frac{1}{6} \sqrt{\frac{5}{2}} \left[-3 \cos\left(\frac{2 n \pi }{3}\right)+\sqrt{3} \sin\left(\frac{2 n \pi }{3}\right)\right].
\end{eqnarray}

\subsubsection{\bf{The linear entropy}}
The single qubit RDM, $\rho_1(n)$, is given as  follows:
\begin{equation}
\rho_1(n) =\frac{1}{2}
\left[
\begin{array}{cc}
 1 & -\frac{i}{9}  \left(1+2 \cos\left(\frac{2 n \pi }{3}\right)\right)^2 \\
 \frac{i}{9}  \left(1+2 \cos\left(\frac{2 n \pi }{3}\right)\right)^2 & 1 \\
\end{array}
\right].
\end{equation}
The eigenvalues are $\left\lbrace\frac{1}{9}\left(3 - 2 \cos\left(\frac{2 n \pi}{3}\right) - \cos\left(\frac{4 n \pi}{3}\right)\right), 
\frac{1}{9}\left(6 + 2 \cos\left(\frac{2 n \pi}{3}\right) + \cos\left(\frac{4 n \pi}{3}\right)\right)\right\rbrace$.
The linear entropy  can be calculated by using the expression $S=1-\mbox{tr}\left[\rho_1^2(n)\right]$ as follows:
\begin{equation}\label{Eq:ap5q84}
S_{(\pi/2,-\pi/2)}^{(5)}(n)=\frac{1}{2}\left[1-\frac{1}{81} \left(1+2 \cos\left(\frac{2 n \pi }{3}\right)\right)^4\right].
\end{equation} 
 From the Eq.(\ref{Eq:ap5q84}), we can see that its entanglement dynamics is periodic in nature with 
 $S_{(\pi/2,-\pi/2)}^{(5)}(n)=S_{(\pi/2,-\pi/2)}^{(5)}(n+3)$. It attains the maximum upper bound value of linear entropy  i.e. $1/2$. 
 This is shown in the (Fig.(1)).
\subsubsection{\bf{Concurrence}}
The two-qubit RDM in this case  is given by,
\begin{equation}
\rho_{12}(n)=\left(
\begin{array}{cccc}
 a_m & b_m^* & b_m^* & c_m \\
 b_m & d_m & d_m & b_m^* \\
 b_m & d_m & d_m & b_m^* \\
 c_m & b_m & b_m & a_m \\
\end{array}
\right),
\end{equation}
here the coefficients are as follows:
\begin{eqnarray} \nonumber 
 a_m&=&{\frac{1}{36} \left[12-\cos\left(\frac{2 n \pi }{3}\right)-2 \cos\left(\frac{4 n \pi }{3}\right)-\sqrt{3} \sin\left(\frac{2
n \pi }{3}\right)+2 \sqrt{3} \sin\left(\frac{4 n \pi }{3}\right)\right]},
b_m={\frac{i}{36}  \left[1+2 \cos\left(\frac{2 n \pi }{3}\right)\right]^2},\\ \nonumber
c_m&=&\frac{1}{36} \left[-3 \cos\left(\frac{2 n \pi }{3}\right)-6 \cos\left(\frac{4 n \pi }{3}\right)+\sqrt{3} \left(\sin\left(\frac{2n \pi }{3}\right)-2 \sin\left(\frac{4 n \pi }{3}\right)\right)\right] \mbox{and}\\ \nonumber
d_m&=&{\frac{1}{36} \left[6+\cos\left(\frac{2 n \pi }{3}\right)+2 \cos\left(\frac{4 n \pi }{3}\right)+\sqrt{3} \sin\left(\frac{2n \pi }{3}\right)-2 \sqrt{3} \sin\left(\frac{4 n \pi }{3}\right)\right]}.
\end{eqnarray}
Time periodicity of all  the coefficients  is $3$ i.e. $\left( a_m(n),b_m(n),c_m(n),d_m(n)\right)=\left( a_{m}(n+3),b_{m}(n+3),c_{m}(n+3),d_{m}(n+3)\right)$. The eigenvalues $\lambda_l$  of $(\sigma_y \otimes \sigma_y)\rho_{12} (\sigma_y \otimes \sigma_y) \rho_{12}^*$ in decreasing order  are given as,\\
$\left\lbrace 0,\frac{1}{8} \left({a_{m}
}^2-8 b_m^2-2 {a_{m}} {c_{m}}+{c_{m}^2}+4 {d_{m}^2}-({a_{m}}-{c_{m}}-2{d_{m}}) \sqrt{{a_{m}^2}-16 {b_{m}^2}-2 {a_{m}} {c_{m}}+{c_{m}^2}+4 {a_{m}} d_{m}-4 {c_{m}} {d_{m}}+4 {d_{m}}^2}\right)\right.$\\
$\left.,\frac{1}{8} \left({a_m^2}-8{b_m^2}-2 {a_{m}} {c_{m}}+{c_m^2}+4 {d_m^2}+({a_{m}}-c_m-2 d_m) \sqrt{a_m^2-16b_m^2-2 a_m c_m+c_m^2+4 a_m d_m-4 c_m d_m+4 d_m^2}\right),\right.$\\ $\left.\frac{1}{4} \left({a_{m}}+{c_{m}}\right)^2\right\rbrace$.\\
% As we discussed earlier due to the periodic nature of the matrix and coefficients its eigenvalue also shows periodic nature having the same periodicity, which we can see by using the simplified eigenvalues of $(\sigma_y \otimes \sigma_y)\rho_{12} (\sigma_y \otimes \sigma_y) \rho_{12}^*$ for  $n$ going from $0$ to $6$  are $\left\lbrace(0, 0, 0, 0), \left(1/4, 1/4, 0, 0\right),\left(1/4, 1/4, 0, 0\right),(0, 0, 0, 0),\left(1/4, 1/4, 0, 0\right),\left(1/4, 1/4, 0, 0\right), (0, 0, 0, 0)\right\rbrace$. These eigenvalues repeat after $n\rightarrow n+3$. By using these  $\lambda_j$ in Eq.~(\ref{Eq:Ap5Q3}) we can easily see that the concurrence vanishes  for each value of $n$, which is shown in  Fig~(\ref{}).
 The eigenvalues $\lambda_l$ of the matrix $(\sigma_y \otimes \sigma_y)\rho_{12} (\sigma_y \otimes \sigma_y) \rho_{12}^*$ exhibit a 
 periodic nature, with the same periodicity as the matrix and the coefficients. By analyzing the simplified eigenvalues $\lambda_l$ 
 for values of $n$ ranging from 0 to 6, we observe the following pattern: $\left\lbrace(0, 0, 0, 0), (1/4, 1/4, 0, 0), (1/4, 1/4, 0, 0),
 (0, 0, 0, 0), (1/4, 1/4, 0, 0), (1/4, 1/4, 0, 0), (0, 0, 0, 0)\right\rbrace$. Notably, these eigenvalues repeat after shifting $n$ 
 by $3$, i.e., $n \rightarrow(n + 3)$. By using these  $\lambda_l$ in Eq.~(\ref{Eq:Ap5Q3}), we can easily see that the concurrence 
 vanishes  for each value of $n$. This is shown in the (Fig.(1)).
 
\section{Exact analytical solution for six-qubits case}
Consider  a six-qubit system confined into a seven-dimensional symmetric subspace. The eigenbasis of operator $\otimes_{l=1}^6\sigma_y$ in which $\mathcal{U}$  is block diagonal are given below:
\begin{eqnarray}
\ket{\phi_0^{\pm}}&=&\frac{1}{\sqrt{2}}(\ket{w_0} \mp \ket{\overline{w_0}}),\\
\ket{\phi_1^{\pm}}&=&\frac{1}{\sqrt{2}}(\ket{w_1} \pm  \ket{\overline{w_1}}),\\
\ket{\phi_2^{\pm}}&=&\frac{1}{\sqrt{2}}(\ket{w_2} \mp  \ket{\overline{w_2}}),\\
\ket{\phi_3^{+}}&=&\frac{1}{\sqrt{20}}\sum_\mathcal{P}\ket{000111} ,
\end{eqnarray}
where $\ket{w_0}=\ket{000000}$,
$\ket{\overline{w_0}}=\ket{111111}$, $\ket{w_1}=\frac{1}{\sqrt{6}} \sum_\mathcal{P} \ket{000001}_\mathcal{P}$, $\ket{\overline {w_1}}=\frac{1}{\sqrt{6}} \sum_\mathcal{P} \ket{011111}_\mathcal{P}$, $\ket{w_2}=\frac{1}{\sqrt{15}} \sum_\mathcal{P} \ket{000011}_\mathcal{P}$ and $\ket{\overline{w_2}}=\frac{1}{\sqrt{15}} \sum_\mathcal{P} \ket{001111}_\mathcal{P}$. The unitary operator $\mathcal{U}$ in this basis are diagonal in $\mathcal{U_+} \left( \mathcal{U_-}\right)$  having dimensions $4\times 4\left(3\times3\right)$ respectively. They are given as follows:
 \begin{equation}
\mathcal{U_+}= \frac{e^{\frac{ i \pi}{4}}}{2\sqrt{2}}\left(
\begin{array}{cccc}
 0 & -\sqrt{3} & 0 & -\sqrt{5} \\
i\sqrt{3} & 0 & i\sqrt{5}& 0 \\
 0 & -\sqrt{5} & 0 & \sqrt{3} \\
 i\sqrt{5} & 0 & -i\sqrt{3} & 0 \\
\end{array}
\right)\mbox{and}
 \end{equation}
 \begin{equation}
\mathcal{U_-}= \frac{e^{\frac{ i \pi}{4}}}{4}{\left(
\begin{array}{ccc}
  1 & 0 &  \sqrt{15}  \\
 0 & -4i & 0 \\
 \sqrt{15}  & 0 & - 1 \\
\end{array}
\right)}.
 \end{equation}
For $6$ qubit, the eigenvalues of the $\mathcal{U_+} \left( \mathcal{U_-}\right)$  is ${\left\{-1,-1,1,1\right\}}\left(\left\{-(-1)^{1/4},-(-1)^{3/4},(-1)^{1/4}\right\}\right)$  and the eigenvectors are  $\left\lbrace\left[\frac{2+2 i}{\sqrt{5}},\sqrt{\frac{3}{5}},0,1\right]^T, \left[\sqrt{\frac{3}{5}},\frac{2-2 i}{\sqrt{5}},1,0\right]^T, \left[-\frac{2+2 i}{\sqrt{5}},\sqrt{\frac{3}{5}},0,1\right]^T, \left[\sqrt{\frac{3}{5}},-\frac{2-2 i}{\sqrt{5}},1,0\right]^T\right\rbrace\left(\left\lbrace\left[-\sqrt{\frac{3}{5}},0,1\right]^T, \left[0,1,0 \right]^T, \left[\sqrt{\frac{5}{3}},0,1\right]^T\right\rbrace\right)$. The $n$th time evolution of the  blocks $\mathcal{U_{\pm}}$ is given as follows:
 \begin{equation}
\mathcal{U}_{+}^n= \frac{1}{8}{\left[
\begin{array}{cccc}
 4 \left(1+e^{i n \pi }\right) & {\sqrt{6}}~ e^{\frac{i  \pi }{4}} \left(-1+e^{i n \pi }\right) & 0 & {\sqrt{10}}~ e^{\frac{i  \pi }{4}} \left(-1+e^{i n \pi }\right) \\
 {\sqrt{6}}~ e^{-\frac{i  \pi }{4}} \left(-1+e^{i n \pi }\right) & 4 \left(1+e^{in \pi }\right) & {\sqrt{10}}~ e^{-\frac{i  \pi }{4}} \left(-1+e^{i n \pi }\right) & 0 \\
 0 & {\sqrt{10}}~ e^{\frac{i  \pi }{4}} \left(-1+e^{i n \pi }\right) &  4\left(1+e^{i n \pi }\right) & -{\sqrt{6}}~ e^{\frac{i  \pi }{4}} \left(-1+e^{i m \pi }\right) \\
\sqrt{10}~ e^{-\frac{i  \pi }{4}} \left(-1+e^{i n \pi } \right) & 0 & -\sqrt{6}~  e^{-\frac{i  \pi }{4}} \left(-1+e^{i n
\pi }\right) &  4\left(1+e^{i n \pi }\right) \\
\end{array}
\right]}~~ \mbox{and}
 \end{equation}
 \begin{equation}
 \mathcal{U}_{-}^n =\frac{e^{\frac{i n \pi }{4}}}{8}{\left[
\begin{array}{ccc}
 5 + 3~e^{i n \pi } & 0 &  \sqrt{15}\left( 1-e^{{ i n \pi }}\right) \\
 0 & 8 ~ e^{\frac{3 i m \pi }{2}} & 0 \\
 \sqrt{15}\left( 1-e^{{ i n \pi }}\right) & 0 & 3 + 5~ e^{i n \pi } \\
\end{array}
\right]}.
 \end{equation}
 The time periodicity of each element of $\mathcal{U}_{+}^n$ and  $\mathcal{U}_{-}^n$  is $2$ and $8$ respectively. Hence the unitary operator $\mathcal{U}^n$ for $6$ qubits has a periodicity of $8$.
 \subsection{Initial state $\otimes^6\ket{0}$= $\ket{\theta_0 =0,\phi_0 =0}$}
The state $\ket{\psi_n}$ can be calculated by $n$th implementations  of  unitary operator $\mathcal{U}$  on this initial state $\otimes^6\ket{0}$ as follows:
\begin{eqnarray}
\ket{\psi_n} =\mathcal{U}^n\ket{000000}&=& \frac{1}{\sqrt{2}}\left(\mathcal{U}_{+}^n\ket{\phi_0^+}+\mathcal{U}_{-}^n\ket{\phi_0^-}\right)\\ 
&=& \frac{1}{\sqrt{2}}\left(\bar{a}_1\ket{\phi_0^+}+\bar{a}_2\ket{\phi_1^+}+\bar{a}_3\ket{\phi_3^+}+\bar{a}_4\ket{\phi_0^-}+\bar{a}_5\ket{\phi_2^-}\right),
\end{eqnarray}
where the coefficients are, $\bar{a}_1=\frac{1}{2} \left(1+e^{i n \pi }\right)$,
 $\bar{a}_2 = \frac{\sqrt{6}}{8} e^{-\frac{i  \pi }{4}} \left(-1+e^{i n \pi }\right)$, $\bar{a}_3=\frac{\sqrt{10}}{8} e^{-\frac{i  \pi }{4}} \left(-1+e^{i n \pi }\right)$, 
 $\bar{a}_4=\frac{1}{8}e^{\frac{i n \pi }{4}}\left({5} +{3} e^{ i n \pi }\right)$ and
   $\bar{a}_5=\frac{\sqrt{15}}{8}e^{\frac{i n \pi }{4}}\left(1 - e^{ i n \pi }\right)$. \\
\subsubsection{\bf{The linear entropy}}
The  single qubit RDM $\rho_1(n)$ is given as follows:
\begin{equation}
\rho_1(n)=\frac{1}{4}{\left[
\begin{array}{cc}
 \ 2+\frac{1}{4} \cos\left(\frac{n \pi }{2}\right) \left(5 \cos\left(\frac{n \pi }{4}\right)+3 \cos\left(\frac{3
n\pi }{4}\right)\right) & G_n \\
 G_n &  \left(2-\cos\left(\frac{n \pi }{4}\right)-\frac{5}{8} \cos\left(\frac{3 n \pi }{4}\right)-\frac{3}{8} \cos\left(\frac{5
n \pi }{4}\right)\right) \\
\end{array}
\right]},
\end{equation}\\
where $G_n={\frac{-1}{{4
\sqrt{2}}} \sin\left(\frac{n \pi }{2}\right) \left[-5 \cos\left(\frac{n \pi }{4}\right)+3 \cos\left(\frac{3 n \pi }{4}\right)+5 \sin\left(\frac{n \pi }{4}\right)+3 \sin\left(\frac{3 n\pi }{4}\right)\right]}$\\

The  expression for linear entropy can be calculated as follows:
\begin{eqnarray}\nonumber
\label{Eq:Ap5Q4}
S_{(0,0)}^{(6)}(n)&=& \frac{1}{512}\left\lbrace\left[ 1 - \cos\left(\frac{n\pi}{2}\right) + \sin\left(\frac{n\pi}{2}\right)\right]^2 \left[284 +229\left( \cos\left(\frac{n\pi}{2}\right)- \sin\left(\frac{n\pi}{2}\right) \right) -96\sin(n\pi) \right.\right.\\  
   && \left.\left. - 9\left( \cos\left(\frac{3n\pi}{2}\right)+  \sin\left(\frac{3n\pi}{2}\right)\right)\right]\right\rbrace.
\end{eqnarray}
By observing (Fig.(1)) and Eq.~(\ref{Eq:Ap5Q4}), we can see that the entanglement dynamics of the system exhibit a periodic pattern. This periodicity is expressed as $S_{(0,0)}^{(6)}(n) = S_{(0,0)}^{(6)}(n+4)$. We also find that the values of linear entropy for consecutive odd \mbox{and} even values of $n$  are identical. It attains the maximum upper bound value i.e. $1/2$.
\subsubsection{\bf{Concurrence}}
The two-qubit RDM is  given as follows:
\begin{equation}
\rho_{12}(n)={\left(
\begin{array}{cccc}
 \tilde{b}_{m} &  \tilde{a}_{m}^* & \tilde{a}_{m}^* & \tilde{d}_{m}^* \\
 \tilde{a}_{m} & \frac{1}{4} \sin^2\left(\frac{n \pi }{2}\right) & \frac{1}{4} \sin^2\left(\frac{n \pi }{2}\right) & \tilde{a}_{m} \\
 \tilde{a}_{m} & \frac{1}{4} \sin^2\left(\frac{n \pi }{2}\right) & \frac{1}{4}  \sin^2\left(\frac{n \pi }{2}\right) & \tilde{a}_{m} \\
 \tilde{d}_{m} & \tilde{a}_{m}^* & \tilde{a}_{m}^*& \tilde{c}_{m}\\
\end{array}
\right)},
\end{equation}
 where  the coefficients are  expressed as  follows:
\begin{eqnarray}\nonumber
\tilde{b}_{m}&=&\frac{1}{32} \left[12+8 \cos\left(\frac{n \pi }{4}\right)+5 \cos\left(\frac{3 n \pi }{4}\right)+4 \cos(n \pi)+3 \cos\left(\frac{5 n \pi }{4}\right)\right],\\ \nonumber
\tilde{c}_{m}&=&\frac{1}{8} \sin^2\left(\frac{n\pi}{8}\right)\left[11+10 \cos\left(\frac{n \pi }{4}\right)+6 \cos\left(\frac{ n \pi }{2}\right)+2\cos\left(\frac{ 3n \pi }{4}\right)+3 \cos\left({ n \pi} \right)\right],\\ \nonumber
\tilde{d}_{m}&=&\frac{1}{32} \left[-4 \cos (n\pi )-i \left(4 i+\sin\left(\frac{3 n \pi }{4}\right)+\sin\left(\frac{5 n \pi }{4}\right)\right)\right] \mbox{and}\\ \nonumber
\tilde{a}_{m}&=&\frac{1}{64 \sqrt{2}}\left[-8 \cos\left(\frac{n \pi }{4}\right)+(5-i) \cos\left(\frac{3 n \pi }{4}\right)+(3+i) \cos\left(\frac{5n \pi }{4}\right)+8 \sin\left(\frac{n \pi }{4}\right)+(5+i) \sin\left(\frac{3 n \pi }{4}\right)\right.\\ &&\nonumber
\left.+(2+2 i) \sin (n\pi) -(3-i) \sin\left(\frac{5n \pi }{4}\right)\right].
\end{eqnarray}
 All these coefficients have periodic nature  of period $4$ i.e $\left(\tilde{b}_{m}(n),\tilde{a}_{m}(n),\tilde{d}_{m}(n),\tilde{c}_{m}(n)\right)=\left(\tilde{b}_{m}(n+4),\tilde{a}_{m}(n+4),\right.$\\$\left.\tilde{d}_{m}(n+4),\tilde{c}_{m}(n+4))\right)$. The  eigenvalues of $(\sigma_y \otimes \sigma_y)\rho_{12} (\sigma_y \otimes \sigma_y) \rho_{12}^*$ for  $n$ = $0$ to $8$  are $\left\lbrace(0, 0, 0, 0), \left(1/4, 1/4, 0, 0\right),\right.$\\$\left.\left(1/4, 1/4, 0, 0\right),(0, 0, 0, 0),(0, 0, 0, 0),\left(1/4, 1/4, 0, 0\right), \left(1/4, 1/4, 0, 0\right), (0, 0, 0, 0),(0, 0, 0, 0)\right\rbrace$. These eigenvalues repeat after every $n\rightarrow n+4$, which shows a periodicity of $4$. By using these eigenvalues in Eq.~(\ref{Eq:Ap5Q3}) we can easily find that the concurrence remains  zero for every value of $n$, which is shown in (Fig.(1)).
\subsection{Initial state $\ket{++++++}$= $\ket{\theta_0 =\pi/2,\phi_0 =-\pi/2}$}
The initial state  can be expressed as $\otimes ^6 {\ket{+}}_y=\frac{1}{4\sqrt{2}} \ket{\phi_0^+}+\frac{i\sqrt{3}}{4}\ket{\phi_1^+}- \frac{\sqrt{15}}{4\sqrt{2}} \ket{\phi_2^+}-\frac{i\sqrt{5}}{4}\ket{\phi_3^+}$. The state $\ket{\psi_n}$ can be calculated by the $n$ iterations  of unitary operator $\mathcal{U}$  on this initial state as follows:
\begin{eqnarray}
\ket{\psi_n}&=&\mathcal{U}_{+}^n\ket{++++++}\\  \nonumber
&=&e^{{\frac{ in \pi }{4}}}\left({\alpha}_{n}^\prime \ket{\phi_0^+}+{\beta}_{n}^\prime  \ket{\phi_1^+}+{\gamma}_{n}^\prime  \ket{\phi_2^+}+\zeta_n  \ket{\phi_3^+}\right).      
\end{eqnarray}
Here the coefficients are  given as follows:
\begin{eqnarray} \nonumber
 {\alpha}_{n}^\prime &=&\left[\left(\frac{1}{16}-\frac{i}{16}\right) \left(-1+e^{i n \pi }\right)+\frac{1+e^{i n\pi }}{8 \sqrt{2}}\right],  ~{\beta}_{n}^\prime =\left[\left(-\frac{1}{8}+\frac{i}{8}\right) \sqrt{\frac{3}{2}} \left(-1+e^{i n \pi }\right)+\frac{1}{8} i \sqrt{3} \left(1+e^{i n \pi }\right)\right],\\ \nonumber
{\gamma}_{n}^\prime &=&\left[\left(-\frac{1}{16}+\frac{i}{16}\right) \sqrt{15} \left(-1+e^{i n \pi }\right)-\frac{1}{8} \sqrt{\frac{15}{2}} \left(1+e^{i n \pi
}\right)\right] ~~\mbox{and}\\ \nonumber~ \zeta_n &=&\left[\left(\frac{1}{8}-\frac{i}{8}\right) \sqrt{\frac{5}{2}} \left(-1+e^{i n \pi }\right)-\frac{1}{8} i \sqrt{5} \left(1+e^{i n \pi }\right)\right]. \\ \nonumber
\end{eqnarray}

\subsubsection{\bf{The linear entropy}}
The single qubit RDM, $\rho_1(n)$, is given as follows:
\begin{equation}
\rho_1(n)=
{\frac{1}{2}\left[
\begin{array}{cc}
 1 & -\frac{i}{2} \left(1+\cos(n \pi)\right) \\
 \frac{i}{2}  \left(1+\cos(n \pi )\right) & 1 \\
\end{array}
\right]}.
\end{equation}
Thus, the expression for linear entropy is given as follows:
\begin{equation}\label{Eq:Ap6Q1}
S_{(\pi/2,-\pi/2)}^{(6)}(n)= \frac{1}{2}\left[1-\frac{1}{4} \left(1+\cos\left( n \pi \right)\right)^2\right]. 
\end{equation}
From the  Eq.({\ref{Eq:Ap6Q1}}), we can see that its entanglement dynamics is periodic in nature with 
$S_{(\pi/2,-\pi/2)}^{(6)}(n)=S_{(\pi/2,-\pi/2)}^{(6)}(n+2)$. It attains the maximum upper bound value i.e.  $1/2$  for odd values of 
$n$ and zero for even values of $n$. This is shown in the(Fig.(1)). 
\subsubsection{\bf{Concurrence}}
The two-qubit  RDM for this state  is given as follows:
\begin{equation}
\rho_{12}(n)={\frac{1}{4}\left[
\begin{array}{cccc}
 1 & -\frac{i}{2}  \left(1+\cos (n \pi )\right) & -\frac{i}{2}  \left(1+\cos (n \pi )\right) & -1 \\
 \frac{i}{2}  \left(1+\cos (n \pi )\right) & 1 & 1 & -\frac{i}{2} \left( 1+\cos (n \pi )\right) \\
 \frac{i}{2}  \left( 1+\cos (n \pi ) \right) & 1 & 1 & -\frac{i}{2} \left( 1+\cos (n \pi )\right) \\
 -1 & \frac{i}{2}  \left(1+\cos (n \pi )\right) & \frac{i}{2}  \left(1+\cos (n \pi )\right) & 1 \\
\end{array}
\right]}.
\end{equation}
 The eigenvalues $\lambda_l$  of $(\sigma_y \otimes \sigma_y)\rho_{12} (\sigma_y \otimes \sigma_y) \rho_{12}^*$ in decreasing order 
 are $\left\lbrace \frac{1}{8} \left(1 - \cos(n \pi)\right),\frac{1}{8} \left(1 -  \cos(n\pi)\right),0,0 \right\rbrace$. 
 From Eq.(\ref{Eq:Ap5Q3}), we can show that the concurrence remains zero for each value of $n$.
\section{Exact analytical solution for seven-qubits case}
For $7$ qubit the permutation symmetric basis is given as follows:
\begin{eqnarray}
\ket{\phi_0^{\pm}}&=&\frac{1}{\sqrt{2}}(\ket{w_0} \mp i \ket{\overline{w_0}}),\\
\ket{\phi_1^{\pm}}&=&\frac{1}{\sqrt{2}}(\ket{w_1} \pm i \ket{\overline{w_1}}),\\
\ket{\phi_2^{\pm}}&=&\frac{1}{\sqrt{2}}(\ket{w_2} \mp i \ket{\overline{w_2}}),\\
\ket{\phi_3^{\pm}}&=&\frac{1}{\sqrt{2}}(\ket{w_3} \pm i \ket{\overline{w_3}}).
\end{eqnarray}\\
All these states are eigenstate of parity operator such that
 $\otimes_{l=1}^7 \sigma_l^y \ket{\phi_j^{\pm}}=\pm \ket{\phi_j^{\pm}}$. Here 
 $\ket{w_0}=\ket{0000000}$,
 $\ket{\overline{w_0}}=\ket{1111111}$,
  $\ket{w_1}=\frac{1}{\sqrt{7}} \sum_\mathcal{P} \ket{0000001}_\mathcal{P}$,
   $\ket{\overline {w_1}}=\frac{1}{\sqrt{7}} \sum_\mathcal{P} \ket{0111111}_\mathcal{P}$, $\ket{w_2}=\frac{1}{\sqrt{21}} \sum_\mathcal{P} \ket{0000011}_\mathcal{P}$, $\ket{\overline{w_2}}=\frac{1}{\sqrt{21}} \sum_\mathcal{P}\ket{0011111}_\mathcal{P}$, $\ket{w_3}=\frac{1}{\sqrt{35}} \sum_\mathcal{P} \ket{0000111}_\mathcal{P}$ and $\ket{\overline{w_3}}=\frac{1}{\sqrt{35}} \sum_\mathcal{P} \ket{0001111}_\mathcal{P}$.
 The unitary operator is block diagonal in $\mathcal{U}_{\pm}$ having dimensions $4\times 4$ respectively. They are given as follows:
\begin{equation}
\mathcal{U}_{+} ={\frac{1}{8}\left(
\begin{array}{cccc}
 -1 & -i \sqrt{7} &  -\sqrt{21}  &- i \sqrt{35} 
\\
 -i \sqrt{7}  & - 5  & - 3i\sqrt{3}  & - \sqrt{5} \\
  \sqrt{21}  &  3 i \sqrt{3}  &  1 & -i\sqrt{15} 
\\
 i \sqrt{35}   &  \sqrt{5}  & -i \sqrt{15}  & -3  \\
\end{array}
\right)}.
\end{equation}
\begin{equation}
\mathcal{U}_{-} ={\frac{1}{8}\left(
\begin{array}{cccc}
 i &  \sqrt{7} &  i\sqrt{21}  & \sqrt{35} 
\\
  \sqrt{7}  &  5i  & 3\sqrt{3}  & i\sqrt{5} \\
  -i\sqrt{21}  &  -3 \sqrt{3}  &  -i & \sqrt{15} 
\\
 -\sqrt{35}   &  -i\sqrt{5}  & \sqrt{15}  & 3i  \\
\end{array}
\right)}.
\end{equation}
For $7$ qubit, the eigenvalues for $\mathcal{U_{+}}$($\mathcal{U_{-}}$) are 
$\left\lbrace-1,-1,e^{\frac{i \pi }{3}},e^{-\frac{i \pi }{3}}\right\rbrace$ $\left(i\left\lbrace1,1,e^{\frac{2i \pi }{3}},e^{-\frac{2i \pi }{3}}\right\rbrace\right)$ and  the eigenvectors are  ${\left\lbrace\left[ \pm i \sqrt{\frac{5}{7}},0,0,1\right]^T, \left[0,\pm i \sqrt{3},1,0 \right]^T,\left[\mp i \sqrt{\frac{7}{5}},\pm i \sqrt{\frac{3}{5}},-\frac{3}{\sqrt{5}},1 \right]^T, \left[\mp i \sqrt{\frac{7}{5}},\mp i \sqrt{\frac{3}{5}},\frac{3}{\sqrt{5}},1 \right]^T \right\rbrace}$. The $n$th time evolution of $\mathcal{U}_{\pm}$ is given as follows:

\begin{equation}
\mathcal{U}_{+}^n=
{e^{i n \pi }\left[
\begin{array}{cccc}
 \frac{1}{12} \left(5+7 \cos\left(\frac{2 n \pi }{3}\right)\right) & \frac{1}{4} i \sqrt{\frac{7}{3}} \sin\left(\frac{2 n \pi }{3}\right)
& \frac{1}{4} \sqrt{7} \sin\left(\frac{2 n \pi }{3}\right) & \frac{1}{6} i \sqrt{35} \sin^2\left(\frac{n \pi }{3}\right) \\
 \frac{1}{4} i \sqrt{\frac{7}{3}} \sin\left(\frac{2 n \pi }{3}\right) & \frac{1}{4} \left(3+\cos\left(\frac{2 n \pi }{3}\right)\right)
& \frac{1}{2} i \sqrt{3} \sin^2\left(\frac{n \pi }{3}\right) & \frac{1}{4} \sqrt{\frac{5}{3}} \sin\left(\frac{2 n \pi }{3}\right) \\
 -\frac{1}{4} \sqrt{7} \sin\left(\frac{2 n \pi }{3}\right) & -\frac{1}{2} i \sqrt{3} \sin^2\left(\frac{n \pi }{3}\right) & \frac{1}{4}
\left(1+3 \cos\left(\frac{2 n \pi }{3}\right)\right) & \frac{1}{4} i \sqrt{5} \sin\left(\frac{2 n \pi }{3}\right) \\
- \frac{1}{6} i \sqrt{35} \sin^2\left(\frac{n \pi }{3}\right) & -\frac{1}{4} \sqrt{\frac{5}{3}} \sin\left(\frac{2 n \pi }{3}\right) &
\frac{1}{4} i \sqrt{5} \sin\left(\frac{2 n \pi }{3}\right) & \frac{1}{12} \left(7+5 \cos\left(\frac{2 n \pi }{3}\right)\right) \\
\end{array}
\right]}\mbox{and}
\end{equation}
\begin{equation}
\mathcal{U}_{-}^n=
e^{\frac{in \pi }{2}}\left[
\begin{array}{cccc}
 \frac{1}{12} \left(5+7 \cos\left(\frac{2 n \pi }{3}\right)\right) & -\frac{i}{4}  \sqrt{\frac{7}{3}} \sin\left(\frac{2 n \pi }{3}\right)
& \frac{\sqrt{7}}{4}  \sin\left(\frac{2 n \pi }{3}\right) & -\frac{1}{6} i \sqrt{35} \sin^2\left(\frac{n \pi }{3}\right) \\
 -\frac{1}{4} i \sqrt{\frac{7}{3}} \sin\left(\frac{2 n \pi }{3}\right) & \frac{1}{4} \left(3+\cos\left(\frac{2 n \pi }{3}\right)\right)
& -\frac{1}{2} i \sqrt{3} \sin^2\left(\frac{n \pi }{3}\right) & \frac{1}{4} \sqrt{\frac{5}{3}} \sin\left(\frac{2 n \pi }{3}\right) \\
 -\frac{1}{4} \sqrt{7} \sin\left(\frac{2 n \pi }{3}\right) & \frac{1}{2} i \sqrt{3} \sin^2\left(\frac{n\pi }{3}\right) & \frac{1}{4}
\left(1+3 \cos\left(\frac{2 n \pi }{3}\right)\right) & -\frac{1}{4} i \sqrt{5} \sin\left(\frac{2 n\pi }{3}\right) \\
 \frac{1}{6} i \sqrt{35} \sin^2\left(\frac{n \pi }{3}\right) & -\frac{1}{4} \sqrt{\frac{5}{3}} \sin\left(\frac{2 n \pi }{3}\right) &
-\frac{1}{4} i \sqrt{5} \sin\left(\frac{2 n \pi }{3}\right) & \frac{1}{12} \left(7+5 \cos\left(\frac{2 n \pi }{3}\right)\right) \\
\end{array}
\right].
\end{equation}
Each element of the  $\mathcal{U}_{+}^n\left(\mathcal{U}_{-}^n\right)$  exhibits a time periodicity of $6(12)$. Therefore, the 
unitary operator $\mathcal{U}^n$ for a system of $7$ qubits shows a periodicity of $12$. 

\subsection{Initial state $\otimes^7\ket{0}$= $\ket{\theta_0 =0,\phi_0 =0}$}
The state $\ket{\psi_n}$ can be evaluated after $n$th iteration of $\mathcal{U}$ on this initial state  as follows:
 \begin{equation}
 \begin{split}
\ket{\psi_n}&=  \mathcal{U}^n |0000000\rangle =
  \frac{1}{\sqrt{2}} \mathcal{U}^n \left( |\phi_{0}^{+} \rangle + |\phi_{0}^{-} \rangle \right) \\ 
& = \frac{1}{\sqrt{2}} \left( \mathcal{U}_{+}^n |\phi_{0}^{+} \rangle + \mathcal{U}_{-}^n |\phi_{0}^{-} \rangle \right)\\
& =\frac{1}{2}e^{{\frac{ in \pi }{2}}}\left\lbrace(1+ i^n)\left(\bar{\alpha_n} \ket{w_0}+i\bar{\beta_n} \ket{\overline{w_1}}+ \bar{\gamma_n} \ket{w_2}+i\bar{\zeta_n} \ket{\overline{w_3}}\right)\right.\\  & \left. +(1-i^n)\left( i\bar{\alpha_n} \ket{\overline{w_0}}-\bar{\beta_n}\ket{w_1}+i \bar{\gamma_n} \ket{\overline{w_2}}-\bar{\zeta_n} \ket{w_3})\right) \right\rbrace,
 \end{split}
 \end{equation}
where the coefficients are:
 \begin{equation}\nonumber
 \bar{\alpha_n}={\frac{1}{12} \left[5+7 \cos\left(\frac{2 n \pi }{3}\right)\right]},~\bar{\beta_n}={\frac{i}{4}  \sqrt{\frac{7}{3}} \sin\left(\frac{2 n \pi }{3}\right)}, ~\bar{\gamma_n}={-\frac{\sqrt{7}}{4}  \sin\left(\frac{2 n \pi }{3}\right)}~\mbox{and}~~ \bar{\zeta_n}={-\frac{i \sqrt{35}}{6}  \sin^2\left(\frac{n \pi }{3}\right)}.
\end{equation}
\subsubsection{\bf{The linear entropy}}
For even $n=2m$, the RDM $\rho_1(2m)$ is diagonal and is given as follows:
 \begin{equation}
 \rho_1 (2m)={\left[
\begin{array}{cc}
 \frac{1}{18} \left(12+5 \cos\left(\frac{4 \pi  m}{3}\right)+\cos\left(\frac{2 \pi  m}{3}\right)\right) & 0 \\
 0 & \frac{1}{9} \left(7+2 \cos\left(\frac{4 \pi  m}{3}\right)\right) \sin^2\left(\frac{2\pi  m}{3}\right) \\
\end{array}
\right]}.
 \end{equation}
The expression  for the linear entropy for  even $n=2m$ is given as follows:
\begin{equation}\label{Eq:Ap5Q5}
 S_{(0,0}^{(7)}(2m)={\frac{1}{81} \left[7+2 \cos\left(\frac{4 m \pi }{3}\right)\right] \left[12+5 \cos\left(\frac{4 m \pi }{3}\right)+\cos\left(\frac{2m \pi }{3}\right)\right] \sin^2\left(\frac{2m \pi }{3}\right)}.
 \end{equation}
For odd $n$= $2m-1$, the RDM $\rho_1(2m-1)$ is not diagonal and is given as follows:
 \begin{equation}
 \rho_1 (2m-1)={\frac{1}{2}\left(
\begin{array}{cc}
 1 & W_n \\
 W_n & 1 \\
\end{array}
\right)},
 \end{equation}
where the value of $W_n$ is written as: 
 \begin{equation}\nonumber
 W_n=\frac{1}{9} \sin\left[\frac{\pi  (2m-1)}{3}\right] \left\lbrace-4 \sqrt{3} \cos\left[\frac{\pi(2m-1)}{3}\right]+\sqrt{3} \cos[\pi  (2m-1)]+6
\sin\left[\frac{\pi(2m-1)}{3}\right]+\sin[(2m-1)\pi]\right\rbrace.
\end{equation}
The expression of   linear entropy for odd  $n$= $2m-1$ is given as,
\begin{eqnarray}\label{Eq:Ap5Q6}
 S_{(0,0}^{(7)}(2m-1)&=&\frac{1}{2}(1-W_n^2)\\ \nonumber
&=& \frac{1}{2}\left\lbrace 1-\frac{1}{81} \sin^2\left(\frac{\pi  (2m-1)}{3}\right) \left[-4 \sqrt{3} \cos\left(\frac{\pi(2m-1)}{3}\right)+\sqrt{3} \cos(\pi
 (2m-1))\right.\right. \\ \nonumber && \left.\left. 
 +\sin((2m-1)\pi)+6 \sin\left(\frac{\pi(2m-1)}{3}\right)\right]^2\right\rbrace.
\end{eqnarray}
 It attains the maximum upper bound value of the linear entropy i.e. $1/2$. From the Eqs.~(\ref{Eq:Ap5Q5}) and (\ref{Eq:Ap5Q6}), 
 we can see that its entanglement dynamics is periodic in nature with $S_{(0,0)}^{(7)}(n)=S_{(0,0)}^{(7)}(n+6)$. We also find that 
 the linear entropy for consecutive odd \mbox{and} even values of $n$  remains same (which is shown in the (Fig.(1))
 i.e.
\begin{equation}
S_{(0,0)}^{(7)}(2m-1)=S_{(0,0)}^{(7)}(2m).
\end{equation} 
\subsubsection{\bf{Concurrence} }
For even $n=2m$, the two qubit RDM, $\rho_{12}(n)$, is an ``$X$ state" \cite{TingYu2007} and is  given as follows:
\begin{equation}
\rho_{12}(2m)={\left(
\begin{array}{cccc}
 s_n& 0 & 0 & y_n \\
 0 & l_n & l_n & 0 \\
 0 & l_n & l_n & 0 \\
 y_n & 0 & 0 & x_n \\
\end{array}
\right)},
\end{equation}
where the coefficients are:
\begin{eqnarray}\nonumber
s_n&=&\frac{1}{12} \left[6+5 \cos\left(\frac{4m \pi  }{3}\right)+\cos\left(\frac{2m \pi  }{3}\right)\right], ~ y_n=\frac{1}{{12 \sqrt{3}}}\left[-5 \sin\left(\frac{4m
\pi  }{3}\right)+\sin\left(\frac{2m \pi  }{3}\right)\right],\\ \nonumber
x_n&=&\frac{1}{18} \left[7+2 \cos\left(\frac{4m
\pi  }{3}\right)\right] \sin^2\left(\frac{2m\pi  }{3}\right)~ \mbox{and}~ l_n=\frac{1}{18} \left[7+2 \cos\left(\frac{4m \pi  }{3}\right)\right] \sin^2\left(\frac{2m\pi  }{3}\right).
\end{eqnarray}
Time periodicity of these coefficients is three i.e. $\left( s_n(m),y_n(m),l_n(m),x_n(m)\right)=\left( s_{n}(m+3),y_{n}(m+3),\right.$\\ $\left.
l_{n}(m+3),x_{n}(m+3)\right)$.
For odd $n=2m-1$,  the two-qubit RDM $\rho_{12}(n)$ is given as follows:
\begin{equation}
\rho_{12}(2m-1)=\frac{1}{2}{\left(
\begin{array}{cccc}
 \tilde{s}_n & \tilde{y}_n & \tilde{y}_n & 2\tilde{x}_n \\
 \tilde{y}_n & 2\tilde{l}_n& 2\tilde{l}_n & \tilde{y}_n \\
 \tilde{y}_n & 2\tilde{l}_n & 2\tilde{l}_n& \tilde{y}_n \\
 2\tilde{x}_n & \tilde{y}_n & \tilde{y}_n & \tilde{s}_n \\
\end{array}
\right)},
\end{equation}
where the coefficients are
\begin{eqnarray}\nonumber 
\tilde{l}_n&=&\frac{1}{18} \left[7+2 \cos\left(\frac{2(2m-1) \pi  }{3}\right)\right] \sin^2\left(\frac{(2m-1)\pi  }{3}\right),
\tilde{x}_n=\frac{1}{{12 \sqrt{3}}}\left[-5 \sin\left(\frac{2(2m-1)
\pi  }{3}\right)+\sin\left(\frac{4(2m-1)\pi  }{3}\right)\right],\\ \nonumber
\tilde{y}_n& =&\frac{1}{18} \sin\left(\frac{(2m-1)\pi  }{3}\right) \left[-4 \sqrt{3} \cos\left(\frac{(2m-1)\pi  }{3}\right)+\sqrt{3} \cos((2m-1)\pi  )+6
\sin\left(\frac{(2m-1)\pi  }{3}\right)+\sin((2m-1)\pi)\right] \mbox{and}\\ \nonumber
\tilde{s}_n&=&\frac{1}{18} \left[12+5 \cos\left(\frac{2(2m-1)
\pi  }{3}\right)+\cos\left(\frac{4(2m-1) \pi  }{3}\right)\right].
\end{eqnarray} 
 Time periodicity of these coefficients is three i.e. $\left(\tilde{l}_n (m),\tilde{s}_n (m),\tilde{y}_n (m),\tilde{x}_n (m)\right)=\left( \tilde{l}_n (m+3),\tilde{s}_n (m+3),\right.$\\$\left.\tilde{y}_n (m+3),\tilde{x}_n (m+3)\right)$. The eigenvalues $ \lambda_ l$ of $(\sigma_y \otimes \sigma_y)\rho_{12} (\sigma_y \otimes \sigma_y) \rho_{12}^*$ for  $n$ = $0$ to $12$ in decreasing order are $\left\lbrace(0, 0, 0, 0), \left(1/4, 1/4, 0, 0\right),\left(1/4, 1/4, 0, 0\right),\left(1/4, 1/4, 0, 0\right) \left(1/4, 1/4, 0, 0\right), (0, 0, 0, 0),(0, 0, 0, 0),\left(1/4, 1/4, 0, 0\right),\right.$\\$\left.\left(1/4, 1/4, 0, 0\right), \left(1/4, 1/4, 0, 0\right),\left(1/4, 1/4, 0, 0\right), (0, 0, 0, 0),(0, 0, 0, 0)\right\rbrace$. These eigenvalues repeat after every $n\rightarrow n+6$. By using these  $\lambda_l$ in Eq.~(\ref{Eq:Ap5Q3}), we can easily see that the concurrence remains  zero for every value of $n$, which is shown in (Fig.(1)).
\subsection{Initial state $\otimes^7\ket{+}$= $\ket{\theta_0 =\pi/2,\phi_0 =-\pi/2}$}
 The initial state  can be expressed as $\otimes ^7 {\ket{+}}_y=\frac{1}{8} \ket{\phi_0^+}+\frac{i\sqrt{7}}{8}\ket{\phi_1^+}- \frac{\sqrt{21}}{8} \ket{\phi_2^+}- i\frac{\sqrt{35}}{8} \ket{\phi_3^+}$. The state $\ket{\psi_n}$ can be obtained by $n$th time evolution of unitary operator $\mathcal{U}$  on this initial state. Thus,
\begin{equation}
\begin{split}
\ket{\psi_n}&=\mathcal{U}_{+}^n \ket{+++++++} \\ 
&= e^{ in \pi }(\bar{b}_{1} \ket{\phi_0^+}+\bar{b}_{2} \ket{\phi_1^+}+ \bar{b}_{3} \ket{\phi_2^+}+\bar{b}_{4}\ket{\phi_3^+}).
\end{split}
\end{equation}
Here  the coefficients are as follows:
 \begin{eqnarray}\nonumber
 \bar{b}_{1}&=&\frac{1}{24} \left[10-7 \cos\left(\frac{2 n \pi }{3}\right)-7 \sqrt{3} \sin\left(\frac{2 n \pi }{3}\right)\right], \bar{b}_{2}=\frac{i \sqrt{7}}{24}  \left[3 \cos\left(\frac{2 n \pi }{3}\right)-\sqrt{3} \sin\left(\frac{2 n\pi }{3}\right)\right]\\
 \nonumber
\bar{b}_{3}&=&\frac{\sqrt{7}}{8}  \left[-\sqrt{3} \cos\left(\frac{2 n \pi }{3}\right)+\sin\left(\frac{2 n \pi }{3}\right)\right]\mbox{and} \;\; \bar{b}_{4}=-\frac{i \sqrt{35} }{24} \left[2+\cos\left(\frac{2 n \pi }{3}\right)+\sqrt{3} \sin\left(\frac{2 n \pi }{3}\right)\right].
 \end{eqnarray}
 \subsubsection{\bf{The linear entropy}}
 The single qubit RDM for this state is given as follows:
\begin{equation}
\rho_1(n)=\left[
\begin{array}{cc}
 1 & -\frac{i}{9}  \left(3+5 \cos\left(\frac{2 n \pi }{3}\right)+\cos\left(\frac{4 n \pi }{3}\right)\right) \\
 \frac{i}{9}  \left(3+5 \cos\left(\frac{2 n\pi }{3}\right)+\cos\left(\frac{4 n\pi }{3}\right)\right) & 1 \\
\end{array}
\right].
\end{equation}
Thus, the expression for  linear entropy is  given as follows:
\begin{equation}\label{Eq:Ap7Q1}
S_{(\pi/2,-\pi/2)}^{(7)}(n)={\frac{1}{2}\left[ 1-\frac{1}{81} \left(3+5 \cos\left(\frac{2 n \pi }{3}\right)+\cos\left(\frac{4 n \pi }{3}\right)\right)^2 \right]}.
\end{equation}
From Eq.({\ref{Eq:Ap7Q1}}), we can see that the linear entropy attains the maximum upper bound value i.e.  
$1/2$. We also find that its entanglement dynamics is periodic in nature with $S_{(\pi/2,-\pi/2)}^{(7)}(n)=
S_{(\pi/2,-\pi/2)}^{(7)}(n+3)$. This is shown in the (Fig.(1)).

\subsubsection{\bf{Concurrence}}
 The two qubit RDM $\rho_{12}(n)$ for this state is given as follows:
\begin{equation}
\rho_{12}(n)={\left(
\begin{array}{cccc}
 f_m & e_m^* & e_m^* & h_m \\
 e_m & g_m & g_m & e_m ^*\\
 e_m & g_m & g_m & e_m^* \\
 h_m & e_m & e_m & f_m \\
\end{array}
\right)}.
\end{equation}
Here the coefficients are as follows:
\begin{eqnarray}\nonumber
h_m&=&\frac{1}{72} \left[-15 \cos\left(\frac{2 n \pi }{3}\right)-3 \cos\left(\frac{4 n \pi }{3}\right)-\sqrt{3} \left(-5
\sin\left(\frac{2 n \pi }{3}\right)+\sin\left(\frac{4 n \pi }{3}\right)\right)\right],\\ \nonumber
g_m&=&\frac{1}{72} \left[12+5 \cos\left(\frac{2 n \pi }{3}\right)+\cos \left(\frac{4 n \pi }{3}\right)+5 \sqrt{3} \sin\left(\frac{2
n \pi }{3}\right)-\sqrt{3} \sin\left(\frac{4 n \pi }{3}\right)\right],\\ \nonumber 
e_m&=&\frac{i}{36}  \left[3+5 \cos\left(\frac{2 n \pi }{3}\right)+\cos\left(\frac{4 n \pi }{3}\right)\right] \mbox{and} ~~f_m=-\frac{1}{72} \left[-5+2 \cos\left(\frac{2 n \pi }{3}\right)\right] \left[5+\cos\left(\frac{2 n \pi }{3}\right)-\sqrt{3}
\sin\left(\frac{2 n \pi }{3}\right)\right].
\end{eqnarray}
Time periodicity of these coefficients is $3$ i.e. $\left( h_m(n),f_m(n),g_m(n),e_m(n)\right)=\left( h_{m}(n+3),f_{m}(n+3),
g_{m}(n+3)\right.$\\$\left.,e_{m}(n+3)\right)$. The eigenvalues $\lambda_l$ of $(\sigma_y \otimes \sigma_y)\rho_{12} (\sigma_y 
\otimes \sigma_y)$ in decreasing order  for  $n$ = $0$ to $6$ are $\left\lbrace(0, 0, 0, 0), \left(1/4, 1/4, 0, 0\right),
\left(1/4, 1/4, 0, 0\right),(0, 0, 0, 0),\left(1/4, 1/4, 0, 0\right), \left(1/4, 1/4, 0, 0\right), (0, 0, 0, 0)\right\rbrace$. 
These eigenvalues repeat after every  $n\rightarrow n+3$, which shows a periodicity of $3$.  By using these  $\lambda_l$ in 
Eq.~(\ref{Eq:Ap5Q3}), we can easily see that the concurrence remains  zero for every value of $n$, which is shown in  
(Fig.(1)).
\section{Exact analytical solution for eight-qubits case}
Here the  eigenbasis for this case  is given as follows:
\begin{eqnarray}
\ket{\phi_0^{\pm}}&=&\frac{1}{\sqrt{2}}(\ket{w_0} \pm \ket{\overline{w_0}}),\\
\ket{\phi_1^{\pm}}&=&\frac{1}{\sqrt{2}}(\ket{w_1} \mp \ket{\overline{w_1}}),\\
\ket{\phi_2^{\pm}}&=&\frac{1}{\sqrt{2}}(\ket{w_2} \pm  \ket{\overline{w_2}}),\\
\ket{\phi_3^{\pm}}&=&\frac{1}{\sqrt{2}}(\ket{w_3} \mp  \ket{\overline{w_3}}),\\
\ket{\phi_4^{+}}&=&\frac{1}{\sqrt{70}}\sum_{\mathcal{P}}\ket{00001111}, 
\end{eqnarray}
where $\ket{w_0}=\ket{00000000}$,~$\ket{\overline{w_0}}=\ket{11111111}$, $\ket{w_1}=\frac{1}{\sqrt{8}}\sum_\mathcal{P} \ket{00000001}_\mathcal{P}$, $\ket{\overline {w_1}}=\frac{1}{\sqrt{8}}\sum_\mathcal{P}\ket{01111111}_\mathcal{P}$,~$\ket{w_2}=\frac{1}{\sqrt{28}}\sum_\mathcal{P} \ket{00000011}_\mathcal{P}$,$\ket{\overline{w_2}}=\frac{1}{\sqrt{28}}\sum_\mathcal{P}\ket{00111111}_\mathcal{P}$,
$\ket{w_3}=\frac{1}{\sqrt{56}}\sum_\mathcal{P} \ket{00000111}_\mathcal{P}$ and $\ket{\overline{w_3}}=\frac{1}{\sqrt{56}} \sum_\mathcal{P} \ket{00011111}_\mathcal{P}$. The unitary operator $\mathcal{U}$ in this basis is block diagonal in $\mathcal{U_+}$ and $\mathcal{U_-}$ having dimensions $5\times 5$ and  $4\times4$ respectively such that:
\begin{equation}
\mathcal{U}_{+}={\frac{1}{8}\left(
\begin{array}{ccccc}
 -1& 0 & -2\sqrt{7}  & 0 & -\sqrt{35}  \\
 0 & -6i & 0 & -2i \sqrt{7}  & 0 \\
-2 \sqrt{7}  & 0 & -4 & 0 & 2\sqrt{5}  \\
 0 & -2i \sqrt{7}  & 0 & 6i & 0 \\
 - \sqrt{35}  & 0 & 2 \sqrt{5}  & 0 & -3 \\
\end{array}
\right)}~ \mbox{and}
\end{equation}
\begin{equation}
\mathcal{U}_{-}={\frac{1}{2 \sqrt{2}}\left(
\begin{array}{cccc}
 0 & 1 & 0 & \sqrt{7}  \\
i & 0 & i\sqrt{7}  & 0 \\
 0 & \sqrt{7}  & 0 & -1 \\
 i \sqrt{7}  & 0 & -i & 0 \\
\end{array}
\right)}.
\end{equation}
For $8$ qubit the eigenvalues of $\mathcal{U_{+}}$ $\left(\mathcal{U_{-}}\right)$ are $\left\lbrace-1,-1,i,-i,1\right\rbrace$ $\left(e^{\frac{i \pi }{4}}\left\lbrace-1,-1,1,1\right\rbrace\right)$ and the eigenvectors are\\
 $\left\lbrace\left[\sqrt{\frac{5}{7}},0,0,0,1 \right]^T, \left[\frac{2}{\sqrt{7}},0,1,0,0 \right]^T,\left[0, -\frac{1}{\sqrt{7}},0,1,0 \right]^T,\left[0,\sqrt{7},0,1,0 \right]^T,\left[-\sqrt{\frac{7}{5}},0,\frac{2}{\sqrt{5}},0,1 \right]^T \right\rbrace \left(\left\lbrace\left[-\frac{2-2 i}{\sqrt{7}},\frac{1}{\sqrt{7}},0,1 \right]^T\right.\right.$\\ $\left.\left.,\left[\frac{1}{\sqrt{7}},-\frac{2+2 i}{\sqrt{7}},1,0 \right]^T,\left[\frac{2-2 i}{\sqrt{7}},\frac{1}{\sqrt{7}},0,1 \right]^T,\left[\frac{1}{\sqrt{7}},\frac{2+2 i}{\sqrt{7}},1,0 \right]^T\right\rbrace \right)$. The $n$th time evolution of $\mathcal{U}_{+}$ and $\mathcal{U}_{-}$ is given as follows:
\begin{equation}
 \mathcal{U}_{+}^n=\frac{1}{16}{\left(
\begin{array}{ccccc}
  \left(7+9 e^{i n \pi }\right) & 0 & 2 \sqrt{7} \left(-1+e^{i n \pi }\right) & 0 &\sqrt{35} \left(-1+e^{i n\pi }\right) \\
 0 & 2~ e^{\frac{i n \pi }{2}} \left(1+7 e^{i n \pi }\right) & 0 & 2 \sqrt{7} e^{\frac{i n \pi }{2}} \left(-1+e^{i n \pi }\right)
& 0 \\
 2 \sqrt{7} \left(-1+e^{i n \pi }\right) & 0 & 4 \left(1+3 e^{i n \pi }\right) & 0 & -2\sqrt{5} \left(-1+e^{i n \pi
}\right) \\
 0 & 2 \sqrt{7} e^{\frac{i n \pi }{2}} \left(-1+e^{i n \pi }\right) & 0 & 2~ e^{\frac{i n \pi }{2}} \left(7+e^{i n \pi }\right)
& 0 \\
  \sqrt{35} \left(-1+e^{i n \pi }\right) & 0 & -2 \sqrt{5} \left(-1+e^{i n \pi }\right) & 0 &  \left(5+11 e^{i
n \pi }\right) \\
\end{array}
\right)} \mbox{and}
\end{equation}
\begin{equation}
 \mathcal{U}_{-}^n=\frac{e^{\frac{i n \pi }{4}}}{2}{\left(
\begin{array}{cccc}
   \left(1+e^{i n \pi }\right) & -\frac{e^{\frac{-i n \pi}{4}}}{2\sqrt{2}} \left(-1+e^{i n \pi
}\right) & 0 &- \frac{e^{-\frac{i n \pi}{4}}}{2\sqrt{2}}\sqrt{7}  \left(-1+e^{i n \pi }\right) \\
- \frac{e^{\frac{i n \pi}{4}}}{2\sqrt{2}} \left(-1+e^{i n \pi }\right) &  \left(1+e^{i n \pi
}\right) & -\frac{e^{\frac{i n \pi}{4}}}{2\sqrt{2}} \sqrt{7}  \left(-1+e^{i n \pi }\right) & 0 \\
 0 &-\frac{e^{\frac{-i n \pi}{4}}}{2\sqrt{2}} \sqrt{7}  \left(-1+e^{i n \pi }\right) & \left(1+e^{i
n \pi }\right) & \frac{e^{\frac{-i n \pi}{4}}}{2\sqrt{2}}  \left(-1+e^{i n \pi }\right) \\
 -\frac{e^{\frac{i n \pi}{4}}}{2\sqrt{2}} \sqrt{7}  \left(-1+e^{i n \pi }\right) & 0 & \frac{e^{\frac{i n \pi}{4}}}{2\sqrt{2}}  \left(-1+e^{i n \pi }\right) &  \left(1+e^{i n \pi }\right) \\
\end{array}
\right)}.
\end{equation}
The time periodicity  of $\mathcal{U}_{+}^n\left(\mathcal{U}_{-}^n\right)$ is $4(8)$. Hence the time periodicity of $\mathcal{U}^n$ is $8$. 
\subsection{Initial state $\otimes^8\ket{0}$= $\ket{\theta_0 =0,\phi_0 =0}$}
The state $\ket{\psi_n}$ can be calculated by $n$th time evolution of unitary operator $\mathcal{U}$ on the initial state $\otimes^8\ket{0}$ as follows:
 \begin{eqnarray}
\ket{\psi_n}&=&\mathcal{U}^{n}\ket{00000000}= \frac{1}{\sqrt{2}}(\mathcal{U}_{+}^n\ket{\phi_0^+}+\mathcal{U}_{-}^n\ket{\phi_0^-}) \\  \nonumber
&=&\frac{1}{\sqrt{2}}\left( {b}_1^\prime\ket{\phi_0^+}+{b}_2^\prime\ket{\phi_2^+}+{b}_3^\prime\ket{\phi_4^+}+{b}_4^\prime\ket{\phi_0^-}+{b}_5^\prime\ket{\phi_1^-}+{b}_6^\prime\ket{\phi_3^-}\right),
\end{eqnarray}
here the coefficients are as follows:
\begin{eqnarray}\nonumber
{b}_1^\prime&=&\frac{1}{16} \left(7+9 e^{i n \pi }\right),\;\; {b}_2^\prime=\frac{\sqrt{7}}{8}  \left(-1+e^{i n \pi }\right), ~{b}_3^\prime=\frac{\sqrt{35}}{16}  \left(-1+e^{i n \pi },\right),~~ {b}_4^\prime =\frac{e^{\frac{i n \pi }{4}}}{2}  \left(1+e^{i n \pi }\right),\\  \nonumber {b}_5^\prime &=&\left(-\frac{1}{8}-\frac{i}{8}\right) e^{\frac{i n \pi }{4}} \left(-1+e^{i n \pi }\right)~\mbox{and}~~ {b}_6^\prime=\left(-\frac{1}{8}-\frac{i}{8}\right) \sqrt{7} e^{\frac{i n \pi }{4}} \left(-1+e^{i n\pi }\right).
\end{eqnarray}
\subsubsection{\bf{The linear entropy}}
 The single qubit RDM, $\rho_1(n)$, is given as follows:
\begin{equation}
\rho_1(n)={\frac{1}{4}\left[
\begin{array}{cc}
  \left(2+\frac{1}{8} \cos\left(\frac{n\pi }{2}\right) \left(9 \cos\left(\frac{n \pi }{4}\right)+7 \cos\left(\frac{3 n \pi }{4}\right)\right)\right)
& Q_n \\
 Q_n &  \left(2-\frac{1}{8} \cos\left(\frac{n \pi }{2}\right) \left(9 \cos\left(\frac{n \pi }{4}\right)+7 \cos\left(\frac{3 n \pi
}{4}\right)\right)\right) \\
\end{array}
\right]},
\end{equation}
where $Q_n=\frac{1}{8\sqrt{2}}\sin\left(\frac{n\pi}{2}\right) \left[9 \cos\left(\frac{n\pi}{4}\right) - 7 \cos\left(\frac{3n\pi}{4}\right) - 
   9 \sin\left(\frac{n\pi}{4}\right) - 7 \sin\left(\frac{3n\pi}{4}\right)\right]$. \\

 The expression for linear entropy is given as:
 \begin{eqnarray}\nonumber\label{Eq:Ap58Q1}
  S_{(0,0)}^{8}(n)&=&\frac{1}{2048}\left\lbrace\left[ 1 - \cos\left(\frac{n\pi}{2}\right) + \sin\left(\frac{n\pi}{2}\right)\right]^2 \left[1212 +1005 \left(\cos\left(\frac{n\pi}{2}\right)-\sin\left(\frac{n\pi}{2}\right)\right)-448\sin(n\pi) 
    \right.\right.\\ 
   && \left.\left. - 49 \left(\cos\left(\frac{3n\pi}{2}\right)+ \sin\left(\frac{3n\pi}{2}\right)\right)\right]\right\rbrace.
 \end{eqnarray}

From the Eq.~(\ref{Eq:Ap58Q1}) we can see that the entanglement dynamics is periodic in nature with 
$S_{(0,0)}^{(8)}(n)=S_{(0,0)}^{(8)}(n+4)$. We also find that the linear entropy for consecutive odd and even values of $n$ 
remain the same. This is shown in the (Fig.(1)).
\subsubsection{\bf{Concurrence}}
 The two-qubit RDM, $\rho_{12}(n)$, is given as follows:
\begin{equation}
 \rho_{12}(n)={\left[
\begin{array}{cccc}
 \tilde{g}_n & \tilde{h}_n & \tilde{h}_n & \tilde{f}_n^* \\
 \tilde{h}_n^* & \frac{1}{4} \sin^2\left(\frac{n \pi }{2}\right) & \frac{1}{4} \sin^2\left(\frac{n \pi }{2}\right) & \tilde{h}_n^* \\
 \tilde{h}_n^*  & \frac{1}{4} \sin^2\left(\frac{n\pi }{2}\right)& \frac{1}{4} \sin^2\left(\frac{n \pi }{2}\right) & \tilde{h}_n^*  \\
 \tilde{f}_n & \tilde{h}_n & \tilde{h}_n & \tilde{e}_n \\
\end{array}
\right]}.
\end{equation}
Here  the coefficients are:
\begin{eqnarray}\nonumber \label{Eq:Ap8Q3}
\tilde{g}_n&=&\frac{1}{64} \left[24+16 \cos\left(\frac{n \pi }{4}\right)+9 \cos\left(\frac{3 n \pi }{4}\right)+8 \cos (n\pi)+7 \cos\left(\frac{5 n \pi }{4}\right)\right],\\ \nonumber
\tilde{f}_n&=&\frac{1}{64} \left[-8 \cos (n\pi )+i \left(-8 i+\sin\left(\frac{3 n \pi }{4}\right)+\sin\left(\frac{5 n \pi }{4}\right)\right)\right],\\ \nonumber
\tilde{e}_n&=&\frac{1}{16} \sin^2\left(\frac{n\pi}{8}\right)\left[23+22 \cos\left(\frac{n \pi }{4}\right)+6 \cos\left(\frac{3 n \pi }{4}\right)+7 \cos (n\pi
)+14\cos\left(\frac{ n \pi }{2}\right)\right]\mbox{and}\\ \nonumber
\tilde{h}_n&=&\frac{1}{{128 \sqrt{2}}}\left[  -16 \cos\left(\frac{n \pi }{4}\right)+(9-i) \cos\left(\frac{3 n \pi }{4}\right)+(7+i) \cos\left(\frac{5
n \pi }{4}\right)+16 \sin\left(\frac{n \pi }{4}\right)+(9+i) \sin\left(\frac{3 n\pi }{4}\right)\right.\\ \nonumber &&\left.(2+2 i)\sin(n\pi) -(7-i) \sin\left(\frac{5
n \pi }{4}\right)\right].
\end{eqnarray}
The time period of all the coefficients is $4$. The eigenvalues $\lambda_l$ in decreasing order of $(\sigma_y \otimes \sigma_y)\rho_{12} (\sigma_y \otimes \sigma_y) \rho_{12}^*$ for  $n$ = $0$ to $8$  are $\left\lbrace(0, 0, 0, 0), \left(1/4, 1/4, 0, 0\right),\left(1/4, 1/4, 0, 0\right),(0, 0, 0, 0),(0, 0, 0, 0),\left(1/4, 1/4, 0, 0\right), \left(1/4, 1/4, 0, 0\right), (0, 0, 0, 0),(0, 0, 0, 0)\right\rbrace$. These eigenvalues repeat after every $n\rightarrow n+4$
, which shows the same periodicity as that of the coefficients i.e. $4$. By using these  $\lambda_l$ in Eq.~(\ref{Eq:Ap5Q3}), 
we can easily see that the concurrence remains  zero for every value of $n$. This is shown in the(Fig.(1)).
\subsection{Initial state $\otimes^8\ket{+}$= $\ket{\theta_0 =\pi/2,\phi_0 =-\pi/2}$}
The initial state can be  expressed as $\otimes ^8 {\ket{+}}_y=\frac{1}{8\sqrt{2}} \ket{\phi_0^+}+\frac{i}{4}\ket{\phi_1^+}- \frac{\sqrt{14}}{8} \ket{\phi_2^+}-\frac{i\sqrt{7}}{4}\ket{\phi_3^+}+\frac{\sqrt{70}}{16}\ket{\phi_4^+}$. The state $\ket{\psi_n}$ can be obtained by $n$th time evolution of unitary operator $\mathcal{U}$ on this initial state. Thus,
\begin{eqnarray}\nonumber
\ket{\psi_n}&=&\mathcal{U}_{+}^n\ket{++++++++}\\ 
&=& {c}_{1}^\prime \ket{\phi_0^+}+ {c}_{2}^\prime \ket{\phi_1^+} +  {c}_{3}^\prime \ket{\phi_2^+}+ {c}_{4}^\prime\ket{\phi_3^+}+ {c}_{5}^\prime\ket{\phi_4^+},
\end{eqnarray}
where $ {c}_{1}^\prime=\frac{1}{8 \sqrt{2}}~ e^{i n \pi }$, ${c}_{2}^\prime=\frac{i}{4}  \; e^{\frac{i n\pi }{2}}$,
$ {c}_{3}^\prime=-\frac{1}{4} \sqrt{\frac{7}{2}}\; e^{i n\pi }$, $ {c}_{4}^\prime=-\frac{i \sqrt{7}}{4} ~ e^{\frac{i n \pi }{2}}$ 
\mbox{and} $ {c}_{5}^\prime=\frac{1}{8} \sqrt{\frac{35}{2}}~ e^{i n \pi }$.\\
\subsubsection{\bf{The linear entropy}}
 The single qubit RDM, $\rho_1(n)$, for this state is given as follows:
\begin{equation}
 \rho_1(n)={\frac{1}{2}\left[
\begin{array}{cc}
 1 & -i \cos\left(\frac{n \pi }{2}\right) \\
 i \cos\left(\frac{n \pi }{2}\right) & 1 \\
\end{array}
\right]}.
\end{equation}
The expression for linear entropy is given as follows:
\begin{equation}\label{Eq:8Ap5}
S_{(\pi/2,-\pi/2)}^{(8)}(n)={\frac{1}{2}\left[1-\cos^2\left(\frac{n \pi }{2}\right)\right]}.
\end{equation}
From the Eq.({\ref{Eq:8Ap5}) we can see that the linear entropy attains the maximum upper bound value i.e. $1/2$ for odd values of 
$n$ and zero for even values of $n$. We also find that its entanglement dynamics is periodic in nature with 
$S_{(\pi/2,-\pi/2)}^{(8)}(n)=S_{(\pi/2,-\pi/2)}^{(8)}(n+2)$. This is shown in the (Fig.(1)).
\subsubsection{\bf{Concurrence}}
 The two qubit RDM, $\rho_{12}(n)$, for this state is given as follows:
\begin{equation}
\rho_{12}={\frac{1}{4}\left[
\begin{array}{cccc}
 1 & - i \cos\left(\frac{n \pi }{2}\right) & - i \cos\left(\frac{n \pi }{2}\right) & -1 \\
  i \cos\left(\frac{n \pi }{2}\right) & 1 & 1 & - i \cos\left(\frac{n \pi }{2}\right) \\
  i \cos\left(\frac{n \pi }{2}\right) & 1 & 1 & - i \cos\left(\frac{n \pi }{2}\right) \\
 -1 &  i \cos\left(\frac{n \pi }{2}\right) &  i \cos\left(\frac{n \pi }{2}\right) & 1 \\
\end{array}
\right]}.
\end{equation}
 The eigenvalues $\lambda_l$  of $(\sigma_y \otimes \sigma_y)\rho_{12} (\sigma_y \otimes \sigma_y) \rho_{12}^*$ in increasing order 
 are $\left\lbrace 0,0,\frac{1}{8} \left(1 - \cos(n \pi)\right),\frac{1}{8} \left(1 - \cos(n\pi)\right) \right\rbrace$. Here we can 
 see that the two eigenvalues are zero and the other two are also same. From the Eq.(\ref{Eq:Ap5Q3}), it can be seen that the 
 concurrence remains zero for every value of $n$. This is shown in the (Fig.(1)).

 \section{Exact analytical solution for nine-qubits case}
The eigenvectors of $ \otimes_{l=1}^9 \sigma_l^y$ operator with eigenvalues $\pm1$
are given by,
\begin{eqnarray}
\ket{\phi_1^{\pm}}&=&\frac{1}{\sqrt{2}}(\ket{w_0} \pm i \ket{\overline{w_0}}),\\
\ket{\phi_2^{\pm}}&=&\frac{1}{\sqrt{2}}(\ket{w_1} \mp i \ket{\overline{w_1}}),\\
\ket{\phi_3^{\pm}}&=&\frac{1}{\sqrt{2}}(\ket{w_2} \pm i \ket{\overline{w_2}}),\\
\ket{\phi_4^{\pm}}&=&\frac{1}{\sqrt{2}}(\ket{w_3} \mp i \ket{\overline{w_3}}),\\
\ket{\phi_5^{\pm}}&=&\frac{1}{\sqrt{2}}(\ket{w_4} \pm i \ket{\overline{w_4}}).
\end{eqnarray}
These are parity eigenstate  such that $ \otimes_{l=1}^9 \sigma_l^y \ket{\phi_j^{\pm}}=\pm \ket{\phi_j^{\pm}}$. Here, $\ket{w_0}=\ket{000000000}$, $\ket{\overline{w_0}}=\ket{111111111}$, $\ket{w_1}=\frac{1}{\sqrt{9}} \sum_\mathcal{P} \ket{000000001}_\mathcal{P}$, $\ket{\overline {w_1}}=\frac{1}{\sqrt{9}} \sum_\mathcal{P} \ket{011111111}_\mathcal{P}$, $\ket{w_2}=\frac{1}{\sqrt{36}} \sum_\mathcal{P} \ket{000000011}_\mathcal{P}$, $\ket{\overline{w_2}}=\frac{1}{\sqrt{36}} \sum_\mathcal{P} \ket{001111111}_\mathcal{P}$, $\ket{w_3}=\frac{1}{\sqrt{84}} \sum_\mathcal{P} \ket{000000111}_\mathcal{P}$, $\ket{\overline{w_3}}=\frac{1}{\sqrt{84}} \sum_\mathcal{P}\ket{000111111}_\mathcal{P}$, $\ket{w_4}=\frac{1}{\sqrt{126}} \sum_\mathcal{P} \ket{000001111}_\mathcal{P}$ and  $\ket{\overline{w_4}}=\frac{1}{\sqrt{126}} \sum_\mathcal{P} \ket{000011111}_\mathcal{P}$. The unitary operator $\mathcal{U}$ in this basis is diagonal with blocks $\mathcal{U_{\pm}}$ each  having dimensions $5\times 5$ respectively such that:
\begin{equation}
\mathcal{U}_{\pm}=\frac {e^ {\frac{{\mp}i \pi  }{4}}}{16}\left(
\begin{array}{ccccc}
1 & \mp 3i~ & {6} & \mp{2i} \sqrt{21}  & {3} \sqrt{14}  \\
 \pm 3 i & -{7}  & \pm {10 i}  & -2\sqrt{21}  & \pm {i}\sqrt{14}  \\
 6  & \mp {10 i} & 8  & 0 & -2{\sqrt{14}} \\
\pm {2i} \sqrt{21} & - 2\sqrt{21} & 0 & 8 & \mp {2i}\sqrt{6} \\
 3\sqrt{14}  & \mp i\sqrt{14}  & -2{\sqrt{14}}  & \pm {2i} \sqrt{6}  & {6}  \\
\end{array}
\right).
\end{equation}
For $9$ qubit, the eigenvalues are $\mathcal{U_{+}}\left(\mathcal{U_{-}}\right)$ is given as ${e^{\frac{3i \pi }{4}}\left\{1,e^{-\frac{2i \pi }{3}},e^{-\frac{2i \pi }{3}},e^{\frac{2i \pi }{3}},e^{\frac{2i \pi }{3}}\right\}}\left(e^{\frac{i \pi }{4}}\left\{-1,e^{\frac{i \pi }{3}},e^{\frac{i \pi }{3}},e^{-\frac{i \pi }{3}},e^{-\frac{i \pi }{3}}\right\}\right)$
and the eigenvectors are $\left\lbrace\left[\frac{3}{\sqrt{14}},\frac{\pm i}{\sqrt{14}},0,0,1 \right]^T,\left[-\frac{3}{\sqrt{14}},\mp \frac{5 i}{\sqrt{14}},2 i \sqrt{\frac{6}{7}},0,1 \right]^T,\left[\pm\frac{1}{\sqrt{7}},-\frac{3 i}{\sqrt{7}},\pm i \sqrt{\frac{3}{7}},1,0 \right]^T,\right.$\\ $\left.\left[-\frac{3}{\sqrt{14}},\mp \frac{5 i}{\sqrt{14}},-2 i \sqrt{\frac{6}{7}},0,1 \right]^T 
,\left[\mp\frac{1}{\sqrt{7}}, \frac{3 i}{\sqrt{7}},\pm i \sqrt{\frac{3}{7}},1,0\right]^T\right\rbrace$. The $n${th} time evolution of  $\mathcal{U}_{\pm}$ is given as follows:
\begin{equation}
 \mathcal{U}_{\pm}^n= e^{\frac{3 i n\pi}{4}}{\left[
\begin{array}{ccccc}
 \frac{1}{8} \left(3+5 \cos \left(\frac{2 n \pi }{3}\right)\right) & \mp\frac{1}{4} i \sin^2 \left(\frac{n \pi }{3}\right) & \frac{1}{4}\sqrt{3} \sin \left(\frac{2 n \pi }{3}\right) & \mp\frac{1}{4} i \sqrt{7} \sin \left(\frac{2 n \pi }{3}\right) & \frac{1}{2} \sqrt{\frac{7}{2}}\sin ^2\left(\frac{n \pi }{3}\right) \\
 \pm \frac{1}{4} i \sin^2 \left(\frac{n \pi }{3}\right) & \frac{1}{24} \left(1+23 \cos\left(\frac{2 n \pi }{3}\right)\right) & \pm {5 i \sin \left(\frac{2n\pi }{3}\right)}/{4 \sqrt{3}} & -\frac{1}{4} \sqrt{7} \sin \left(\frac{2 n \pi }{3}\right) & \pm\frac{1}{6} i \sqrt{\frac{7}{2}} \sin^2\left(\frac{n\pi }{3}\right) \\
 -\frac{1}{4} \sqrt{3} \sin \left(\frac{2 n \pi }{3}\right) & \pm {5 i \sin \left(\frac{2 n \pi }{3}\right)}/{4 \sqrt{3}} & \cos\left(\frac{2n \pi }{3}\right) & 0 & \frac{1}{2} \sqrt{\frac{7}{6}} \sin \left(\frac{2 n \pi }{3}\right) \\
 \mp \frac{1}{4} i \sqrt{7} \sin \left(\frac{2 n \pi }{3}\right) & \frac{1}{4} \sqrt{7} \sin \left(\frac{2 n \pi }{3}\right) & 0 & \cos \left(\frac{2n \pi }{3}\right) & \pm {i \sin \left(\frac{2 n \pi }{3}\right)}/{2 \sqrt{2}} \\
 \frac{1}{2} \sqrt{\frac{7}{2}} \sin^2 \left(\frac{n \pi }{3}\right) & \mp\frac{1}{6} i \sqrt{\frac{7}{2}} \sin^2 \left(\frac{n \pi }{3}\right)& -\frac{1}{2} \sqrt{\frac{7}{6}} \sin \left(\frac{2 n \pi }{3}\right) & \pm {i \sin \left(\frac{2 n \pi }{3}\right)}/{2 \sqrt{2}} & \frac{1}{12}\left(7+5 \cos \left(\frac{2 n \pi }{3}\right)\right) \\
\end{array}
\right]}.
\end{equation}
It can be shown that each element of the  $\mathcal{U}_{\pm}^n$  matrix exhibits a time periodicity of $24$. Therefore, the unitary operator $\mathcal{U}^n$  possesses a periodicity of $24$.
\subsection{Initial state $\otimes^9\ket{0}$= $\ket{\theta_0 =0,\phi_0 =0}$}
The state $\ket{\psi_n}$ can be evaluated after  the $n$th iteration of $\mathcal{U}$ on this initial state and is  given as follows:
\begin{equation}
 \begin{split}
\ket{\psi_n}&=\mathcal{U}^n |000000000\rangle =\frac{1}{\sqrt{2}} \mathcal{U}^n \left( |\phi_{0}^{+} \rangle + |\phi_{0}^{-} \rangle \right) \\ 
&=\frac{1}{\sqrt{2}} \left( \mathcal{U}_{+}^n |\phi_{0}^{+} \rangle + \mathcal{U}_{-}^n |\phi_{0}^{-} \rangle \right) \\
&=\frac{1}{2} e^{{\frac{ 3in \pi }{4}}}\left\lbrace(1+ i^n)\left(\delta_n \ket{w_0}-i\epsilon_n \ket{\overline{w_1}}+ \eta_n \ket{w_2}+-i\omega_n \ket{\overline{w_3}}+\chi_n \ket{w_4}\right)+\right.  \\ & \left.(1-i^n)\left(i\delta_n \ket{\overline{w_0}}+\epsilon_n\ket{w_1}+i \eta_n \ket{\overline{w_2}}+\omega_n \ket{w_3}+i \chi_n \ket{\overline{w_4}}\right)\right\rbrace,
 \end{split}
 \end{equation}
 where  the coefficients are  $\delta_n= \frac{1}{8} \left[3+5 \cos \left(\frac{2 n \pi }{3}\right)\right]$, $\epsilon_n= \frac{i}{4}  \sin^2 \left(\frac{n \pi }{3}\right)$, $\eta_n=-\frac{\sqrt{3}}{4}  \sin \left(\frac{2 n \pi }{3}\right)$, $\omega_n= -\frac{i \sqrt{7}}{4}  \sin \left(\frac{2 n \pi }{3}\right)$ \mbox{and}
$ \chi_n=\frac{1}{2} \sqrt{\frac{7}{2}} \sin^2 \left(\frac{n \pi }{3}\right)$.\\
\subsubsection{\bf{The linear entropy}}
For even $n=2m$ the single qubit RDM, $\rho_1(2m)$, is diagonal and is given as follows:
\begin{equation}
\rho_1(2m)=\left[
\begin{array}{cc}
 \frac{1}{9} \left(6+2 \cos\left(\frac{4 m \pi }{3}\right)+\cos\left(\frac{2 m \pi }{3}\right)\right) & 0 \\
0 & \frac{4}{9} \left(2+\cos\left(\frac{4 m \pi }{3}\right)\right) \sin^2 \left(\frac{2m \pi }{3}\right)\\
\end{array}
\right].
\end{equation}
Thus, the expression for linear entropy is given as follows:
\begin{equation}
S_{(0,0)}^{(9)}(2m)={\frac{8}{81} \left[2+\cos\left(\frac{4 m \pi }{3}\right)\right] \left[6+2 \cos\left(\frac{4 m \pi }{3}\right)+\cos\left(\frac{2
m \pi }{3}\right)\right] \sin^2\left(\frac{2m\pi }{3}\right)}.
\end{equation}
For odd $n$= $2m-1$ the single qubit RDM, $\rho_1(2m-1)$, is not diagonal and is given as follows:
\begin{equation}\label{Eq:Ap9Q2}
\rho_1(2m-1)={\frac{1}{2}\left(
\begin{array}{cc}
 1 & \bar{r}_n \\
 \bar{r}_n & 1 \\
\end{array}
\right)},
\end{equation}
 where 
\begin{equation}\nonumber
\bar{r}_n=\frac{4}{9} \sin^2\left[\frac{(2m-1) \pi }{3}\right] \left[2+\cos\left(\frac{2 (2m-1)\pi }{3}\right)-\sqrt{3} \sin\left(\frac{2 (2m-1) \pi
}{3}\right)\right].\\
\end{equation}
Thus, the expression for linear entropy is given as follows:
\begin{equation}\label{Eq:Ap9Q1}
S_{(0,0}^{(9)}(2m-1)= {\frac{1}{2}-\frac{8}{81} \sin^4\left(\frac{(2m-1)\pi }{3}\right) \left[2+\cos\left(\frac{2 (2m-1) \pi }{3}\right)-\sqrt{3} \sin\left(\frac{2
(2m-1) \pi }{3}\right)\right]^2}.
\end{equation}
From the Eqs.~(\ref{Eq:Ap9Q2}) and (\ref{Eq:Ap9Q1}), we can see that the linear entropy is periodic in $m$ having periodicity $3$, which implies that its entanglement dynamics is periodic in $n$ having period $6$ i.e. $S_{(0,0)}^{(9)}(n)=S_{(0,0)}^{(9)}(n+6)$. We also seen that the  linear entropy for consecutive odd \mbox{and} even values of $n$  remains same, (which is shown in (Fig.(1)) i.e.
\begin{equation}
S_{(0,0)}^{(9)}(2m-1)=S_{(0,0)}^{(9)}(2m).
\end{equation} 

\subsubsection{\bf{Concurrence}}
 The two qubit RDM, $\rho_{12}(n)$, for even value  $n=2m$ is given as follows:
\begin{equation}
 \rho_{12}(2m)={\left(
\begin{array}{cccc}
 z_n & 0 & 0 & q_n \\
 0 & {v_n} & {v_n} & 0 \\
 0 & {v_n} & {v_n} & 0 \\
 q_n & 0 & 0 & t_n \\
\end{array}
\right)},
\end{equation}
where the coefficients are:
\begin{eqnarray}\nonumber
z_n&=&\frac{1}{72} \left[36+25 \cos\left(\frac{4m \pi }{3}\right)+11 \cos\left(\frac{2m \pi }{3}\right)\right], q_n={\frac{1}{{8\sqrt{3}}}\left[-3 \sin\left(\frac{4m\pi }{3}\right)+\sin\left(\frac{2m\pi }{3}\right)\right]}\\ \nonumber
t_n&=&\frac{1}{36}
\left[17+10 \cos \left(\frac{4m\pi }{3}\right)\right]\sin^2\left(\frac{2m\pi }{3}\right) \mbox{and}~~ {v_n}=\frac{1}{12} \left[5+2 \cos\left(\frac{4m \pi }{3}\right)\right] \sin^2\left(\frac{2m \pi }{3}\right).
\end{eqnarray}
Time periodicity of all the  coefficients is three i.e. $\left( z_n(m),q_n(m),t_n(m),v_n(m)\right)=\left( z_{n}(m+3),q_{n}(m+3),
\right.$\\ $\left.t_{n}(m+3),v_{n}(m+3)\right)$. The $\rho_{12}(n)$ for odd values  $n=2m-1$ is given as follows:
 \begin{equation}
  \rho_{12}(2m-1)={\frac{1}{2}\left(
\begin{array}{cccc}
 \tilde{z}_n & \tilde{t}_n & \tilde{t}_n & 2\tilde{q}_n \\
 \tilde{t}_n& 2\tilde{v}_n & 2\tilde{v}_n & \tilde{t}_n \\
 \tilde{t}_n & 2\tilde{v}_n & 2\tilde{v}_n & \tilde{t}_n\\
 2\tilde{q}_n & \tilde{t}_n& \tilde{t}_n & \tilde{z}_n \\
\end{array}
\right)}.
\end{equation}
Here the coefficients are as follows:
\begin{eqnarray} \nonumber
\tilde{z}_n&=&\frac{1}{12} \left[8+3 \cos \left(\frac{2 (2m-1) \pi }{3}\right)+\cos\left(\frac{4 (2m-1) \pi }{3}\right)\right] , \\ \nonumber
 \tilde{t}_n&=&\frac{2}{9} \sin^2\left[\frac{(2m-1) \pi }{3}\right] \left[2+\cos\left(\frac{2 (2m-1)\pi }{3}\right)-\sqrt{3} \sin \left(\frac{2(2m-1) \pi }{3}\right)\right],\\ \nonumber
 \tilde{v}_n& =&\frac{1}{12} \left[5+2 \cos\left(\frac{2(2m-1) \pi }{3}\right)\right] \sin^2\left(\frac{(2m-1) \pi }{3}\right)
\mbox{and}~~ \tilde{q}_n={\frac{1}{{8\sqrt{3}}}\left[-3 \sin\left(\frac{2(2m-1)\pi }{3}\right)+\sin\left(\frac{(2m-1)\pi }{3}\right)\right]}. 
\end{eqnarray}
Time periodicity of all the coefficients is 3 i.e.  $\left( \tilde{z}_n(m),\tilde{t}_n(m),\tilde{v}_n(m),\tilde{q}_n(m)\right)=\left(\tilde{z}_n(m+3),\tilde{t}_n(m+3),\tilde{q}_n(m+3)\right.$\\$\left.,\tilde{v}_n(m+3)\right)$. The simplified  eigenvalues $\lambda_l$  of $(\sigma_y \otimes \sigma_y)\rho_{12} (\sigma_y \otimes \sigma_y) \rho_{12}^*$ in decreasing order  for  $n$ = $0$ to $12$  are $\left\lbrace(0, 0, 0, 0), \left(1/4, 1/4, 0, 0\right),\left(1/4, 1/4, 0, 0\right),\left(1/4, 1/4, 0, 0\right), \left(1/4, 1/4, 0, 0\right), (0, 0, 0, 0),(0, 0, 0, 0),\left(1/4, 1/4, 0, 0\right),\right.$\\$\left.\left(1/4, 1/4, 0, 0\right), \left(1/4, 1/4, 0, 0\right),\left(1/4, 1/4, 0, 0\right), (0, 0, 0, 0),(0, 0, 0, 0)\right\rbrace$. These eigenvalues repeat after every $n\rightarrow n+6$. By using these  $\lambda_l$ in Eq.~(\ref{Eq:Ap5Q3}), we can easily see that the concurrence remains  zero for every value of $n$, which is shown in (Fig.(1)).
\subsection{Initial state $\otimes^9\ket{+}$= $\ket{\theta_0 =\pi/2,\phi_0 =-\pi/2}$}
The  initial state can be  expressed as, $\otimes ^9 {\ket{+}}_y=(\frac{1}{16} \ket{\phi_0^+}+\frac{3 i}{16}\ket{\phi_1^+}- \frac{3}{8} \ket{\phi_2^+}-\frac{i\sqrt{21}}{8} \ket{\phi_3^+}+\frac{3\sqrt{14}}{16} \ket{\phi_4^+})$. The state $\ket{\psi_n}$ can be obtained by the $n$th time evolution of unitary operator $\mathcal{U}$ on this initial state. Thus,
\begin{eqnarray}
\ket{\psi_n}&=& \mathcal{U}_{+}^n\ket{+++++++++}\\ \nonumber
            &=& e^{ \frac{3 i n \pi}{4} }({d}_1^\prime \ket{\phi_0^+}+{d}_2^\prime \ket{\phi_1^+}+ {d}_3^\prime \ket{\phi_2^+}+{d}_4^\prime \ket{\phi_3^+}+{d}_5^\prime \ket{\phi_4^+}).
\end{eqnarray}
Here the coefficients are as follows:
\begin{eqnarray}\nonumber
{d}_1^\prime &=&\frac{1}{16} \left[6-5 \cos \left(\frac{2 n \pi }{3}\right)-5 \sqrt{3} \sin \left(\frac{2 n \pi }{3}\right)\right],~ {d}_2^\prime =\frac{i}{16}  \left[2+\cos \left(\frac{2 n \pi }{3}\right)+\sqrt{3} \sin \left(\frac{2 n \pi }{3}\right)\right],\\  \nonumber
{d}_3^\prime &=&\frac{1}{16} \left[-6 \cos \left(\frac{2 n \pi }{3}\right)+2 \sqrt{3} \sin \left(\frac{2 n \pi }{3}\right)\right],~ {d}_4^\prime =\frac{1}{16} \left[-2 i \sqrt{7} \left(\sqrt{3} \cos\left(\frac{2 n \pi }{3}\right)-\sin\left(\frac{2 n \pi }{3}\right)\right)\right] \mbox{and} \\  \nonumber
{d}_5^\prime &=&\frac{\sqrt{14}}{16}  \left[2+\cos\left(\frac{2 n \pi }{3}\right)+\sqrt{3} \sin \left(\frac{2 n \pi }{3}\right)\right].
 \end{eqnarray}
 \subsubsection{\bf{The linear entropy}}
 The one qubit RDM for this state is given as follows:
\begin{equation}
 \rho_1(n)={\frac{1}{2}\left[
\begin{array}{cc}
 1 & -\frac{i}{9}  \left(1+2 \cos\left(\frac{2 n \pi }{3}\right)\right)^2 \\
 \frac{i}{9}  \left(1+2 \cos\left(\frac{2 n \pi }{3}\right)\right)^2 & 1 \\
\end{array}
\right]}.
\end{equation}
Thus, the expression for linear entropy is given as follows:
\begin{equation}\label{Eq:Ap9Q3}
S_{(\pi/2,-\pi/2)}^{(9)}(n)=\frac{1}{2}\left[1-\frac{1}{81} \left(1+2 \cos\left(\frac{2 n \pi }{3}\right)\right)^4\right].
\end{equation}
From the Eq.({\ref{Eq:Ap9Q3}), we can see that its entanglement dynamics is periodic in nature with 
$S_{(\pi/2,-\pi/2)}^{(9)}(n)=S_{(\pi/2,-\pi/2)}^{(9)}(n+3)$. We also find that it attains the maximum upper bound value i.e. 
$1/2$ which is shown in (Fig.(1)).
\subsubsection{\bf{Concurrence}}
The two-qubit RDM is obtained as follows:
\begin{equation}
\rho_{12}(n)=\left(
\begin{array}{cccc}
 {z}_n^\prime& {{t}_n^\prime}^* & {{t}_n^\prime}^*  & {q}_n^\prime \\
 {t}_n^\prime & {v}_n^\prime & {v}_n^\prime& {{t}_n^\prime}^* \\
 {t}_n^\prime & {v}_n^\prime & {v}_n^\prime & {{t}_n^\prime}^*  \\
 {q}_n^\prime & {t}_n^\prime & {t}_n^\prime & {z}_n^\prime \\
\end{array}
\right),
\end{equation}
where the coefficients are as follows:
\begin{eqnarray}\nonumber
{z}_n^\prime &=&\frac{1}{48} \left[16-3 \cos\left(\frac{2 n \pi }{3}\right)-\cos\left(\frac{4 n \pi }{3}\right)-3 \sqrt{3} \sin\left(\frac{2n \pi }{3}\right)+\sqrt{3} \sin\left(\frac{4 n \pi }{3}\right)\right], {t}_n^\prime =\frac{i}{36}  \left[1+2 \cos\left(\frac{2 n \pi }{3}\right)\right]^2 ,\\ \nonumber
{q}_n^\prime &=&\frac{1}{48} \left[-9 \cos\left(\frac{2 n \pi }{3}\right)-3 \cos \left(\frac{4 n \pi }{3}\right)-\sqrt{3} \left(-3 \sin\left(\frac{2
n \pi }{3}\right)+\sin \left(\frac{4 n \pi }{3}\right)\right)\right]\mbox{and}\\ \nonumber
{v}_n^\prime &=&\frac{1}{48} \left[8+3 \cos \left(\frac{2 n \pi }{3}\right)+\cos\left(\frac{4 n \pi }{3}\right)+3 \sqrt{3} \sin\left(\frac{2
n \pi }{3}\right)-\sqrt{3} \sin\left(\frac{4 n \pi }{3}\right)\right].
\end{eqnarray}
Time periodicity of all the coefficients is $3$ i.e. $( {z}_n^\prime(n),{t}_n^\prime(n),{q}_n^\prime(n),{v}_n^\prime(n))=( {z}_n^\prime(n+3),{t}_n^\prime(n+3),{q}_n^\prime(n+3),{v}_n^\prime(n+3))$.
 The eigenvalues $\lambda_l$ of $(\sigma_y \otimes \sigma_y)\rho_{12} (\sigma_y \otimes \sigma_y) \rho_{12}^*$ in decreasing order for  $n$ = $0$ to $6$ are $\left\lbrace(0, 0, 0, 0), \left(1/4, 1/4, 0, 0\right),\left(1/4, 1/4, 0, 0\right),(0, 0, 0, 0),\left(1/4, 1/4, 0, 0\right), \left(1/4, 1/4, 0, 0\right), (0, 0, 0, 0)\right\rbrace$. These eigenvalues repeat after every $n\rightarrow n+3$, which shows the periodicity of $3$. By using these  $\lambda_l$ in Eq.~(\ref{Eq:Ap5Q3}), we can easily see that the concurrence remains  zero for every value of $n$, which is shown in (Fig.(1)).

 \section{Exact analytical solution for ten-qubits case}
Considering a $10$ qubit system that is confined in an eleven-dimensional symmetric subspace. The eigenbasis  in which the unitary operator becomes block diagonal is given as follows:
\begin{eqnarray}
\ket{\phi_0^{\pm}}&=&\frac{1}{\sqrt{2}}(\ket{w_0} \mp \ket{\overline{w_0}}),\\
\ket{\phi_1^{\pm}}&=&\frac{1}{\sqrt{2}}(\ket{w_1} \pm  \ket{\overline{w_1}}),\\
\ket{\phi_2^{\pm}}&=&\frac{1}{\sqrt{2}}(\ket{w_2} \mp  \ket{\overline{w_2}}),\\
\ket{\phi_3^{\pm}}&=&\frac{1}{\sqrt{2}}(\ket{w_3} \pm  \ket{\overline{w_3}}),\\
\ket{\phi_4^{\pm}}&=&\frac{1}{\sqrt{2}}(\ket{w_4} \mp  \ket{\overline{w_4}}),\\
\ket{\phi_5^{+}}&=&\frac{1}{\sqrt{252}}\sum_\mathcal{P}\ket{0000011111}, 
\end{eqnarray}
here $\ket{w_0}=\ket{0000000000}$, $\ket{\overline{w_0}}=\ket{1111111111}$, $\ket{w_1}=\frac{1}{\sqrt{10}} \sum_\mathcal{P} \ket{0000000001}_\mathcal{P}$, $\ket{\overline {w_1}}=\frac{1}{\sqrt{10}} \sum_\mathcal{P} \ket{0111111111}_\mathcal{P}$, $\ket{w_2}=\frac{1}{\sqrt{45}} \sum_\mathcal{P} \ket{0000000011}_\mathcal{P}$, $\ket{\overline{w_2}}=\frac{1}{\sqrt{45}} \sum_\mathcal{P} \ket{0011111111}_\mathcal{P}$, $\ket{w_3}=\frac{1}{\sqrt{120}}\sum_\mathcal{P} \ket{0000000111}_\mathcal{P}$, $\ket{\overline{w_3}}=\frac{1}{\sqrt{120}} \sum_\mathcal{P} \ket{0001111111}_\mathcal{P}$, $\ket{w_4}=\frac{1}{\sqrt{210}}\sum_\mathcal{P} \ket{0000001111}_\mathcal{P}$, $\ket{\overline{w_4}}=\frac{1}{\sqrt{210}} \sum_\mathcal{P} \ket{0000111111}_\mathcal{P}$. As we discussed earlier in this set of basis $\mathcal{U}$ is block diagonal in $\mathcal{U}_{+}$ and $\mathcal{U}_{-}$ having dimensions $6\times6$ and $5\times5$ respectively. The block $\mathcal{U}_{+}$ is written in positive parity basis $\left\lbrace\phi_0^+,\phi_1^+,\phi_2^+,\phi_3^+,\phi_4^+,\phi_5^+ \right\rbrace $ as follows: 
\begin{equation}
\mathcal{U}_{+}=\frac{1}{8\sqrt{2}}\left(
\begin{array}{cccccc}
 0 & - \sqrt{{5}}~ e^{\frac{3 i \pi}{4}} & 0 & -2 \sqrt{{15}}~ e^{\frac{3 i \pi}{4}} & 0 & -{3}\sqrt{{7}}~
e^{\frac{3 i \pi}{4}} \\
  \sqrt{{5}} ~e^{\frac{-3 i \pi}{4}} & 0 & 9~ e^{\frac{-3 i \pi}{4}} & 0 &  \sqrt{42} ~e^{\frac{-3 i \pi}{4}} & 0 \\
 0 & -9~ e^{\frac{3 i \pi}{4}} & 0 & -2\sqrt{{3}}~ e^{\frac{3 i \pi}{4}} & 0 &  \sqrt{{35}}
~e^{\frac{3 i \pi}{4}} \\
 2 \sqrt{{15}}~ e^{\frac{-3 i \pi}{4}} & 0 &  2\sqrt{3} e^{\frac{-3 i \pi}{4}} & 0 & - 2\sqrt{14}~ e^{\frac{-3 i \pi}{4}} & 0 \\
 0 & -\sqrt{42}~ e^{\frac{3 i \pi}{4}} & 0 & 2\sqrt{14}~ e^{\frac{3 i \pi}{4}} & 0 & - \sqrt{30}~ e^{\frac{3 i \pi}{4}}
\\
 {3}\sqrt{{7}}~ e^{\frac{-3 i \pi}{4}} & 0 & - \sqrt{{35}}~ e^{\frac{-3 i \pi}{4}} & 0 &  \sqrt{30}~ e^{\frac{-3 i \pi}{4}} & 0 \\
\end{array}
\right),
\end{equation}
whereas the block $\mathcal{U}_{-}$  written in negative parity basis $\left\lbrace\phi_0^-,\phi_1^-,\phi_2^-,\phi_3^-,\phi_4^-\right\rbrace$ is as follows: 
\begin{equation}
\mathcal{U}_{-}={ \frac{1}{16}\left(
\begin{array}{ccccc}
 e^{\frac{3 i \pi}{4}} & 0 & {3} \sqrt{5}~ e^{\frac{3 i \pi}{4}} & 0 &  \sqrt{{210}}~ e^{\frac{3 i \pi}{4}} \\
 0 & -8~ e^{-\frac{3 i \pi}{4}} & 0 & -8 \sqrt{3}~ e^{-\frac{3 i \pi}{4}} & 0 \\
 {3} \sqrt{5}~ e^{\frac{3 i \pi}{4}} & 0 & {13}~ e^{\frac{3 i \pi}{4}} & 0 & - \sqrt{{42}}~ e^{\frac{3 i \pi}{4}}
\\
 0 & -8 \sqrt{3} ~e^{-\frac{3 i \pi}{4}} & 0 & 8~ e^{-\frac{3 i \pi}{4}} & 0 \\
 \sqrt{{210}}~ e^{\frac{3 i \pi}{4}} & 0 & -\sqrt{{42}}~ e^{\frac{3 i \pi}{4}} & 0 & 2~ e^{\frac{3 i \pi}{4}} \\
\end{array}
\right)}.
\end{equation}
 The eigenvalues for $\mathcal{U_{+}}\left(\mathcal{U_{-}}\right)$ are ${\{i,i,i,-i,-i,-i\}}\left(\left\{(-1)^{1/4},-(-1)^{3/4},(-1)^{3/4},(-1)^{3/4},-(-1)^{1/4}\right\}\right)$ and the eigenvectors are $\left\lbrace\left[-\frac{1+i}{\sqrt{7}},-\sqrt{\frac{5}{7}},(1+i) \sqrt{\frac{5}{7}},0,0,1 \right]^T,\left[\sqrt{\frac{5}{42}},(-4+4 i) \sqrt{\frac{2}{21}},3 \sqrt{\frac{3}{14}},0,1,0 \right]^T,\right.$\\ $\left. \left[\left(-\frac{1}{2}-\frac{i}{2}\right) \sqrt{\frac{5}{3}},\frac{2}{\sqrt{3}},\left(-\frac{1}{2}-\frac{i}{2}\right) \sqrt{3},1,0,0 \right]^T,
\left[\frac{1+i}{\sqrt{7}},-\sqrt{\frac{5}{7}},(-1-i) \sqrt{\frac{5}{7}},0,0,1 \right]^T,\left[\sqrt{\frac{5}{42}},(4-4 i) \sqrt{\frac{2}{21}},3 \sqrt{\frac{3}{14}},0,1,0 \right]^T,\right.$\\ $\left.\left[\left(\frac{1}{2}+\frac{i}{2}\right) \sqrt{\frac{5}{3}},\frac{2}{\sqrt{3}},\left(\frac{1}{2}+\frac{i}{2}\right) \sqrt{3},1,0,0 \right]^T \right\rbrace \left(\left\lbrace\left[0, \sqrt{3},0,1,0\right]^T,\left[-\sqrt{\frac{15}{14}},0,\sqrt{\frac{3}{14}},0,1\right]^T, \left[\sqrt{\frac{14}{15}},0,0,0,1\right]^T,\left[\frac{1}{\sqrt{5}},0,1,0,0\right]^T,\right.\right.$\\ $\left.\left.\left[0,-\frac{1}{\sqrt{3}},0,1,0\right]^T \right\rbrace\right)$.  The $n$th time evolution  of $\mathcal{U}_{\pm}$ is  given  as follows :
\begin{equation}
{\mathcal{U}_{+}^n=\left[
\begin{array}{cccccc}
 \cos\left(\frac{n \pi }{2}\right) & \bar{a}_1 \left(\frac{1}{16}-\frac{i }{16}\right) \sqrt{5}  & 0 & \bar{a}_1  \left(\frac{1}{8}-\frac{i}{8}\right) \sqrt{15}
& 0 & \bar{a}_1 \left(\frac{3}{16}-\frac{3 i}{16}\right) \sqrt{7} \\
\bar{a}_1  \left(-\frac{1}{16}-\frac{i}{16}\right) \sqrt{5}  & \cos\left(\frac{n \pi }{2}\right) & \bar{a}_1 \left(-\frac{9}{16}-\frac{9 i}{16}\right)  & 0 &
-\frac{\bar{a}_1}{8} (-1)^{1/4} \sqrt{21}  & 0 \\
 0 & \bar{a}_1\left(\frac{9}{16}-\frac{9 i}{16}\right)  & \cos\left(\frac{n \pi }{2}\right) & \bar{a}_1\left(\frac{1}{8}-\frac{i}{8}\right) \sqrt{3} & 0 &
\bar{a}_1 \left(-\frac{1}{16}+\frac{i}{16}\right) \sqrt{35} \\
 \bar{a}_1 \left(-\frac{1}{8}-\frac{i}{8}\right) \sqrt{15}  & 0 & \bar{a}_1 \left(-\frac{1}{8}-\frac{i}{8}\right) \sqrt{3}  & \cos\left(\frac{n \pi }{2}\right)
& \bar{a}_1 \left(\frac{1}{4}+\frac{i}{4}\right) \sqrt{\frac{7}{2}}  & 0 \\
 0 &  \bar{a}_1\left(\frac{1}{8}-\frac{i}{8}\right) \sqrt{\frac{21}{2}} & 0 & \frac{\bar{a}_1 }{4} (-1)^{3/4} \sqrt{7} & \cos \left(\frac{n \pi }{2}\right)&\bar{a}_1  \left(\frac{1}{8}-\frac{i}{8}\right) \sqrt{\frac{15}{2}} \\
\bar{a}_1  \left(-\frac{3}{16}-\frac{3 i}{16}\right) \sqrt{7}  & 0 & \bar{a}_1 \left(\frac{1}{16}+\frac{i}{16}\right) \sqrt{35}  & 0 &  \frac{ -\bar{a}_1}{8} (-1)^{1/4} \sqrt{15}
 & \cos \left(\frac{n \pi }{2}\right) \\
\end{array}
\right]},
\end{equation}
\begin{equation}
\mathcal{U}_{-}^n={ \frac{e^{\frac{ i n \pi }{4}}}{32} \left[
\begin{array}{ccccc}
  e^{\frac{ i n \pi }{2}} \left(17+15 e^{i n \pi }\right) & 0 & -{3\bar{a}_2} \sqrt{5}~ e^{\frac{ i n \pi }{2}}  & 0 & -2{\bar{a}_2}
\sqrt{\frac{105}{2}}~ e^{\frac{i n \pi }{2}}  \\
 0 & 8~ e^{\frac{i n \pi }{4}} \left(3+e^{i n \pi }\right) & 0 & -8{\bar{a}_2} \sqrt{3}   & 0 \\
 -{3\bar{a}_2} \sqrt{5}~ e^{\frac{i n \pi }{2}}  & 0 & e^{\frac{ i n \pi }{2}} \left(29+3 e^{i n \pi }\right) & 0 & 2{\bar{a}_2}
\sqrt{\frac{21}{2}} ~e^{\frac{ i n \pi }{2}}  \\
 0 & -8{\bar{a}_2} \sqrt{3}  & 0 & 8 \left(1+3 e^{i n \pi }\right) & 0 \\
 -2{\bar{a}_2} \sqrt{\frac{105}{2}} e^{\frac{ i n \pi }{2}}  & 0 & 2{\bar{a}_2} \sqrt{\frac{21}{2}} ~e^{\frac{ i n \pi }{2}}  & 0 & 2~
e^{\frac{ i n \pi }{2}} \left(9+7 e^{i n \pi }\right) \\
\end{array}
\right]},
\end{equation}
where \({\bar{a}_1=\sin\left(\frac{n \pi }{2}\right)}\) and \({\bar{a}_2=\left(-1+e^{i n \pi }\right)}\). The time periodicity of each element of $\mathcal{U}_{+}^n\left(\mathcal{U}_{-}^n\right)$ is $4(8)$. Hence the time periodicity of $\mathcal{U}^n$ is $8$.

\subsection{Initial state $\otimes^{10}\ket{0}$= $\ket{\theta_0 =0,\phi_0 =0}$}
The state $\ket{\psi_n}$ can be obtained after the  $n$ implementations of the unitary operator $\mathcal{U}$ on this initial state. Thus,
\begin{eqnarray}
\ket{\psi_n}&=&\mathcal{U}^{n}\ket{0000000000}= \frac{1}{\sqrt{2}}(\mathcal{U}_{+}^n\ket{\phi_0^+}+\mathcal{U}_{-}^n\ket{\phi_0^-}) \\  \nonumber
&=&\frac{1}{\sqrt{2}}\left( {r}_1^\prime\ket{\phi_0^+}+r_2^\prime\ket{\phi_1^+}+r_3^\prime\ket{\phi_3^+}+r_4^\prime\ket{\phi_5^+}+r_5^\prime\ket{\phi_0^-}+r_6^\prime\ket{\phi_2^-} +r_7^\prime\ket{\phi_4^-}\right),
\end{eqnarray}
where the coefficients are:
\begin{eqnarray}\nonumber
r_1^\prime=\cos\left(\frac{n \pi }{2}\right),~ r_2^\prime=-\frac{\sqrt{10}~e^{\frac{ i  \pi }{4}}}{16}  \sin\left(\frac{n \pi }{2}\right),~ {r_3}^\prime=-\frac{\sqrt{30}~e^{\frac{ i  \pi }{4}}}{8} \sin\left(\frac{n \pi }{2}\right),~ r_4^\prime=-\frac{3\sqrt{14}~e^{\frac{ i  \pi }{4}}~}{16}\sin\left(\frac{n \pi }{2}\right),\\ \nonumber
r_5^\prime=\frac{e^{\frac{3 i n \pi }{4}}}{32}  \left(17+15 e^{i n \pi }\right), ~r_6^\prime=-\frac{3\sqrt{5} ~e^{\frac{3 i n \pi }{4}}}{32}  \left(-1+e^{i n \pi }\right)~\mbox{and}~ r_7^\prime =-\frac{e^{\frac{3 i n \pi }{4}}}{16} \sqrt{\frac{105}{2}}  \left(-1+e^{i n \pi }\right).
\end{eqnarray}
\subsubsection{\bf{The linear entropy}}
The single qubit RDM for this state is given as follows:
\begin{equation}
\rho_1(n)={\frac{1}{4}\left(
\begin{array}{cc}
 2+\frac{17}{32} \cos\left(\frac{n \pi }{4}\right)+\cos\left(\frac{5 n \pi }{4}\right)+\frac{15}{32} \cos\left(\frac{9 n \pi }{4}\right)
& \bar{z}_n \\
 \bar{z}_n  & 2-\frac{17}{32} \cos\left(\frac{n \pi }{4}\right)-\cos\left(\frac{5 n \pi }{4}\right)-\frac{15}{32} \cos\left(\frac{9 n \pi
}{4}\right) \\
\end{array}
\right)},
\end{equation}
where $\bar{z}_n ={\frac{1}{{16\sqrt{2}}}\sin \left(\frac{n \pi }{2}\right) \left[-17 \cos \left(\frac{3 n \pi}{4}\right)+15 \cos \left(\frac{7 n \pi }{4}\right)-17 \sin \left(\frac{3 n \pi }{4}\right)+15 \sin \left(\frac{7 n \pi }{4}\right)\right]}$.\\

The expression for linear entropy of a single-qubit reduced state is given by,
\begin{eqnarray}\label{Eq:ap10Q1}
S_{(0,0)}^{(10)}(n)=&&1-\dfrac{1}{8} \left\lbrace 4+\left[\frac{17}{32} \cos\left(\frac{n \pi }{4}\right)+\cos\left(\frac{5 n \pi }{4}\right)+\frac{15}{32}
\cos\left(\frac{9 n \pi }{4}\right)\right]^2\right\rbrace-\dfrac{1}{4096}\sin^2\left(\frac{n \pi }{2}\right)\\ \nonumber
&& \left[17 \left(\cos\left(\frac{3 n \pi }{4}\right)+\sin\left(\frac{3 n \pi }{4}\right)\right)-15\left(\cos\left(\frac{7 n \pi }{4}\right)+\sin\left(\frac{7 n \pi }{4}\right)\right)\right]^2.
\end{eqnarray}
With the help of Eq.(\ref{Eq:ap10Q1}), we can evaluate the entanglement dynamics for this state,  which we found periodic in nature having periodicity $4$ i.e. $S_{(0,0)}^{(10)}(n)=S_{(0,0)}^{(10)}(n+4)$. Notably, the linear entropy attains its maximum upper bound value of $1/2$ and remains unchanged for consecutive odd and even values of $n$, which is shown in (Fig.(1)). %The value of linear entropy attains its maximum upper bound value i.e. $1/2$ and it remains the same for consecutive odd and even values of $n$ for the 10 qubit system as well.\ref fig.
\subsubsection{\bf{Concurrence}}
 The two qubit RDM $\rho_{12}(n)$ for this state is given as follows:
\begin{equation}
\rho_{12}(n)={\left[
\begin{array}{cccc}
\bar{w}_n& \bar{y}_n & \bar{y}_n& \bar{x}_n^*\\
 \bar{y}_n^* & \frac{1}{4} \sin^2 \left(\frac{n \pi }{2}\right) & \frac{1}{4} \sin^2\left(\frac{n \pi }{2}\right) & \bar{y}_n^* \\
 \bar{y}_n^* & \frac{1}{4} \sin^2 \left(\frac{n \pi }{2}\right) & \frac{1}{4} \sin^2 \left(\frac{n \pi }{2}\right)& \bar{y}_n^* \\
 \bar{x}_n & \bar{y}_n & \bar{y}_n & \bar{s}_n\\
\end{array}
\right]}.
\end{equation}
Here all the coefficients are:
\begin{eqnarray}\nonumber
\bar{w}_n&=&\frac{1}{128} \left[17 \cos \left(\frac{n \pi }{4}\right)+16 \left(3+\cos (n\pi )+2 \cos\left(\frac{5n \pi }{4}\right)\right)+15 \cos\left(\frac{9 n \pi }{4}\right)\right],\\ \nonumber
\bar{s}_n&=&\frac{1}{128} \left[-17 \cos\left(\frac{n \pi }{4}\right)+16 \left(3+\cos( n\pi )-2 \cos \left(\frac{5
n \pi }{4}\right)\right)-15 \cos\left(\frac{9 n \pi }{4}\right)\right],\\ \nonumber
\bar{x}_n&=&\frac{1}{128} \left[16-16 \cos(n \pi )+i \sin \left(\frac{n \pi }{4}\right)-i \sin \left(\frac{9 n \pi
}{4}\right)\right]\mbox{and}\\ \nonumber
\bar{y}_n &=&-\frac{e^{\frac{-i\pi}{4}}}{32} \sin\left(\frac{n \pi }{2}\right) \left[2 \cos\left(\frac{n \pi }{2}\right)+(9+8i) \cos\left(\frac{3 n \pi }{4}\right)-(7+8 i) \cos\left(\frac{7 n \pi }{4}\right)+(8+9 i) \sin\left(\frac{3 n \pi }{4}\right)\right.\\ \nonumber &&\left. -(8+7i) \sin\left(\frac{7 n \pi }{4}\right)\right].
\end{eqnarray}
All the coefficients are periodic having periodicity $4$ i.e.
 $\left(\bar{w}_n(n),\bar{s}_n(n),\bar{x}_n(n),\bar{y}_n(n)\right)=\left(\bar{s}_n(n+4),\bar{x}_n(n+4),\bar{y}_n(n+4),
 \bar{w}_n(n+4)\right)$. The eigenvalue $\lambda_l$ of $(\sigma_y \otimes \sigma_y)\rho_{12} (\sigma_y \otimes \sigma_y) \rho_{12}^*$  
 for $n=0$ to $8$  are $\left\lbrace(0, 0, 0, 0), \left(1/4, 1/4, 0, 0\right),\left(1/4, 1/4, 0, 0\right),(0, 0, 0, 0),(0, 0, 0, 0),
 \left(1/4, 1/4, 0, 0\right), \left(1/4, 1/4, 0, 0\right), (0, 0, 0, 0),(0, 0, 0, 0)\right\rbrace$, which shows period of $4$.  
 By using these  $\lambda_l$ in Eq.~(\ref{Eq:Ap5Q3}), we find that the concurrence vanishes for every value of $n$. 
 This is shown in the (Fig.(1)).

\subsection{Initial state $\otimes^{10}\ket{+}$= $\ket{\theta_0 =\pi/2,\phi_0 =-\pi/2}$}
The initial state can be written as, $\otimes ^{10} {\ket{+}}_y=\frac{1}{16\sqrt{2}}\left( \ket{\phi_0^+}+{i \sqrt{10}}\ket{\phi_1^+}- {3\sqrt{5}} \ket{\phi_2^+}-{2i\sqrt{30}}\ket{\phi_3^+}+{\sqrt{210}}\ket{\phi_4^+}\right.$\\$\left.+{3 i\sqrt{14}}\ket{\phi_5^+}\right)$. The state $\ket{\psi_n}$ can be obtained by operating unitary operator $n$ times on this initial state. Thus,
\begin{eqnarray}
\ket{\psi_n}&=& \mathcal{U}_{+}^n\ket{++++++++++}\\  \nonumber
            &=&r_1 \ket{\phi_0^+}+r_2 \ket{\phi_1^+}+ r_3 \ket{\phi_2^+}+r_4 \ket{\phi_3^+}+r_5 \ket{\phi_4^+}+r_6 \ket{\phi_5^+},
\end{eqnarray}
where  the coefficients are:
\begin{eqnarray} \nonumber
 r_1&=&\frac{1}{32} \left[\sqrt{2} \cos\left(\frac{n \pi }{2}\right)+(1+i) \sin\left(\frac{n \pi }{2}\right)\right], ~r_3=-\frac{3\sqrt{5} }{32} \left[\sqrt{2} \cos\left(\frac{n \pi }{2}\right)+(1+i) \sin\left(\frac{n \pi }{2}\right)\right],\\  \nonumber
r_2&=&\frac{i}{16} \sqrt{\frac{5}{2}} \left[\sqrt{2} \cos\left(\frac{n \pi }{2}\right)-(1-i) \sin\left(\frac{n \pi }{2}\right)\right), ~r_4=\frac{\sqrt{15}}{8}  \left(-i \cos \left(\frac{n \pi }{2}\right)+{\exp}(i \pi /4) \sin\left(\frac{n \pi }{2}\right)\right],\\ \nonumber
r_5&=&\frac{\sqrt{105}}{16}  \left[\cos\left(\frac{n \pi }{2}\right)+{\exp}(i \pi /4)\sin\left(\frac{n \pi }{2}\right)\right] \mbox{and}~ r_6=\frac{3i}{16}  \sqrt{\frac{7}{2}} \left[\sqrt{2} \cos \left(\frac{n \pi }{2}\right)-(1-i) \sin\left(\frac{n \pi }{2}\right)\right].
\end{eqnarray}
\subsubsection{\bf{The linear entropy}}
 The RDM for a single qubit is given as follows:
\begin{equation}
  \rho_1(n)={\frac{1}{2}\left[
\begin{array}{cc}
 1 & -\frac{i}{2}  (1+\cos (n \pi )) \\
 \frac{i}{2} (1+\cos (n \pi )) & 1 \\
\end{array}
\right]}.
\end{equation}
Thus, the   linear entropy is given as follows:
\begin{equation}\label{eq:ap10Q5}
S_{(\pi/2,-\pi/2)}^{(10)}(n)=\frac{1}{2}\left[1-\frac{1}{4} (1+\cos(n \pi ))^2\right].
\end{equation}
From the Eq.(\ref{eq:ap10Q5}) we can see that its entanglement dynamics is  periodic in nature having periodicity $2$. We also find that it attains the maximum upper bound value i.e. 1/2 for odd values of $n$ and zero for even values of $n$.
\subsubsection{\bf{Concurrence}}
 The two-qubit  RDM for this state is given as follows,
\begin{equation}
\rho_{12}(n)={\frac{1}{4}\left[
\begin{array}{cccc}
 1 & -\frac{i}{2}  (1+\cos (n \pi )) & -\frac{i}{2} (1+\cos(n \pi )) & -1 \\
 \frac{i}{2} (1+\cos (n \pi )) & 1 & 1 & -\frac{i}{2}  (1+\cos(n \pi )) \\
 \frac{i}{2}  (1+\cos (n \pi )) & 1 & 1 & -\frac{i}{2}  (1+\cos (n \pi )) \\
 -1 & \frac{i}{2}  (1+\cos (n \pi )) & \frac{i}{2}  (1+\cos (n \pi )) & 1 \\
\end{array}
\right]}.
\end{equation}
 The eigenvalues $\lambda_l$  of $(\sigma_y \otimes \sigma_y)\rho_{12} (\sigma_y \otimes \sigma_y) \rho_{12}^*$  in decreasing order are 
$\left\lbrace \frac{1}{8} \left(1 - \cos(n \pi)\right),\frac{1}{8} \left(1 -  \cos(n\pi)\right),0,0\right\rbrace$. 
By using these eigenvalues in the eq.(\ref{Eq:Ap5Q3}), we can see that its concurrence vanishes for all values of $n$. 
This is shown in the (Fig.(1)).

\section{Exact analytical solution for eleven-qubits case}
Considering a $11$ qubit  system which is confined in a twelve-dimensional symmetric subspace. The eigenbasis  in which the unitary operator becomes block diagonal is given as follows:
\begin{eqnarray}
\ket{\phi_0^{\pm}}&=&\frac{1}{\sqrt{2}}(\ket{w_0} \mp i \ket{\overline{w_0}}),\\
\ket{\phi_1^{\pm}}&=&\frac{1}{\sqrt{2}}(\ket{w_1} \pm i \ket{\overline{w_1}}),\\
\ket{\phi_2^{\pm}}&=&\frac{1}{\sqrt{2}}(\ket{w_2} \mp i \ket{\overline{w_2}}),\\
\ket{\phi_3^{\pm}}&=&\frac{1}{\sqrt{2}}(\ket{w_3} \pm i \ket{\overline{w_3}}),\\
\ket{\phi_4^{\pm}}&=&\frac{1}{\sqrt{2}}(\ket{w_4} \mp i \ket{\overline{w_4}}),\\
\ket{\phi_5^{\pm}}&=&\frac{1}{\sqrt{2}}(\ket{w_5} \pm i \ket{\overline{w_5}}).
\end{eqnarray}
These are parity eigenstate  such that $ \otimes_{l=1}^{11} \sigma_l^y \ket{\phi_j^{\pm}}=\pm \ket{\phi_j^{\pm}}$. 
Here $\ket{w_0}=\ket{00000000000}$, $\ket{\overline{w_0}}=\ket{11111111111}$, $\ket{w_1}=\frac{1}{\sqrt{11}} \sum_\mathcal{P} 
\ket{00000000001}_\mathcal{P}$, $\ket{\overline {w_1}}=\frac{1}{\sqrt{11}} \sum_\mathcal{P} \ket{011111111}_\mathcal{P}$, 
$\ket{w_2}=\frac{1}{\sqrt{55}} \sum_\mathcal{P} \ket{00000000011}_\mathcal{P}$, $\ket{\overline{w_2}}=\frac{1}{\sqrt{55}} 
\sum_\mathcal{P} \ket{00111111111}_\mathcal{P}$,$\ket{w_3}=\frac{1}{\sqrt{165}} \sum_\mathcal{P} \ket{00000000111}_\mathcal{P}$, 
$\ket{\overline{w_3}}=\frac{1}{\sqrt{165}} \sum_\mathcal{P} \ket{00011111111}_\mathcal{P}$, $\ket{w_4}=\frac{1}{\sqrt{330}} 
\sum_\mathcal{P} \ket{00000001111}_\mathcal{P}$,$\ket{\overline{w_4}}=\frac{1}{\sqrt{330}} \sum_\mathcal{P} \ket{00001111111}_
\mathcal{P}$, $\ket{w_5}=\frac{1}{\sqrt{462}} \sum_\mathcal{P} \ket{00000011111}_\mathcal{P}$,  $\ket{\overline{w_5}}_
\mathcal{P}=\frac{1}{\sqrt{462}} \sum_\mathcal{P}\ket{00000111111}_\mathcal{P}$. 
The $\mathcal{U}$ is block diagonal in $\mathcal{U}_{+}$ and $\mathcal{U}_{-}$ each having dimension $6\times 6$ respectively. They are given as follows:
\begin{equation}
\mathcal{U}_{+}={\frac{1}{32}} \left(
\begin{array}{cccccc}
 i & -\sqrt{11}  & i\sqrt{55}  & -\sqrt{165}  &  i\sqrt{330} &-\sqrt{462}  \\
  - \sqrt{11}  & 9i  & -7 \sqrt{5}  & 5i \sqrt{15}  &
- 3\sqrt{30}  & i\sqrt{42}  \\
 -i\sqrt{55}  &  7  \sqrt{5}  & -19i  &  7 \sqrt{3} &
i\sqrt{6}  & -\sqrt{210}  \\
   \sqrt{165}  & - 5i \sqrt{15}  & 7 \sqrt{3}  & 5i  &- 11 \sqrt{2}
 & i\sqrt{70}  \\
  i\sqrt{330} &-  3 \sqrt{30}  & -i\sqrt{6}  &  11\sqrt{2} & - 6i  & -2\sqrt{35}  \\
- \sqrt{462}  & i\sqrt{42}  &  \sqrt{210}  & -i\sqrt{70}  &  - 2
\sqrt{35}  & 10i  \\
\end{array}
\right)   \mbox{and}
\end{equation}
\begin{equation}
\mathcal{U}_{-}={\frac{1}{32}} \left(
\begin{array}{cccccc}
 1 & -i\sqrt{11}  & \sqrt{55}  & -i\sqrt{165}  &  \sqrt{330} &-i\sqrt{462}  \\
  - i\sqrt{11}  & 9  & -7i \sqrt{5}  & 5 \sqrt{15}  &
- 3i\sqrt{30}  & \sqrt{42}  \\
 -\sqrt{55}  &  7i  \sqrt{5}  & -19  &  7i \sqrt{3} &
\sqrt{6}  & -i\sqrt{210}  \\
   i\sqrt{165}  & - 5 \sqrt{15}  & 7i \sqrt{3}  & 5  &- 11 i\sqrt{2}
 & \sqrt{70}  \\
  \sqrt{330} &-3i \sqrt{30}  & -\sqrt{6}  &  11i\sqrt{2} & - 6  & -2i\sqrt{35}  \\
- i\sqrt{462}  & \sqrt{42}  &  i\sqrt{210}  & -\sqrt{70}  &  - 2i\sqrt{35}  & 10  \\
\end{array}
\right).
\end{equation}
For $11$ qubit , the eigenvalues for $\mathcal{U_{+}}\left(\mathcal{U_{-}}\right)$ is given as ${\left\{i\left(-1,-1,e^{\frac{i \pi }{3}},e^{\frac{i \pi }{3}},e^{-\frac{i \pi }{3}},e^{-\frac{i \pi }{3}}\right)\right\}}\left(\left\{-1,-1,e^{\frac{i \pi }{3}},e^{\frac{i \pi }{3}},e^{-\frac{i \pi }{3}},e^{-\frac{i \pi }{3}}\right\}\right)$ and eigenvectors are $\left\lbrace\left[0,-\sqrt{\frac{7}{6}},\mp i \sqrt{\frac{35}{6}},0,0,1 \right]^T,\left[-\sqrt{\frac{15}{22}},0,0,\mp \frac{i}{\sqrt{2}},1,0 \right]^T,\left[\pm\sqrt{\frac{11}{14}},-2 \sqrt{\frac{2}{21}},\pm i \sqrt{\frac{10}{21}},-i \sqrt{\frac{15}{14}},0,1 \right]^T \right.$\\$\left.,\left[0,\pm\sqrt{\frac{5}{2}},-\frac{i}{\sqrt{2}},\pm i \sqrt{2},1,0 \right]^T,\left[\mp\sqrt{\frac{11}{14}},-2 \sqrt{\frac{2}{21}},\pm i \sqrt{\frac{10}{21}},i \sqrt{\frac{15}{14}},0,1 \right]^T,\left[0,\mp\sqrt{\frac{5}{2}},\frac{i}{\sqrt{2}},\pm i \sqrt{2},1,0 \right]^T\right\rbrace$. The $n$th time evolution of $\mathcal{U}_{\pm}$ is given as follows:
\begin{equation}
\mathcal{U}_{+}^n={e^{\frac{i n\pi }{2}}}\left(
\begin{array}{cccccc}
 \frac{1}{16} (5+11 a_3^\prime) & -\frac{i a_1^\prime}{16}\sqrt{\frac{11}{3}}  & -\frac{a_1^\prime}{16} \sqrt{\frac{55}{3}}  & -\frac{i a_2^\prime}{8} \sqrt{\frac{55}{3}} 
& -\frac{a_2^\prime}{4} \sqrt{\frac{55}{6}}  & -\frac{i a_1^\prime}{8} \sqrt{\frac{77}{2}}  \\
 -\frac{i a_1^\prime}{16} \sqrt{\frac{11}{3}} & \frac{1}{48} (7+41a_3^\prime) & -\frac{7i a_2^\prime}{24} \sqrt{5}  & -\frac{5}{16} \sqrt{5} a_1^\prime & -\frac{3a_1^\prime}{8} i \sqrt{\frac{5}{2}}
 & -\frac{a_2^\prime}{4} \sqrt{\frac{7}{6}}  \\
 \frac{a_1^\prime}{16} \sqrt{\frac{55}{3}}  & \frac{7i a_2^\prime}{24} \sqrt{5}  & \frac{1}{48} (35+13 a_3^\prime) & \frac{7}{16} i a_1^\prime & -\frac{a_1^\prime}{8 \sqrt{2}} & -\frac{i a_2^\prime}{4}
\sqrt{\frac{35}{6}}  \\
 \frac{i a_2^\prime}{8}\sqrt{\frac{55}{3}}  & \frac{5a_1^\prime}{16} \sqrt{5}  & \frac{7}{16} i a_1^\prime  & \frac{1}{48} (11+37 a_3^\prime) & -\frac{11 i a_2^\prime}{12 \sqrt{2}}
& -\frac{a_1^\prime}{8} \sqrt{\frac{35}{6}}  \\
 -\frac{a_2^\prime}{4} \sqrt{\frac{55}{6}}  & -\frac{3a_1^\prime}{8} i \sqrt{\frac{5}{2}}  & \frac{a_1^\prime}{8 \sqrt{2}} & \frac{11 i a_2^\prime}{12 \sqrt{2}} & \frac{1}{24} (11+13
a_3^\prime) & -\frac{i a_1^\prime}{8} \sqrt{\frac{35}{3}} \\
 -\frac{i a_1^\prime}{8} \sqrt{\frac{77}{2}}  & -\frac{a_2^\prime}{4} \sqrt{\frac{7}{6}}  & \frac{1}{4} i a_2^\prime \sqrt{\frac{35}{6}} 
& \frac{a_1^\prime}{8} \sqrt{\frac{35}{6}}  & -\frac{i a_1^\prime}{8} \sqrt{\frac{35}{3}}  & \frac{1}{8} (1+7 a_3^\prime) \\
\end{array}
\right).
\end{equation}
\begin{equation}
 \mathcal{U}_{-}^n={e^{i n \pi} }\left(
\begin{array}{cccccc}
 \frac{1}{16} (5+11 a_3^\prime) & \frac{i a_1^\prime}{16}\sqrt{\frac{11}{3}}  & -\frac{a_1^\prime}{16} \sqrt{\frac{55}{3}}  & \frac{i a_2^\prime}{8} \sqrt{\frac{55}{3}}  &
-\frac{a_2^\prime}{4} \sqrt{\frac{55}{6}}  & \frac{i a_1^\prime}{8} \sqrt{\frac{77}{2}}  \\
 \frac{i a_1^\prime}{16} \sqrt{\frac{11}{3}} & \frac{1}{48} (7+41a_3^\prime) & \frac{7i a_2^\prime}{24} \sqrt{5}  & -\frac{5}{16} \sqrt{5} a_1^\prime & \frac{3a_1^\prime}{8} i \sqrt{\frac{5}{2}}
 & -\frac{a_2^\prime}{4} \sqrt{\frac{7}{6}}  \\
 \frac{a_1^\prime}{16} \sqrt{\frac{55}{3}}  & -\frac{7i a_2^\prime}{24} \sqrt{5}  & \frac{1}{48} (35+13 a_3^\prime) & -\frac{7}{16} i a_1^\prime & -\frac{a_1^\prime}{8 \sqrt{2}} & \frac{i a_2^\prime}{4}
\sqrt{\frac{35}{6}}  \\
 -\frac{i a_2^\prime}{8} \sqrt{\frac{55}{3}}  & \frac{5a_1^\prime}{16} \sqrt{5}  & -\frac{7}{16} i a_1^\prime & \frac{1}{48} (11+37 a_3^\prime) & \frac{11 i a_2^\prime}{12 \sqrt{2}}
& -\frac{a_1^\prime}{8} \sqrt{\frac{35}{6}}  \\
 -\frac{a_2^\prime}{4} \sqrt{\frac{55}{6}}  & \frac{3a_1^\prime}{8} i \sqrt{\frac{5}{2}}  & \frac{a_1^\prime}{8 \sqrt{2}} & -\frac{11 i a_2^\prime}{12 \sqrt{2}} & \frac{1}{24} (11+13
a_3^\prime) & \frac{i a_1^\prime}{8} \sqrt{\frac{35}{3}} \\
 \frac{i a_1^\prime}{8} \sqrt{\frac{77}{2}}  & -\frac{a_2^\prime}{4} \sqrt{\frac{7}{6}}  & -\frac{1}{4} ia_1^\prime \sqrt{\frac{35}{6}}
& \frac{a_1^\prime}{8} \sqrt{\frac{35}{6}}  & \frac{i a_1^\prime}{8} \sqrt{\frac{35}{3}} & \frac{1}{8} (1+7 a_3^\prime) \\
\end{array}
\right),
\end{equation}
where the coefficients are $a_1^\prime=\sin \left({2 n\pi }/{3}\right)$, $a_2^\prime=\sin^2 \left({n\pi }/{3}\right)$ \mbox{and} $a_3^\prime=\cos \left({2 n \pi }/{3}\right)$. Time periodicity of $\mathcal{U_{+}}\left(\mathcal{U_{-}}\right)$  is $12(6)$. Hence the periodicity of $\mathcal{U}$ is $12$.

\subsection{Initial state $\otimes^{11}\ket{0}$= $\ket{\theta_0 =0,\phi_0 =0}$}
The state $\ket{\psi_n}$  is obtained by the $n$th time evolution of unitary operator $\mathcal{U}$ on this initial state and is given as follows:
\begin{equation}
\begin{split}
\ket{\psi_n}&=\mathcal{U}^n |000000000000\rangle =\frac{1}{\sqrt{2}}~ \mathcal{U}^n \left( |\phi_{0}^{+} \rangle + |\phi_{0}^{-} \rangle \right) \\ 
&=\frac{1}{\sqrt{2}} \left( \mathcal{U}_{+}^n |\phi_{0}^{+} \rangle + \mathcal{U}_{-}^n |\phi_{0}^{-} \rangle \right) \\ 
&=\frac{1}{2}e^{{ i n \pi }}\left\lbrace(1+ i^n)(\tilde{\delta}_n \ket{w_0}+i\tilde{\epsilon}_n \ket{\overline{w_1}}+ \tilde{\xi}_n \ket{w_2}+i\tilde{\omega}_n \ket{\overline{w_3}}+\tilde{\chi}_n \ket{w_4}+ i \tilde{\eta}_n \ket{\overline{w_5}})\right.\\&\left.+(1-i^n)(i\tilde{\delta}_n\ket{\overline{w_0}}-\tilde{\epsilon}_n\ket{w_1}+i \tilde{\xi}_n \ket{\overline{w_2}}-\tilde{\omega}_n\ket{w_3}+i\tilde{\chi}_n \ket{\overline{w_4}}-\tilde{\eta}_n \ket{w_5})\right\rbrace,
\end{split}
\end{equation}
where \(\tilde{\delta}_n=\frac{1}{16} \left[5+11 \cos\left(\frac{2 n\pi }{3}\right)\right]\),~ \(\tilde{\epsilon}_n=-\frac{i}{16}  \sqrt{\frac{11}{3}} \sin\left(\frac{2 n \pi }{3}\right)\),~ \(\tilde{\xi}_n=\frac{1}{16} \sqrt{\frac{55}{3}} \sin\left(\frac{2 n \pi }{3}\right)\),~ \(\tilde{\omega}_n=\frac{i}{8}  \sqrt{\frac{55}{3}} \sin^2\left(\frac{n\pi }{3}\right)\), \\ \(\tilde{\chi}_n=-\frac{1}{4} \sqrt{\frac{55}{6}} \sin^2\left(\frac{n\pi }{3}\right)\)~ and~~ \(\tilde{\eta}_n=-\frac{i}{8}  \sqrt{\frac{77}{2}} \sin\left(\frac{2 n\pi }{3}\right)\).\\
\subsubsection{\bf{The linear entropy}}
The single qubit  RDM for even $n=2m$ is given as follows:
\begin{equation}
\rho_1(2m)={\left[
\begin{array}{cc}
 \frac{1}{24} \left(16+5 \cos\left(\frac{4 m \pi }{3}\right)+3 \cos\left(\frac{2 m \pi }{3}\right)\right) & 0 \\
 0 & \frac{1}{12} \left(11+6 \cos\left(\frac{4 m \pi }{3}\right)\right) \sin^2\left(\frac{2m \pi }{3}\right)\\
\end{array}
\right]}.
\end{equation}
Thus, the expression for linear entropy is given as follows:
\begin{equation}\label{Eq:ap11q1}
S_{(0,0)}^{(11)}(2m)=\frac{1}{144} \left(11+6 \cos\left(\frac{4 m \pi }{3}\right)\right) \left(16+5 \cos\left(\frac{4 m \pi }{3}\right)
+3 \cos\left(\frac{2m \pi }{3}\right)\right) \sin^2\left(\frac{2m \pi }{3}\right). 
\end{equation}
 The single qubit RDM for odd $n=2m-1$ is given as follows:
\begin{equation}
\rho_1(2m-1)={\frac{1}{2}\left(
\begin{array}{cc}
 1 & E_m \\
 E_m & 1 \\
\end{array}
\right)}.
\end{equation}
where
\begin{equation} \nonumber
 E_m=\frac{1}{12} \sin\left[\frac{(2m-1) \pi }{3}\right] \left[-2 \sqrt{3} \cos\left(\frac{(2m-1) \pi }{3}\right)+3 \sqrt{3} \cos((2m-1)
\pi )+8 \sin \left(\frac{(2m-1) \pi }{3}\right)+3 \sin((2m-1)\pi)\right].
\end{equation}
Thus, the expression for linear entropy is given as follows:
\begin{eqnarray} \label{Eq:ap11q2}
S_{0,0}^{(11)}(2m-1)&=&\frac{1}{2}\left[1-E_m^2 \right]\\ \nonumber
&=&\frac{1}{2}\left[1-\frac{1}{144} \sin^2\left(\frac{(2m-1) \pi }{3}\right) \left(-2 \sqrt{3} \cos\left(\frac{(2m-1) \pi }{3}\right)+3 \sqrt{3}
\cos((2m-1) \pi )\right.\right.\\ \nonumber && \left.\left.
+3\sin((2m-1)\pi)+8 \sin\left(\frac{(2m-1) \pi }{3}\right)\right)^2\right].
\end{eqnarray}
From the Eqs.({\ref{Eq:ap11q1}) and (\ref{Eq:ap11q2}}), we found that linear entropy remains the same for consecutive odd and even 
values and its entanglement dynamics is periodic in nature having periodicity $6$. This is shown in the (Fig.(1)).
\subsubsection{\bf{Concurrence }}
The two qubit RDM, $\rho_{12}(2m)$, for  even  values  $n=2m$ on this state is given as follows:
\begin{equation}
\rho_{12}(2m)={\left(
\begin{array}{cccc}
h_n^\prime & 0 & 0 & w_n^\prime \\
 0 & v_n^\prime & v_n^\prime & 0 \\
 0 & v_n^\prime & v_n^\prime& 0 \\
 w_n^\prime & 0 & 0 & y_n^\prime \\
\end{array}
\right)},
\end{equation}
here  the coefficients are: 
\begin{eqnarray}\nonumber
 v_n^\prime=\frac{1}{24} \left[11+6 \cos\left(\frac{4m \pi }{3}\right)\right] \sin^2\left(\frac{2m \pi }{3}\right),~~ h_n^\prime=\frac{1}{16} \left[8+5 \cos \left(\frac{4m\pi }{3}\right)+3 \cos \left(\frac{2m \pi }{3}\right)\right], \\ \nonumber
w_n^\prime=\frac{1}{16 \sqrt{3}}\left[-5 \sin \left( \frac{4m \pi }{3}\right)+3 \sin \left(\frac{2m \pi }{3}\right)\right]~\mbox{and}~ y_n^\prime=\frac{1}{24} \left[11+6 \cos\left(\frac{4m \pi }{3}\right)\right] \sin^2\left(\frac{2m\pi }{3}\right).
\end{eqnarray}
Similarly  for odd values  $n=2m-1$  we get:
\begin{equation}
\rho_{12}(2m-1)={\left(
\begin{array}{cccc}
 l_n^\prime & j_n^\prime & j_n^\prime & 2p_n^\prime \\
j_n^\prime & 2k_n^\prime& 2k_n^\prime & j_n^\prime \\
 j_n^\prime& 2k_n^\prime & 2k_n^\prime & j_n^\prime \\
 2p_n^\prime & j_n^\prime & j_n^\prime & l_n^\prime \\
\end{array}
\right)},
\end{equation}
here the coefficients are 
 \begin{eqnarray} \nonumber
  j_n^\prime&=&\frac{1}{24} \sin\left[\frac{(2m-1)\pi }{3}\right] \left[-2 \sqrt{3} \cos\left(\frac{(2m-1) \pi }{3}\right)+3 \sqrt{3} \cos
((2m-1)\pi )+8 \sin\left(\frac{(2m-1)\pi }{3}\right)+3\sin((2m-1)\pi)\right], \\  \nonumber
l_n^\prime&=&\frac{1}{24} \left(16+5 \cos\left(\frac{2 (2m-1) \pi }{3}\right)+3 \cos\left(\frac{4 (2m-1) \pi }{3}\right)\right),
k_n^\prime=\frac{1}{24} \left[11+6 \cos\left(\frac{2(2m-1)  \pi }{3}\right)\right] \sin^2\left(\frac{(2m-1)  \pi }{3}\right) ~~\mbox{and}\\ \nonumber
p_n^\prime&=&\frac{1}{16 \sqrt{3}}\left[-5 \sin \left( \frac{2(2m-1) \pi }{3}\right)+3 \sin \left(\frac{(2m-1) \pi }{3}\right)\right].
\end{eqnarray}
All the coefficients are periodic in nature having periodicity three. The simplified  eigenvalues of 
$(\sigma_y \otimes \sigma_y)\rho_{12} (\sigma_y \otimes \sigma_y) \rho_{12}^*$ 
in decreasing order  for  $n = 0$ to $12$ are 
$\left\lbrace(0, 0, 0, 0),\left(1/4, 1/4, 0, 0\right),\left(1/4, 1/4, 0, 0\right),\left(1/4, 1/4, 0, 0\right), \left(1/4, 1/4, 0, 0\right), \right.$\\$\left.(0, 0, 0, 0),(0, 0, 0, 0),\left(1/4, 1/4, 0, 0\right),\left(1/4, 1/4, 0, 0\right), \left(1/4, 1/4, 0, 0\right),\left(1/4, 1/4, 0, 0\right), (0, 0, 0, 0),(0, 0, 0, 0)\right\rbrace$.
By using these values in the Eq.(\ref{Eq:Ap5Q3}), we find that its concurrence vanishes for all values of $n$.
\subsection{Initial state $\otimes^{11}\ket{+}$= $\ket{\theta_0 =\pi/2,\phi_0 =-\pi/2}$}
The initial  state can be expressed as, $\otimes ^{11} {\ket{+}}_y=\frac{1}{32}\left( \ket{\phi_0^+}+{ i \sqrt{11}}\ket{\phi_1^+}- {\sqrt{55}} \ket{\phi_2^+}-i\sqrt{165}\ket{\phi_3^+}+{\sqrt{330}} \ket{\phi_4^+}\right.$\\$\left.+{i\sqrt{462}} \ket{\phi_5^+}\right)$. The state $\ket{\psi_n}$ can be obtained by $n$th  time evolution of unitary operator $\mathcal{U}$ on the initial state. Thus,
\begin{eqnarray}
\ket{\psi_n}&=&\mathcal{U}_+^n \ket{+++++++++++}\\ \nonumber
            &=&e^{ \frac{i n \pi}{2} }({\delta}_n^\prime \ket{\phi_0^+}+{\zeta}_n^\prime \ket{\phi_1^+}+ {\xi}_n^\prime \ket{\phi_2^+}+{\eta}_n^\prime \ket{\phi_3^+}+{\chi}_n^\prime \ket{\phi_4^+}+{\omega}_n^\prime \ket{\phi_5^+}),
\end{eqnarray}
where the coefficients are as follows:
\begin{eqnarray}\nonumber
{\delta}_n^\prime &=&\frac{1}{32} \left[-10+11 \cos\left(\frac{2 n \pi }{3}\right)+11 \sqrt{3} \sin\left(\frac{2 n \pi }{3}\right)\right], ~{\zeta}_n^\prime=\frac{i\sqrt{11}}{96}  \left[3 \cos\left(\frac{2 n \pi }{3}\right)-\sqrt{3} \sin\left(\frac{2 n\pi }{3}\right)\right],\\ \nonumber
{\xi}_n^\prime&=&\frac{\sqrt{55}}{96}  \left[-3 \cos\left(\frac{2 n \pi }{3}\right)+\sqrt{3} \sin\left(\frac{2 n \pi }{3}\right)\right], ~{\eta}_n^\prime=-\frac{i \sqrt{55}}{96}  \left[2 \sqrt{3}+\sqrt{3} \cos\left(\frac{2 n\pi }{3}\right)+3 \sin\left(\frac{2 n\pi }{3}\right)\right],\\ \nonumber
{\chi}_n^\prime &=&\frac{1}{48} \sqrt{\frac{55}{2}} \left[2 \sqrt{3}+\sqrt{3} \cos\left(\frac{2 n \pi }{3}\right)+3 \sin \left(\frac{2
n \pi }{3}\right)\right]\mbox{and}~ {\omega}_n^\prime=\frac{i}{16}  \sqrt{\frac{77}{2}}  \left[\sqrt{3} \cos\left(\frac{2 n\pi }{3}\right)-\sin\left(\frac{2 n\pi }{3}\right)\right].
\end{eqnarray}

\subsubsection{\bf{The linear entropy}}
 The one-qubit  RDM $\rho_1(n)$ in this case is given as follows:
\begin{equation}
\rho_1(n)={\frac{1}{2}\left[
\begin{array}{cc}
 1 & -\frac{i}{12}\left(4+5 \cos\left(\frac{2 n \pi }{3}\right)+3 \cos\left(\frac{4 n \pi }{3}\right)\right) \\
 \frac{i}{12} \left(4+5 \cos\left(\frac{2 n \pi }{3}\right)+3 \cos\left(\frac{4 n \pi }{3}\right)\right) & 1 \\
\end{array}
\right]}.
\end{equation}
Thus, the expression for the linear entropy is given as follows:
\begin{equation}\label{eqap11}
S_{(\pi/2,-\pi/2)}^{(11)}(n)=\frac{1}{2}{\left[1-\frac{1}{144} \left(4+5 \cos\left(\frac{2 n\pi }{3}\right)+3 \cos\left(\frac{4 n \pi }{3}\right)\right)^2\right]}.
\end{equation}
From the Eq.(\ref{eqap11}), we can see that its linear entropy is periodic with period $3$. It attains the maximum upper bound value i.e. $1/2$.
\subsubsection{\bf{Concurrence}}
 The two-qubit RDM in this case is given by,
\begin{equation}
\rho_{12}(n)=\left(
\begin{array}{cccc}
 \bar{j}_n& \bar{l}_n^* & \bar{l}_n^* & \bar{k}_n \\
 \bar{l}_n& \bar{p}_n & \bar{p}_n& \bar{l}_n^*\\
 \bar{l}_n & \bar{p}_n & \bar{p}_n & \bar{l}_n^*\\
 \bar{k}_n & \bar{l}_n & \bar{l}_n& \bar{j}_n\\
\end{array}
\right),
\end{equation}
 where the coefficients are as follows:
 \begin{eqnarray}\nonumber
\bar{j}_n&=&\frac{1}{96} \left[32-5 \cos \left(\frac{2 n \pi }{3}\right)-3 \cos \left(\frac{4 n \pi }{3}\right)-5 \sqrt{3} \sin\left(\frac{2
n \pi }{3}\right)+3 \sqrt{3} \sin\left(\frac{4 n \pi }{3}\right)\right],\\ \nonumber
\bar{l}_n&=&\frac{i}{48}  \left[4+5 \cos \left(\frac{2 n \pi }{3}\right)+3 \cos\left(\frac{4 n\pi }{3}\right)\right],\\  \nonumber
\bar{k}_n&=&\frac{1}{96} \left[-15 \cos\left(\frac{2 n \pi }{3}\right)-9 \cos\left(\frac{4 n \pi }{3}\right)+\sqrt{3} \left(5 \sin\left(\frac{2
n \pi }{3}\right)-3 \sin\left(\frac{4 n \pi }{3}\right)\right)\right] ~~\mbox{and}\\ \nonumber
\bar{p}_n&=&\frac{1}{96} \left[16+5 \cos\left(\frac{2 n \pi }{3}\right)+3 \cos \left(\frac{4 n \pi }{3}\right)+5 \sqrt{3} \sin\left(\frac{2
n\pi }{3}\right)-3 \sqrt{3} \sin\left(\frac{4 n \pi }{3}\right)\right].
\end{eqnarray}
Thus, all these coefficients and the matrix itself are periodic of period $3$. 
The eigenvalues $\lambda_l$ of $(\sigma_y \otimes \sigma_y)\rho_{12} (\sigma_y \otimes \sigma_y) \rho_{12}^*$ 
in decreasing order  for  
$n$ = $0$ to $6$ are $\left\lbrace(0, 0, 0, 0), \left(1/4, 1/4, 0, 0\right),\left(1/4, 1/4, 0, 0\right),(0, 0, 0, 0),
\left(1/4, 1/4, 0, 0\right), \left(1/4, 1/4, 0, 0\right), (0, 0, 0, 0)\right\rbrace$. By using these values in Eq.(\ref{Eq:Ap5Q3}), 
we found that its concurrence vanishes for all $n$. This is shown in the (Fig.(1)).

% % % % % % % % % % % % % % % % % % % % % % % % % % % % % % % % % % % % % % % % % 
\section{Proof of the Commutation Relation }
\label{Sec:ProofCommutation}
We now prove the following relation from the Ref.\cite{ExactDogra2019}:
\begin{equation}
\Bigr[\mathcal{U},\otimes_{l=1}^{N} \sigma^y_l\Bigr]=0,
\end{equation} 
\begin{equation}\label{uni1}
\mbox{where}~{\mathcal U}=\exp\left(-i \frac{\pi}{4}  \sum_{ l< l'=1}^{N} \sigma^z_{l} \sigma^z_{l'}\right)
\exp\left( -i \frac{\pi}{4} \sum_{l=1}^{N}\sigma^y_l \right).
\end{equation}

To simplify, let us write $\mathcal{U}$ as: $\mathcal{U}$=$\mathcal{U_A} \mathcal{U_B}$, where 
$\mathcal{U_A}$=$\exp\left( \frac{-i\pi}{4}  \sum_{ l< l'=1}^{N} \sigma^z_{l} \sigma^z_{l'}\right)$ and $\mathcal{U_B}$=$\exp\left(  \frac{-i\pi}{4} \sum_{l=1}^{N}\sigma^y_l \right)$.
Thus, the commutation relation simplyfies to:
\begin{equation} \label{eq5}
\Bigr[\mathcal{U_A}~ \mathcal{U_B},\otimes_{l=1}^{N} \sigma^y_l\Bigr]=\mathcal{U_A}\Bigr[ \mathcal{U_B},\otimes_{l=1}^{N} \sigma^y_l\Bigr]+\Bigr[\mathcal{U_A} ,\otimes_{l=1}^{N} \sigma^y_l\Bigr]\mathcal{U_B}.
\end{equation}
It can be seen easily  that $\sigma_y$ commutes with $\otimes_{l=1}^{N} \sigma^y_l$.  So the commutator $\Bigr[ \mathcal{U_B},\otimes_{l=1}^{N} \sigma^y_l\Bigr]=0$.
Now  solving for the second part of commutator in the Eq.(\ref{eq5}) by  simplifying the commutation relation as follows:
\begin{eqnarray} \label{eq6}\nonumber
\Bigr[\frac{-i\pi}{4}\sum_{l<{l^\prime}=1}^{2j}\sigma^z_{l} \sigma^z_{l'},\otimes_{l=1}^{N} \sigma^y_l\Bigr]&= & \frac{-i\pi}{4}\Bigr[\sigma^z_{1}\sigma^z_{2}+ \sigma^z_{1}\sigma^z_{3}+\sigma^z_{2}\sigma^z_{3}+..........+\sigma^z_{2j-1}\sigma^z_{2j},\otimes_{l=1}^{N} \sigma^y_l\Bigr]\\ 
&=& \frac{-i\pi}{4}\left(\Bigr[\sigma^z_{1}\sigma^z_{2},\otimes_{l=1}^{N} \sigma^y_l\Bigr]+\Bigr[\sigma^z_{1}\sigma^z_{3},\otimes_{l=1}^{N} \sigma^y_l\Bigr]+..........\Bigr[\sigma^z_{N-1}\sigma^z_{N},\otimes_{l=1}^{N} \sigma^y_l\Bigr]\right).
\end{eqnarray}
Let us solve the following:
\begin{eqnarray} \label{eq7}\nonumber
\Bigr[\sigma^z_{1}\sigma^z_{2},\otimes_{l=1}^{N} \sigma^y_l\Bigr]&=&\Bigr[\sigma^z_{1}\sigma^z_{2}, \sigma^y_1 \sigma^y_2 \otimes_{l=3}^{N}\sigma^y_{l}\Bigr] \\ \nonumber 
&=& \Bigr[\sigma^z_{1}\sigma^z_{2},\sigma^y_{1}\sigma^y_2\Bigr]\otimes_{l=3}^{N} \sigma^y_l  + \sigma^y_1\sigma^y_{2}\Bigr[\sigma^z_{1}\sigma^z_{2},\otimes_{l=3}^{N} \sigma^y_{l}\Bigr] \\ \nonumber 
&=& \sigma^z_{1}\left(\Bigr[\sigma^z_{2},\sigma^y_{1}\Bigr]\sigma_2^y + \sigma^y_{1} \Bigr[\sigma^z_{2},\sigma^y_{2}\Bigr]\right) \otimes_{l=3}^{N} \sigma^y_l +\left(\Bigr[\sigma^z_{1},\sigma^y_{1}\Bigr]\sigma^y_{2}+\sigma^y_{1}\Bigr[\sigma^z_{1},\sigma^y_{2}\Bigr]\right)\sigma^z_{2}\otimes_{l=3}^{N} \sigma^y_l\\  && + ~\sigma^y_{1}\sigma^y_2 \left(\sigma^z_{1}\Bigr[\sigma^z_{2}, \otimes_{l=3}^{N}\sigma^y_{l}\Bigr]+\Bigr[\sigma^z_{1}, \otimes_{l=3}^{N}\sigma^y_{l}\Bigr]\sigma^z_{2}\right).
\end{eqnarray}
The commutation of Pauli operator  with different  indices is zero i.e.  
$[\sigma_{1}^z,\otimes_{l=3}^{N}\sigma_{l}^y]=$ $[\sigma_{2}^z,\otimes_{l=3}^{N}\sigma_{l}^y]=$ $[\sigma_{1}^z,\sigma_{2}^y]=0$ 
But with the same indices  they don't commute, then  by  using all the commutation  relations in the Eq.(\ref{eq7}) we get:
\begin{eqnarray}
\Bigr[\sigma^z_{1}\sigma^z_{2},\otimes_{l=1}^{N} \sigma^y_l\Bigr]&=&\left(-2i\sigma^x_{1}\times -i\sigma^x_{2}  + (-2i\sigma^x_{1})\times i\sigma^x_{2} \right)\otimes_{l=3}^{N} \sigma^y_l
\\ \nonumber &=&\left(-2~\sigma^x_{1} \sigma^x_{2} +2~\sigma^x_{1} \sigma^x_{2}\right)\otimes_{l=3}^{N} \sigma^y_l\\
 &=& 0.
\end{eqnarray}
Similarly, other commutators in the Eq.(\ref{eq6}) commutes with $\otimes_{l=1}^{N} \sigma^y_l$. So the commutator 
$\Bigr[\frac{-i\pi}{4}\sum_{l<{l^\prime}=1}^{N}\sigma^z_{l} \sigma^z_{l'},\otimes_{l=1}^{N} \sigma^y_l\Bigr]=0$. We know that 
if the commutation of  $[A,B]=0$, then the commutation relation  of $\Bigr[e^{A},B\Bigr]$ is also zero. Thus, we can 
say that $\Bigr[e^{\frac{-i\pi}{4}\sum_{l<{l^\prime}=1}^{N}\sigma^z_{l} \sigma^z_{l'}},\otimes_{l=1}^{N} \sigma^y_l\Bigr]=0$. By 
putting all the values in the Eq.(\ref{eq5}) we can conclude that:

\begin{equation}
\Bigr[\mathcal{U},\otimes_{l=1}^{N} \sigma^y_l\Bigr]=0.
\end{equation}

\end{document}